\documentclass[
reprint,
superscriptaddress,
 amsmath,amssymb,
 aps,
prapplied,
floatfix,
longbibliography
]{revtex4-2}

\newcommand{\topic}[1]{}
\newcommand{\hide}[1]{}

\newcommand{\panel}[1]{(#1)}

\usepackage{amsmath,amssymb}
\usepackage{graphicx}
\usepackage{dcolumn}
\usepackage{bm}
\usepackage[colorlinks]{hyperref}
\hypersetup{
    plainpages=true,
    breaklinks=true,
    hypertexnames=false,
    pageanchor=true,
    colorlinks=true,
    linkcolor={blue},
    citecolor={blue},
    urlcolor={blue},
    anchorcolor={black}
}
\usepackage{multirow}
\usepackage{hhline}
\usepackage{braket, siunitx}
\usepackage{bbold}
\DeclareMathOperator\arccosh{arccosh}
\usepackage{makecell}
\usepackage[many]{tcolorbox} 
\newtcolorbox{boxA}{
    boxrule = 1.pt,
    colframe = black
}

\begin{document}

\preprint{APS/123-QED}

\title{Pulse Design of Baseband Flux Control for \linebreak
Adiabatic Controlled-Phase Gates in Superconducting Circuits}

\def\EECSaffil{Department of Electrical Engineering and Computer Science, Massachusetts Institute of Technology, Cambridge, MA 02139, USA}
\def\RLEaffil{Research Laboratory of Electronics, Massachusetts Institute of Technology, Cambridge, MA 02139, USA}
\def\LLaffil{MIT Lincoln Laboratory, Lexington, MA 02421, USA}
\def\Physaffil{Department of Physics, Massachusetts Institute of Technology, Cambridge, MA 02139, USA}
\def\MERLaffil{Mitsubishi Electric Research Laboratories, Cambridge, MA 02139, USA}
\def\affilAQ{Atlantic Quantum, Cambridge, MA 02139}

\author{Qi~Ding}
\email{qding@mit.edu}
\affiliation{\EECSaffil}
\affiliation{\RLEaffil}

\author{Alan~V.~Oppenheim}
\affiliation{\EECSaffil} 
\affiliation{\RLEaffil}

\author{Petros~T.~Boufounos}
\affiliation{\MERLaffil} 

\author{Simon~Gustavsson} 
\altaffiliation[Present address: ]{\affilAQ}
\affiliation{\RLEaffil} 

\author{Jeffrey~A.~Grover}
\affiliation{\RLEaffil} 

\author{Thomas~A.~Baran}
\affiliation{\RLEaffil} 

\author{William~D.~Oliver}
\email{william.oliver@mit.edu}
\affiliation{\EECSaffil} 
\affiliation{\RLEaffil} 
\affiliation{\Physaffil}

\date{\today}

\begin{abstract}
Despite progress towards achieving low error rates with superconducting qubits, error-prone two-qubit gates remain a bottleneck for realizing large-scale quantum computers. Therefore, a systematic framework to design high-fidelity gates becomes imperative. One type of two-qubit gate in superconducting qubits is the controlled-phase (CPHASE) gate, which utilizes a conditional interaction between higher energy levels of the qubits controlled by a baseband flux pulse on one of the qubits or a tunable coupler. In this work, we study an adiabatic implementation of CPHASE gates and formulate the design of the control trajectory for the gate as a pulse-design problem. We show in simulation that the Chebyshev-based trajectory can, in certain cases, enable gates with gate infidelity lower by an average of $23.3\%$ when compared to the widely used Slepian-based trajectory. 
\end{abstract}

\maketitle

\section{Introduction}

High-fidelity entangling gates are one of the fundamental requirements in the pursuit of large-scale fault-tolerant quantum computing~\cite{Zajac2018,Wright2019BenchmarkingA1,He2019ATG,PhysRevLett.125.240503,krantz_quantum_2019}. Over the past decades, superconducting qubits have emerged as a leading platform for quantum computing, with several advances in terms of gate fidelity, extensibility, and use-case demonstrations~\cite{arute_quantum_2019,doi:10.1146/annurev-conmatphys-031119-050605,Youngseok_2023}. These improvements have enabled superconducting quantum computing platforms to begin tackling significant challenges, including the implementation of prototype quantum error correction (QEC) protocols~\cite{Riste_2014, Maika2016, Christian_2019, Rajeev_2023}. Numerous variants of superconducting qubits and their architecture have been proposed and experimentally demonstrated~\cite{Nakamura1999, Ioffe1999, Mooij1999, Koch2007, Manucharyan2009,Yan2016, Gyenis2021}, along with a variety of different schemes for realizing entangling gates~\cite{Majer2007,Rigetti2010,Poletto2012,Chow_2013,martinis_fast_2014,kelly_prl_2014, fei_tunable_2018, sung_realization_2021, Zhang2021, Ding2023}. Despite these significant advances, two-qubit gate performance continues to limit the development of future fault-tolerant quantum computing systems.

Two-qubit entangling gates employed in superconducting qubits can be broadly classified into two categories. The first category encompasses capacitively coupled, fixed-frequency qubits, in some cases mediated by a resonator, where the implementation of two-qubit gates relies on all-microwave control~\cite{Maika2016,Chow2012,Chow_2013,Gambetta2015,Paik2016, Krinner2020,Mitchell2021,Kandala2021}. Fixed-frequency qubits typically have longer coherence times and no baseband control lines. However, frequency crowding and potential collisions become increasingly challenging as the system size grows. The second category involves frequency-tunable qubits, where the frequency of the qubits can be adjusted using baseband magnetic flux~\cite{Neeley2010,Chen2014,Barends2019,Caldwell2018,sung_realization_2021}. These qubits can be coupled directly or through frequency-tunable coupling elements. In this architecture, two-qubit gates are typically achieved by applying local baseband magnetic-flux pulses to tune the frequencies of the qubits and/or couplers. It is worth noting that this approach results in increased hardware complexity and susceptibility to flux-related noise, thereby exacerbating experimental calibration challenges. Nevertheless, it mitigates the frequency collision issues, and gates relying on baseband flux control generally exhibit faster operation speed compared to all-microwave-activated gates. Additionally, there exist alternative architectural designs and gate schemes that seek to amalgamate features from both of these categories~\cite{dipaolo2022extensible}. In this work, we focus on baseband flux control gates with tunable qubits. More specifically, we study controlled-phase (CPHASE) gates and in particular the controlled-Z (CZ) gate, which corresponds to a conditional phase accumulation of $\pi$.

The fidelity of CPHASE gates depends heavily on the specific pulse shape of the baseband flux, as deviations can cause phase errors and leakage to undesired states. In this work, we first formulate the problem of baseband flux control design as a pulse design problem. In this way, we are able to design the gate by leveraging tools from the signal processing community. Second, we propose a Chebyshev-based trajectory as an alternative to the widely used Slepian-based trajectory. We analytically study the Chebyshev-based trajectory using a two-level system abstraction. Finally, we compare the performance of the Chebyshev-based and Slepian-based trajectories by simulating a CZ gate applied to two capacitively coupled transmon qubits. Simulation results show that the Chebyshev-based trajectory can be designed to induce lower leakage error while maintaining smaller pulse duration, in certain cases. In addition, we show that the proposed Chebyshev-based trajectory can be readily implemented in state-of-the-art hardware by considering practical hardware constraints in simulation.

The manuscript is organized as follows. In Section~\ref{sec: Problem formulation}, we formulate the pulse-design problem using a two-level system abstraction and state explicitly the criterion to be investigated. In Section~\ref{sec: Finite-length discrete-time pulses}, we introduce the definition and examples of finite-length, discrete-time pulses that will be exploited. Then, in Section~\ref{sec: Optimal solutions}, we propose the Chebyshev-based trajectory as an alternative solution compared to the Slepian counterpart. In Section~\ref{sec: Simulation results}, we present time-domain simulation results and demonstrate the advantage of the Chebyshev-based trajectory when implementing a CZ gate in two directly coupled transmon qubits. We also study the effect of realistic hardware limitations on these trajectories. We conclude and discuss outlook in Section~\ref{sec: Conclusion and outlook}.

\section{Problem formulation} \label{sec: Problem formulation}
\subsection{The CPHASE gate in tunable transmon qubits} \label{sec:physics_background}

The general approach to implementing a CPHASE gate in tunable transmon qubits using baseband flux control is summarized in Appendix~\ref{app: CPHASE gate}.

Two important factors in this design are the leakage error and gate duration. Leakage error refers to unwanted qubit population of the total qubit population outside of the computational subspace after the gate operation. In this case, the dominant leakage is from $\ket{11}$ to $\ket{20}$, since they are intentionally brought into resonance. On resonance, these states will hybridize and open an avoided crossing. Therefore, a trajectory towards the avoided crossing must be sufficiently slow in order for the leakage error to be sufficiently small in the adiabatic implementation of the CPHASE gate. On the other hand, for coherence-limited qubits like superconducting qubits, faster trajectories directly translate to higher fidelity. In other words, the process should be ``fast and adiabatic.'' Furthermore, as these two factors are intrinsically contradictory, a design of the trajectory should be made to achieve best performance. In this work, as we will explain in more detail in Section~\ref{sec:problem_statement}, we refer to this problem as the pulse design or control trajectory design problem.

Considerable efforts have been made to the development and experimental validation of a high-fidelity baseband flux-controlled CZ gate. The Slepian-based trajectory is the current standard to implement an adiabatic CZ gate~\cite{martinis_fast_2014}. The control trajectory is based on the Slepian window from the use of optimal window functions. This approach is experimentally demonstrated to reach a CZ gate fidelity up to $99.4\%$~\cite{barends_superconducting_2014}. Rol et al.~\cite{rol_fast_2019} appended two Slepian-based trajectories together to form a bipolar flux pulse named the Net Zero (NZ) pulse, which is more robust to long time distortions in the control line compared to unipolar ones. The latter is experimentally demonstrated to reach a CZ gate fidelity up to $99.7\%$~\cite{PhysRevX.12.011005}. Building upon the NZ pulse, Negîrneac et al.~\cite{negirneac_high-fidelity_2021} develop a variation named the sudden net-zero (SNZ) CZ gates, which simplifies the pulse calibration. The Slepian-based trajectory is also utilized to implement a non-adiabatic CZ gate with fidelity $99.76\pm0.07\%$ in a more sophisticated system consisting of two transmon qubits coupled with a tunable coupler~\cite{sung_realization_2021}.  A flat-top Gaussian pulse has been employed to implement CPHASE gates in a similar qubit-coupler-qubit architecture~\cite{Collodo2020}. Chu et al.~\cite{chu_coupler-assisted_2021} also study the CZ gate in a system with a tunable coupler and propose a modified control trajectory by adding prefactor weights to the Slepian-based trajectory. Another general approach~\cite{kelly_prl_2014} is to perform repeated experiments using closed-loop feedback to evaluate the current gate performance according to some metrics, and then numerically optimize the pulse, starting from a heuristically decent Slepian-based trajectory~\cite{martinis_fast_2014}. 

\subsection{Two-level system abstraction} \label{sec:two_level}

The primary error channel is leakage from $\ket{11}$ to $\ket{20}$ during the gate. In this section, we therefore focus on a simpler two-level abstraction of the problem that couples the diabatic states $\ket{11}$ and $\ket{20}$ to form eigenstates $\ket{\psi_{-}}$ (``ground state'') and $\ket{\psi_{+}}$ (``excited state'').

\begin{figure}[tb]
\begin{center}
\includegraphics[width=0.5\textwidth]{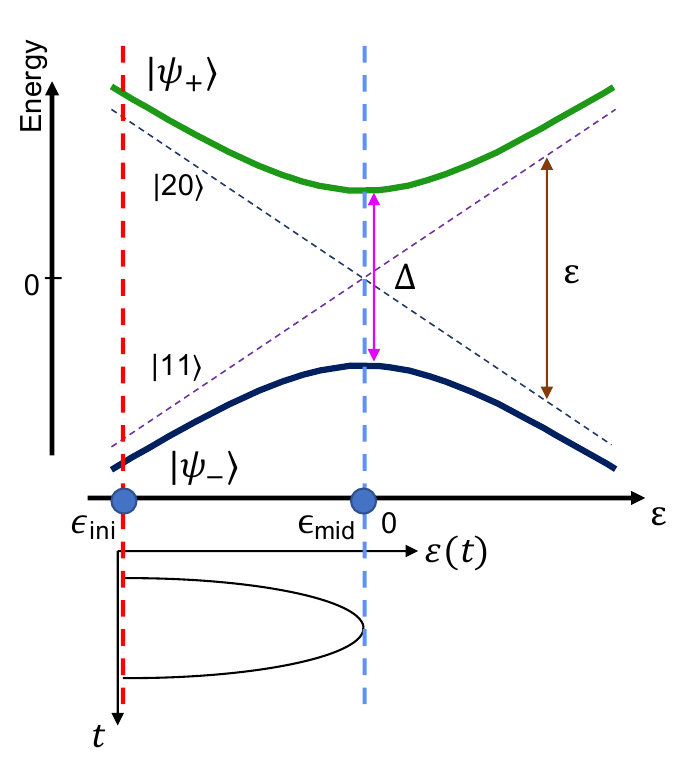}
\caption{Eigenenergies of the two-level system with Hamiltonian $H$ as a function of $\varepsilon$. In the upper plot, the thinner dashed lines represent the diabatic states $\ket{11}$ and $\ket{20}$; the solid lines represent the eigenstates $\ket{\psi_{-}}$ and $\ket{\psi_{+}}$ that result from the coupling strength $\Delta$. The lower plot shows the schematic of an example trajectory for $\varepsilon(t)$ to implement a CPHASE gate. The thicker dashed lines across the upper and lower plots indicate the initial and middle points of $\varepsilon(t)$.}
\label{fig:p2}
\end{center}
\end{figure}

Consider a two-level system whose Hamiltonian is
    \begin{equation} \label{hamiltonian}
        H = \frac{\varepsilon(t)}{2} \sigma_z + \frac{\Delta}{2} \sigma_x = \frac{1}{2}\begin{bmatrix} \varepsilon(t) & \Delta \\ \Delta & -\varepsilon(t) \end{bmatrix},
    \end{equation}
where $\Delta$ is a constant denoting the coupling strength that hybridizes the diabatic states $\ket{11}$ and $\ket{20}$, and $\varepsilon(t)$ is a function of time, which dictates the energy difference between the diabatic states. Of particular interest are the two eigenstates $\ket{\psi_{-}}$, $\ket{\psi_{+}}$ and the corresponding eigenenergies $E_{-}$, $E_{+}$ of this system. Solving the eigen-problem for $H$ yields
    \begin{subequations} \label{eq:eigen_states}
    \begin{equation} \label{eigenvector_excited}
        \ket{\psi_{-}} = \begin{bmatrix} -\sin(\theta(t)/2) \\ \cos(\theta(t)/2) \end{bmatrix} \ \text{and}\  \ket{\psi_{+}} = \begin{bmatrix} \cos(\theta(t)/2) \\ \sin(\theta(t)/2) \end{bmatrix},
    \end{equation}
    \begin{equation} \label{eigenenergy_ground}
        E_{-} = -\frac{1}{2}\sqrt{\varepsilon(t)^2+\Delta^2} \ \text{and}\  E_{+} = \frac{1}{2}\sqrt{\varepsilon(t)^2+\Delta^2} \ ,
    \end{equation}
    \end{subequations}
where $\theta(t)$ is defined as
\begin{equation} \label{eq:theta_def}
    \theta(t) = \arctan \frac{\Delta}{\varepsilon(t)} \ .
\end{equation}
In Appendix~\ref{app: Derived formula for leakage error and its approximation}, we review the geometric interpretation of $\theta(t)$ on the Bloch sphere. In Section~\ref{sec:problem_statement}, $\theta(t)$ is considered an intermediate control variable whose trajectory is to be designed. Fig.~\ref{fig:p2} depicts the eigenenergies of the two-level system with Hamiltonian $H$ as a function of $\varepsilon \in [-\infty,+\infty]$.

In this abstracted two-level system, the CPHASE gate problem is transformed into preparing the system in the initial state $\ket{\psi_{-}}$ and designing $\varepsilon(t)$ to vary the instantaneous energy as depicted in the lower plot of Fig.~\ref{fig:p2}. In particular, $\varepsilon(t)$ varies from $\epsilon_{\textrm{ini}}$ to $\epsilon_{\textrm{mid}}$ and returns to $\epsilon_{\textrm{ini}}$ where $\epsilon_{\textrm{mid}} \approx 0$. Correspondingly, $\theta(t)$ varies from $\theta_{\textrm{ini}}=\arctan(\Delta/\epsilon_{\textrm{ini}})$ to $\theta_{\textrm{mid}}=\arctan(\Delta/\epsilon_{\textrm{mid}})$ and returns to $\theta_{\textrm{ini}}$.

\subsection{Formula for leakage error} \label{sec: Formula for leakage error}

The formula for the leakage error $P_e$ from $\ket{11}$ to $\ket{20}$ after the pulse is implemented is (see Appendix~\ref{app: Derived formula for leakage error and its approximation} for details)
\begin{equation} \label{eq:pe_approx}
    P_e = \frac{\displaystyle \bigg|\int \frac{\textrm{d}\theta}{\textrm{d}t}e^{-i\int^t \omega(t')\textrm{d}t'} \textrm{d}t \bigg|^2}{4} \ ,
\end{equation}
where $\omega(t')$ is the time-dependent frequency difference between eigenstates $\ket{\psi_{-}}$ and $\ket{\psi_{+}}$ due to the trajectory. A nonlinear time-frame transformation is introduced (see Appendix~\ref{app: Nonlinear time frame transformation}), i.e., $\omega_\tau \textrm{d}\tau = \omega(t)\textrm{d}t$, which effectively accounts for the time-dependence of $\omega(t')$, making it a time-independent frequency $\omega_\tau$. We express the leakage error $P_e$ in this new time frame $\tau$
\begin{equation} \label{eq:leakage_nonlinear_time_frame}
    P_e = \frac{\displaystyle \bigg|\int \frac{\textrm{d}\tilde\theta}{\textrm{d}\tau}e^{-i\omega_\tau \tau} \textrm{d}\tau \bigg|^2}{4} \ ,
\end{equation}
where $\omega_\tau$ is a constant, $\tau=\tau(t)$ is a nonlinear function of $t$, and $\tilde\theta(\tau)$ is the transformation of $\theta(t)$ in the new time frame $\tau$. In this time frame, we have obtained a simpler form of the leakage error $P_e$, which can be interpreted as a function of the Fourier transform of $\textrm{d}\tilde\theta/\textrm{d}\tau$ evaluated at $\omega_\tau$.
After designing $\textrm{d}\tilde\theta /\textrm{d}\tau$, we transform back to the original time frame $t$ and obtain the corresponding $\textrm{d}\theta /\textrm{d}t$ using the technique described in Appendix~\ref{app: Nonlinear time frame transformation}. We discuss the validity of Eq.~\ref{eq:leakage_nonlinear_time_frame} along with the nonlinear time-frame transformation in Appendix~\ref{app: validity}. $\omega_\tau$ can take any constant value, however, as explained in Appendix~\ref{app: Nonlinear time frame transformation}, we set $\omega_\tau = \Delta$ for convenience.

\subsection{Problem statement} \label{sec:problem_statement}

We propose a problem statement with explicit requirement: we desire the shortest pulse given a specified allowable leakage error. Then, taking advantage of the time and frequency scaling property of the Fourier transform, we transform the problem into a requirement on frequency. The novelty in our problem formulation enables a straightforward comparison between different pulses. Finally, we comment on the continuous time to discrete time transformation. 

\subsubsection{Nomenclature}

In this section, we more formally define the pulse design or control-trajectory design that is the focus of this work. We consider the case where two qubits are capacitively coupled, one of which is flux-tunable (QB1). Our ultimate goal is to design a baseband flux pulse that changes the external magnetic flux $\Phi_{\text{ext}}(t)$ that threads the SQUID loop of QB1 to change its qubit frequency $\omega_1(t)$ so that a CPHASE gate with desired characteristics is obtained. In the abstracted two-level system discussed in Section~\ref{sec:two_level}, the goal is the design of $\varepsilon(t)$. Then, an intermediate control variable $\theta(t)$ is defined such that it has a one-to-one correspondence to $\varepsilon(t)$. Therefore, the goal of designing $\varepsilon(t)$ is equivalent to designing $\theta(t)$. Finally, in Eq.~\ref{eq:leakage_nonlinear_time_frame}, the formula for the leakage error is written in terms of the Fourier transform of $\textrm{d}\tilde\theta/\textrm{d}\tau$ in the nonlinear time frame $\tau$. Designing a flux pulse further transforms into finding a trajectory $\textrm{d}\tilde\theta/\textrm{d}\tau$, which we refer to as ``control trajectory design.'' In the following sections, we denote $\tilde g(\tau) := \textrm{d}\tilde\theta/\textrm{d}\tau$ for brevity. The discrete form of $\tilde g(\tau)$ is denoted as $\tilde g[n]$.

The design pipeline is presented below. We first design $\tilde g(\tau)$ and obtain $\tilde \theta(\tau)$. Second, we compute $\theta(t)$ through the inverse of the nonlinear time-frame transformation. Third, we obtain $\varepsilon(t)$ according to Eq.~\ref{eq:theta_def} with proper values of $\epsilon_\textrm{ini} (\theta_\textrm{ini})$ and $\epsilon_\textrm{mid} (\theta_\textrm{mid})$. Then, we convert $\varepsilon(t)$ to $\omega_{1}(t)$ and finally to $\Phi_{\text{ext}}(t)$ according to Eq.~\ref{eq:flux_frequency}:
\begin{equation} \label{eq:design_pipeline}
     \tilde g(\tau) \rightarrow \tilde \theta(\tau) \rightarrow \theta(t) \rightarrow \varepsilon(t) \rightarrow \omega_{1}(t) \rightarrow \Phi_{\text{ext}}(t) \ .
\end{equation}

\subsubsection{Constraint on duration} \label{opt_over_length}

Denoting $\tilde g(\tau) = \textrm{d}\tilde\theta/\textrm{d}\tau$, we can rewrite Eq.~\ref{eq:leakage_nonlinear_time_frame} as
\begin{equation} \label{eq: the_leakage_equation}
    P_e = \frac{\displaystyle \bigg|\int\tilde g(\tau)e^{-i\Delta \tau} \textrm{d}\tau \bigg|^2}{4}
    = \frac{\displaystyle \big|G(i\Delta) \big|^2}{4} \ ,
\end{equation}
where $G(i\Delta)$ is the Fourier transform of $\tilde g(t)$ evaluated at $\omega_\tau = \Delta$.

With the goal of implementing a high-fidelity CPHASE gate, there are three key quantities: phase accumulation, leakage error, and gate duration. Phase accumulation is directly related to $\theta(t)$ and thus $\tilde \theta(\tau)$, as discussed in Appendix~\ref{sec:accumulated phase}. Given a particular shape of $\tilde \theta(\tau)$, we can always obtain a desired phase accumulation by fine-tuning the amplitude. For now we assume a normalized amplitude as will be explained later in this section for the purpose of analysis, and we will generalize in simulation. We thus focus on the two remaining interrelated factors: leakage error and gate duration. We are interested in designing a $\tilde g(\tau)$ with as short a duration as possible, given some acceptable leakage error threshold. Here the parameters to be designed are the pulse shape and duration of $\tilde g(\tau)$. In addition, in cases where $\tilde g(\tau)$ has the same shape but a longer duration, we desire that the leakage error remain below the threshold. This makes intuitive sense, because a longer $\tilde g(\tau)$ corresponds to a slower evolution and therefore should induce no more (and often less) leakage error. 

Let $\tau_d$ be the duration of $\tilde g(\tau)$. We consider time-symmetric trajectories $\tilde \theta(\tau)=\tilde\theta(\tau_d-\tau)$ such that $\tilde \theta(\tau)$ starts from some initial value $\tilde\theta_\textrm{ini}$, evolves to some intermediate value $\tilde\theta_\textrm{mid}$, and then returns to the initial value $\tilde\theta_\textrm{ini}$. As the derivative of $\tilde\theta(\tau)$, $\tilde g(\tau) = \textrm{d}\tilde \theta/\textrm{d}\tau$ is anti-symmetric in time $\tau$, i.e., $\tilde g(\tau) = -\tilde g(\tau_d-\tau)$. Since $\tilde\theta(\tau)$ is symmetric, we have $\tilde\theta(\tau_d/2)=\tilde\theta_\textrm{mid}$, and $\int_0^{\tau_d/2}\tilde g(\tau)\textrm{d}\tau = \tilde\theta_\textrm{mid}-\tilde\theta_\textrm{ini} =-\int_{\tau_d/2}^{\tau_d}\tilde g(\tau)\textrm{d}\tau$. We further impose a normalization constraint on the control trajectory so that $\int_0^{\tau_d/2}\tilde g(\tau)\textrm{d}\tau = -\int_{\tau_d/2}^{\tau_d}\tilde g(\tau)\textrm{d}\tau =1$.

Now, $\tilde g(\tau)$ is time-limited to the interval $[0,\tau_d]$, i.e., $\tilde g(\tau)=0 \text{ when} \  \tau<0 \ \text{or} \  \tau>\tau_d$. We consider the leakage error in Eq.~\ref{eq: the_leakage_equation}, paying specific attention to $|G_{\tau_d}(i\Delta)|$, which is the magnitude of the Fourier transform of $\tilde g(\tau)$ of duration $\tau_d$ evaluated at $\Delta$. The problem can be stated as follows:
\begin{boxA}
{\bf Statement 1:} \\
Given an error threshold $P_e \leq \gamma^2/4$, i.e., $|G(i\Delta)| \leq \gamma$, find the $\tilde g(\tau)$ of duration $\tau_d^*$ with $\tau_d^* = \min(\tau_{dc})$, where $\tau_{dc}$ is defined such that for any $\tau_d \geq \tau_{dc}$, $|G_{\tau_d}(i\Delta)| \leq \gamma$ is satisfied. 
\end{boxA}

\subsubsection{\texorpdfstring{Time-frequency transformation: constraint on frequency}{Time-frequency transformation: constraint on frequency}} \label{opt_over_freq}

We transform the problem statement into one that is more readily addressable, utilizing the time and frequency scaling property of the Fourier transform. We first introduce a set $\mathcal{A}$ with uncountably many elements corresponding to an infinite number of control trajectory shapes and use $\tilde g^{a}(\tau)$ with $a \in \mathcal{A}$ to denote a control trajectory confined in time to the interval $[0,1]$, i.e., $\tilde g^{a}(\tau)=0, \text{when} \  \tau<0 \ \text{or} \  \tau>1$. We further impose an additional symmetry requirement on the pulse, such that we have $\int_0^{1/2} \tilde g^{a}(\tau) \textrm{d}\tau = - \int_{1/2}^1 \tilde g^{a}(\tau) \textrm{d}\tau = 1$. We refer to $\tilde g^{a}(\tau)$ as a normalized pulse shape labelled by $a$. 

For any $\tilde g(\tau)$ of duration $\tau_d$, there must exist some $a \in \mathcal{A}$ such that $\tilde g(\tau) = \tilde g^{a}(\tau/\tau_d)/\tau_d$. With the time and frequency scaling property of the Fourier transform, we have $G_{\tau_d}(i\Delta) = G^a(i\Delta \tau_d)$, where $G^a(i\Delta \tau_d)$ is the Fourier transform of $\tilde g^{a}(\tau)$ evaluated at $\Delta \tau_d$. We denote $\omega = \Delta \tau_d$. Note that in the expression $G^a(i\Delta \tau_d)$, $\Delta$ and $\tau_d$ are nominally on an equivalent footing. Therefore, it would be equivalent to construct the problem with $\tau_d$ given and $\Delta$ varied and to be minimized, instead of fixing $\Delta$ and minimizing over $\tau_d$. For convenience, we set $\tau_d=1$ without loss of generality. In this way, we reformulate Statement 1 into: 
\begin{boxA}
{\bf Statement 2:} \\
Given an error threshold $P_e \leq \gamma^2/4$, i.e., $|G^a(i\omega)| \leq \gamma$, find a trajectory shape $\tilde g^{a}(\tau)$ such that $\omega^*$ is minimized, where $\omega^*$ is defined as the minimum frequency such that for any $\omega \geq \omega^*$, $|G^a(i\omega)| \leq \gamma$ is satisfied.
\end{boxA}

\subsubsection{Continuous time to discrete time transformation} \label{ctdt}

The problem formulation so far has been stated in continuous time. However, the pulses are represented in discrete time when we perform simulation. Also, in experiments, a discrete-time pulse needs to be specified for the digital controller of the pulse-generation hardware (e.g., arbitrary waveform generator), followed by a certain interpolation scheme in order to output a continuous-time pulse. Therefore, we consider the design in discrete time. Let $F_s= 1/T_s$ be the sampling frequency and $T_s$ be the sampling period. Then we have $\tilde g[n] = \tilde g(nT_s)$ for $n=0,1,\dots,N-1$ where $N-1 =\lfloor \tau_d/T_s \rfloor$ with $\lfloor x \rfloor$ denoting the greatest integer less than or equal to $x$. We refer to $N$ as the length of $\tilde g[n]$. In this work, $\tilde g(\tau)$ is time-limited by definition. Fortunately, as we show in the Section~\ref{sec: Optimal solutions}, the frequency spectrum of $\tilde g(\tau)$ of interest tends to zero relatively quickly as frequency increases. Therefore, with high enough sampling frequency, the problem of aliasing can be maintained at a minimal level. 

We recast the problem formulation in discrete time in a complete form as follows:
\begin{boxA}
{\bf Statement 3:} \\
    Determine an anti-symmetric control trajectory $\tilde g[n]$ of length $N$, where $\tilde g[n]$ is normalized:
        \begin{enumerate}
            \item $\tilde g[n]=-\tilde g[N-n]$, specifically, $\tilde g[(N-1)/2]=0$ for odd $N$,
            \item $\tilde g[n]=0$ when $n<0$ or $n>N-1$,
            \item $\sum_0^{N/2-1}\tilde g[n]=1=-\sum_{N/2}^{N-1}\tilde g[n]$ for even $N$, or $\sum_0^{(N-1)/2}\tilde g[n]=1=-\sum_{(N-1)/2}^{N-1}\tilde g[n]$ for odd $N$,
        \end{enumerate}
    such that $\omega^*$ is minimized, where $\omega^*$ is defined as $\omega^* = \min(\omega_c)$ such that
        \begin{equation}
            \text{for any } \omega \geq \omega_c, \ |G(e^{i\omega})| \leq \gamma,
            \nonumber
        \end{equation}
    where $G(e^{i\omega})$ is the discrete time Fourier transform of $\tilde g[n]$ and $\gamma$ is given.
\end{boxA}

\section{Finite-length, discrete-time pulses} \label{sec: Finite-length discrete-time pulses}

In order to establish some background information on the pulse design problem, we introduce the definition and notation of finite-length, discrete-time pulses. We first review the Slepian pulses. We further introduce another set of pulses designed using the weighted Chebyshev approximation (WCA), which are referred to as the Chebyshev pulses II.

\subsection{Definition and notation} \label{sec:review_sp}

In this work, we focus on the design of finite-length, discrete-time pulses, which we will refer to as pulses for brevity going forward. The pulses take on certain values over some chosen finite-length, discrete-time interval and are zero-valued outside the interval, defined mathematically as
\begin{equation}
    w[n]=\begin{cases}
        \hat{w}[n], & \ \ 0\leq n\leq N-1 \\
        0, & \ \ \text{otherwise}
    \end{cases} ,
\end{equation}
where $\hat{w}[n]$ denotes the values over the interval $[0,N-1]$, and $N$ is a finite positive integer. The discrete-time Fourier transform of the pulse $w[n]$ is
\begin{equation}
    W(e^{i\omega})= \sum_{n=0}^{N-1}w[n]e^{-i\omega n} \ .
\end{equation}
\subsection{Examples of common finite-length, discrete-time pulses} \label{sec:pulse_window}

In this section, we briefly summarize two examples of pulses---the Slepian pulses and the Chebyshev pulses that we will compare. Appendix~\ref{app: Slepian and Chebyshev} contains more details about the mathematical structure of these pulses.

The Slepian pulses are a set of orthogonal functions that are optimized to have maximum energy concentration in the frequency or time domains. As we discussed in Section~\ref{sec:physics_background}, Slepian-based pulses are commonly used in superconducting circuits to perform high-fidelity baseband flux controlled CZ gates~\cite{martinis_fast_2014}. We denote the Slepian pulses by $\{v_n^{(k)}(N,W), \ k=0,1,\dots,N-1\}$, where $n=0,1,\dots,N-1$ is the index of the pulse, $k$ is the order of each pulse, and $N$ and $W$ are parameters referred to as the length and mainlobe width of the pulse, respectively. In our problem formulation, we are especially interested in the second Slepian pulse ($k=1$), because it is an anti-symmetric pulse by definition and has the largest energy concentration among all anti-symmetric Slepian pulses. We denote the second Slepian pulse ($k=1$) as $w_{\textrm{sl2}}^{NW}[n]$. We omit $NW$ going forward for brevity when it is given in context.

\begin{figure*}[ht]
   \centering
   \includegraphics[width=1\textwidth]{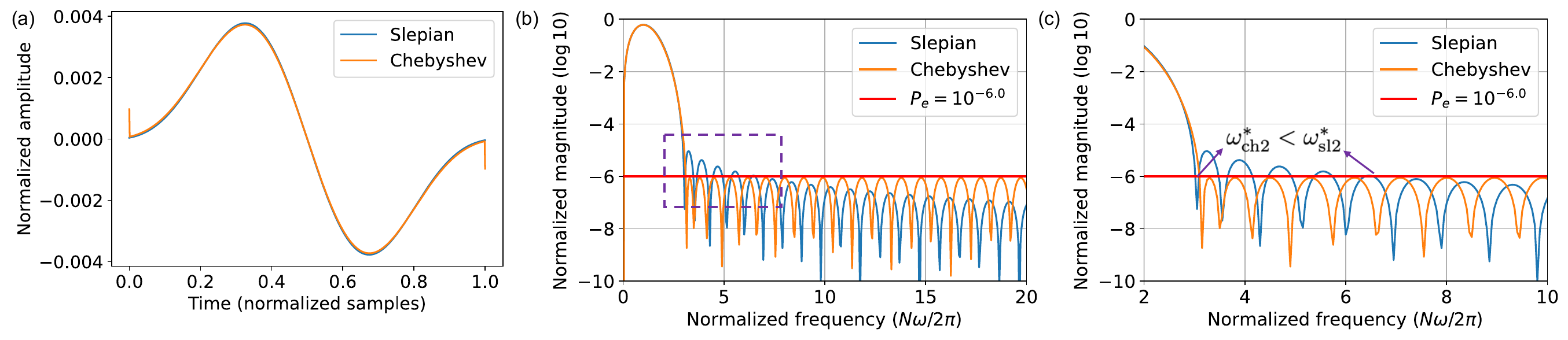}
   \caption{The (a) time-domain and (b) frequency-domain representations and (c) zoom-in version of the Slepian-based trajectory for $N=1001$ and $NW=2.9$ and the Chebyshev-based trajectory for $N=1001$ whose sidelobe amplitude $\gamma_{ch}$ is designed so that $\gamma_{ch}^2/4 \leq \gamma^2/4=10^{-6.0}$, which is also the given error threshold. We find that $\omega^*_{\textrm{ch2}}<\omega^*_{\textrm{sl2}}$.}
   \label{fig:p3}
\end{figure*}

Chebyshev pulses minimize the mainlobe width given a specified sidelobe amplitude. They are characterized by their ability to provide a trade-off between mainlobe width and sidelobe amplitude, making them useful in applications such as filter design and spectrum analysis. Referred to as Chebyshev pulses I and denoted as $w_{\text{ch1}}[n]$, these pulses are symmetric in time. In our exploration, we introduce a complementary, anti-symmetric variation named Chebyshev pulses II, denoted as $w_{\text{ch2}}^{\beta}[n]$. This variation is derived through weighted Chebyshev approximation (WCA), a technique for optimizing a polynomial approximation of a given function. Here, $\beta$ serves as the parameter input for this approximation process. For brevity, we omit $\beta$ going forward. Notably, Chebyshev pulses II share the equiripple sidelobe amplitude characteristic and exhibit only one ripple in the passband, mirroring the traits of Chebyshev pulses I. Further insights about the weighted Chebyshev approximation are summarized in Appendix~\ref{app: Weighted Chebyshev approximation}. In the rest of the paper, we focus on the second Slepian pulses ($k=1$) and the Chebyshev pulses II, and refer to them as the Slepian pulses and the Chebyshev pulses respectively.

\section{Comparison between Chebyshev- and Slepian-based trajectories} \label{sec: Optimal solutions}

We have formulated the CPHASE gate design problem into a pulse design problem and further transformed it into the design of a control trajectory $\tilde g[n]$, where the frequency-domain characteristics are emphasized. Special attention is paid to the comparisons between the Chebyshev-based trajectories and the Slepian-based trajectories that will be defined in this section.

The control trajectory $\tilde g[n]$ needs to satisfy a normalization constraint, and also is specified to be anti-symmetric according to our discussion in {\bf Statement 3}. The finite length constraint is naturally satisfied by finite-length, discrete-time pulses. The anti-symmetry constraint $\tilde g[n]=-\tilde g[N-n]$ is also satisfied using the Slepian pulses $w_{\text{sl2}}[n]$ and the Chebyshev pulses $w_{\text{ch2}}[n]$. Therefore, the control trajectories can be defined through a straightforward normalization of $w_{\text{sl2}}[n]$ and $w_{\text{ch2}}[n]$. We denote the corresponding control trajectories by $\tilde g_{\textrm{sl2}}[n]$ and $\tilde g_{\textrm{ch2}}[n]$.

To obtain the shortest pulse given a specified allowable leakage error, we need to minimize the frequency cutoff $\omega^*$ (see definition in {\bf Statement 3}) while keeping amplitude of the frequency components beyond $\omega^*$ below some constant determined by the allowable leakage error. We propose the Chebyshev-based trajectory $\tilde g_{\textrm{ch2}}[n]$ as an alternative to the Slepian-based trajectory $\tilde g_{\textrm{sl2}}[n]$. While $\tilde g_{\textrm{sl2}}[n]$ can fulfill the desired features because it is originally designed to obtain a maximized energy concentration within some frequency band, it features a decreasing amplitude envelop of sidelobe frequency components, which diminishes its capability to push $\omega^*$ even smaller. On the other hand, $\tilde g_{\textrm{ch2}}[n]$, by design, features a flat amplitude envelop of sidelobe frequency components, which allows $\omega^*$ to be even smaller. The argument is that we can allow higher sidelobe amplitudes in larger frequency components as long as they stay below a specified threshold, and therefore we can in turn decrease the concentration in smaller frequency components. In this way, $\tilde g_{\textrm{ch2}}[n]$ can be designed to obtain a smaller $\omega^*$ while maintaining a sidelobe amplitude below some threshold. Here, $\omega^*$ denotes the smallest attainable frequency under certain constraints given some leakage threshold, as we specify in {\bf Statement 3}. This eventually leads to the fact that the Chebyshev-based trajectory can be designed to be shorter than the Slepian-based trajectory.

We show an example of designing the Chebyshev-based  trajectory and the Slepian-based trajectory according to the given leakage error threshold. First, we determine the length $N=1001$ for both trajectories to be compared. Then we choose a half bandwidth $NW=2.9$ to determine the Slepian-based trajectory $\tilde g_{\textrm{sl2}}[n]$ as a benchmark pulse. We then design $\tilde g_{\textrm{ch2}}[n]$ so that its sidelobe amplitude $\gamma_{ch}$ is such that $\gamma_{ch} \leq \gamma$, where $\gamma^2/4=10^{-6.0}$ is the given leakage error threshold. As depicted in Figs.~\ref{fig:p3}\panel{a}-\panel{b}, we compare the time-domain and frequency-domain representations of $\tilde g_{\textrm{ch2}}[n]$ and $\tilde g_{\textrm{sl2}}[n]$. The dashed blue box area in Fig.~\ref{fig:p3}\panel{b} (expanded in Fig.~\ref{fig:p3}\panel{c}) shows that $\omega^*_{\textrm{ch2}}<\omega^*_{\textrm{sl2}}$. While satisfying the restriction on sidelobe amplitude, $\tilde g_{\textrm{ch2}}[n]$ outperforms $\tilde g_{\textrm{sl2}}[n]$ and features a smaller $\omega^*$. Note that there exist impulses at both endpoints of the $\tilde g_{\textrm{ch2}}[n]$ in Fig.~\ref{fig:p3}\panel{a}, which is a feature that contributes to the equiripple property in the frequency domain. The effect of these impulses will be filtered through numerical integration and interpolation when we transform $\tilde g_{\textrm{ch2}}[n]$ to $\varepsilon_{\textrm{ch2}}(t)$.

\section{Simulation results} \label{sec: Simulation results}

We show time-domain simulation results of a CZ gate, utilizing the Slepian-based trajectories and the Chebyshev-based trajectories as defined in Section~\ref{sec: Optimal solutions}.

\subsection{Setting and procedure} \label{sec:setting_procedure}

We consider two capacitively coupled transmon qubits, one of which is flux-tunable (QB1) and the other having a fixed frequency (QB2). The system Hamiltonian is
\begin{equation}
    H = \sum_{i=1,2}\bigg(\omega_i a^\dagger_i a_i+ \displaystyle\frac{\alpha_i}{2}a^\dagger_i a^\dagger_i a_i a_i\bigg) +g(a^\dagger_1 + a_1)(a^\dagger_2 + a_2) \ ,
\end{equation}
where $a^\dagger_i$, $a_i$ are the raising (creation) and lowering (annihilation) operators in the eigenbasis of the corresponding qubit, $\omega_i$ is the qubit frequency of QB$i$, and $\alpha_i$ is the anharmonicity of QB$i$.

In our simulation, we choose the parameters $\omega_2/2\pi = 4.7$ GHz, $\alpha_1/2\pi=\alpha_2/2\pi=-300$ MHz, and $g/2\pi=10\sqrt{2}\approx14.142$ MHz. We set $\omega_1/2\pi=5.8$ GHz initially and tune $\omega_1$ to perform a CZ gate. These parameters form a typical parameter set for transmon qubits and the operation of a CZ gate. In order to perform a CZ gate, we detune $\omega_1$ so that $\omega_1+\omega_2 \approx 2\omega_1+\alpha_1$. If we move exactly to the degeneracy point of diabatic states $\ket{11}$ and $\ket{20}$, then we have $\omega_1/2\pi = 5.0$ GHz. We vary $\omega_1/2\pi$ from $5.8$ GHz to approximately $5.0$ GHz depending on the amplitude of the control pulse, and then back to $5.8$ GHz.

The procedure of our simulation is as follows:
\begin{enumerate}
    \item Determine a control trajectory $\tilde g_i[n]$. 
    \item For each desired amplitude of the pulse, compute the corresponding control pulse $\varepsilon_i[n]$ for $\tilde g_i[n]$ and use an interpolation of $\varepsilon_i[n]$ as a control pulse $\varepsilon_i(t)$ to detune $\omega_1$.
    \item Simulate a CPHASE gate using QuTiP~\cite{JOHANSSON20131234} for a range of desired amplitude and duration of $\varepsilon_i(t)$. Calculate the phase accumulation and leakage error as a function of pulse duration and amplitude.
    \item Collect the duration and amplitude pairs that obtain a phase accumulation $\phi=\pi$. This is to ensure the implementation of a CZ gate. Determine the corresponding leakage error $P_e$ as a function of pulse duration.
\end{enumerate}

\subsection{Simulation examples} \label{sec:Simulation examples}

In the following simulation examples, we define a normalized amplitude $A = |\epsilon_\textrm{mid}-\epsilon_\textrm{ini}|/(5.8-5.0)$, where $A=1$ indicates that we go exactly to the degeneracy point of diabatic states $\ket{11}$ and $\ket{20}$, while $A=0$ indicates that we stay at the starting point. Here, we take $0 \leq A \leq 1$.

\subsubsection{A comparison example}\label{sec:example1}

\begin{figure}[tb]
   \centering
   \includegraphics[width=0.49\textwidth]{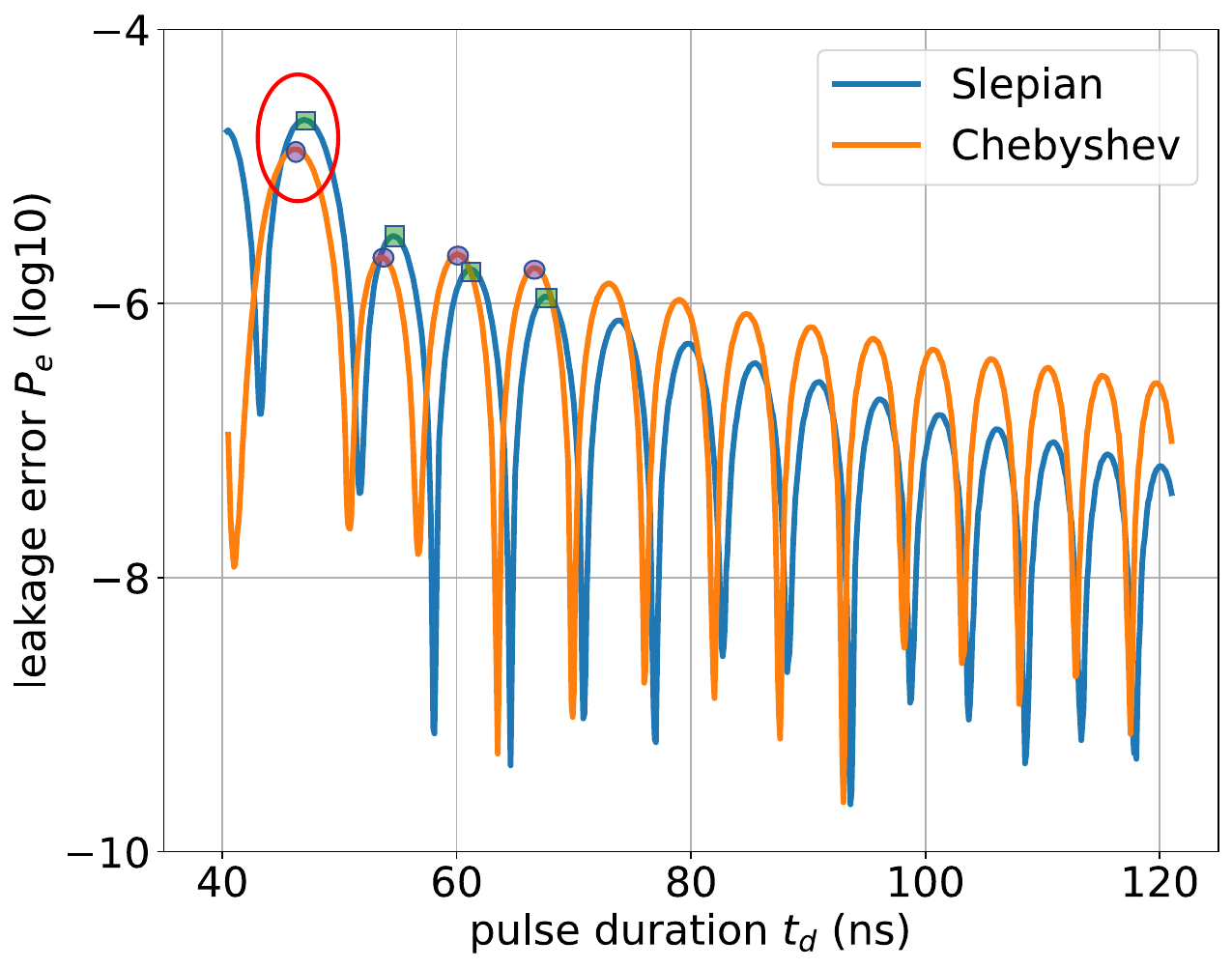}
   \caption{A comparison example. Leakage error $P_e$ as a function of pulse duration $t_d$ for a CZ gate, using $\tilde g_{\textrm{sl2}}[n]$ and $\tilde g_{\textrm{ch2}}[n]$ in Section~\ref{sec: Optimal solutions}. Potential operating points are marked in green squares and purple dots (only four for each $\tilde g[n]$ are shown). The best operating points with shortest pulse duration are indicated by the red circle.}
   \label{fig:p4}
\end{figure}

Figure~\ref{fig:p4} shows the leakage error $P_e$ for a CZ gate as a function of pulse duration $t_d$. Note that the amplitude $A$ of the control pulse is adjusted to ensure an exact phase accumulation of $\phi=\pi$. Figures of phase accumulation and leakage error for a range of control pulse duration and amplitude can be found in Appendix~\ref{app: Phase accumulation and leakage error of example}. As we observe in Fig.~\ref{fig:p4}, the leakage error $P_e$ corresponding to $\tilde g_{\textrm{ch2}}[n]$ appears generally below that of $\tilde g_{\textrm{sl2}}[n]$ for $t_d<\mathtt{\sim}60$ ns. Since the general trend the latter decreases is faster than the former as a function of pulse duration $t_d$, we observe the leakage error $P_e$ corresponding to $\tilde g_{\textrm{sl2}}[n]$ begins to appear below that of $\tilde g_{\textrm{ch2}}[n]$ for $t_d>\mathtt{\sim}60$ ns. This is true for all the simulation examples shown in Appendix~\ref{app: Additional simulation results}. This feature agrees with what we see in theoretical analysis from Figs.~\ref{fig:p3}\panel{c}-\panel{d}, except that the general trend of the leakage error $P_e$ corresponding to $\tilde g_{\textrm{ch2}}[n]$ also decreases slowly rather than remaining exactly flat. We attribute this difference between simulation and analysis to the numerical integration and interpolation in the process of transforming $\tilde g[n]$ into $\varepsilon(t)$. This reduces the impact of impulses at both ends of $\tilde g_{\textrm{ch2}}[n]$ within $\varepsilon_{\textrm{ch2}}(t)$, because the impulses significantly contribute to the equiripple sidelobe characteristic of $\tilde g_{\textrm{ch2}}[n]$.

Potential operating points are considered to be the insensitive points to pulse duration, i.e., the points of the leakage error lobes as shown by the green squares and purple dots in Fig.~\ref{fig:p4}. In addition, if there is a slight deviation in pulse duration at those points, we would be almost surely have a lower leakage error regardless of the direction of the deviation. We determine the best operating point to be the one with the smallest pulse duration, indicated by the red circle in Fig.~\ref{fig:p4}. Since $\tilde g_{\textrm{ch2}}[n]$ generally pushes the leakage error lower in the range of relatively smaller pulse duration, we are able to achieve an operating point with lower leakage error while also maintaining a similar or even smaller pulse duration compared to $\tilde g_{\textrm{sl2}}[n]$. The operating points for $\tilde g_{\textrm{sl2}}[n]$ and $\tilde g_{\textrm{ch2}}[n]$ are $t_d = 47.0$ ns with $P_e = 10^{-4.66}$ and $t_d = 46.1$ ns with $P_e = 10^{-4.72}$, respectively. We further compute the average gate fidelity as described in Appendix~\ref{app: Average gate fidelity}, and find that the infidelity $1-F_g = 5.5\times 10^{-6}$ for $\tilde g_{\textrm{sl2}}[n]$ and $1-F_g = 6.8\times 10^{-6}$ for $\tilde g_{\textrm{ch2}}[n]$ respectively. In this case, the average gate infidelity of the Slepian-based trajectory is lower than that of the Chebyshev-based trajectory although the leakage error of the former is greater than that of the latter. This is mainly because the unwanted population exchange between $\ket{01}$ and $\ket{10}$ within the computational subspace contributes to the average gate infidelity.

\subsubsection{An aggregate of comparison examples}

We conduct more comparisons between various pairs of benchmark Slepian-based trajectories and designed Chebyshev-based trajectories. The main difference is that the leakage error threshold are specified differently when designing the trajectories. We summarize the main results in this section. More details about the additional comparisons are shown in Appendix~\ref{app: Additional simulation results}.

We observe similar results in terms of leakage error to that in Section~\ref{sec:example1} for a majority of the comparison examples. We find that $\tilde g_{\textrm{ch2}}[n]$ can be designed to achieve a lower leakage error than its counterpart $\tilde g_{\textrm{sl2}}[n]$ by roughly $6.0\%$ in certain cases, while also maintaining smaller pulse duration by an average of $0.6$ ns. In several unusual cases, where there appears an unusually small leakage lobe before the first main leakage lobe, $\tilde g_{\textrm{sl2}}[n]$ in fact induces lower leakage error with smaller pulse duration. This indicates that simulation verification is important once a control trajectory is designed because there can be discrepancies between analysis and simulation due to some approximation. When it comes to gate infidelity, for most of the comparison examples ((a)-(g)), we find that the Chebyshev-based trajectories have lower gate infidelity than the Slepian-based trajectories. In example (h), the Slepian-based trajectory has lower gate infidelity because the leakage error is significantly smaller, due to the fact that there appears a significantly small leakage lobe before the first main leakage lobe. On average, the Chebyshev-based trajectories achieve a $23.3\%$ reduction in gate infidelity. This can also be interpretated as an increase in gate fidelity by $0.0009\%$. We note that this small number in terms of increase in gate fidelity results from the fact that the Slepian-based trajectories have already reached a relatively high fidelity (close to $1$). In other words, the percentage is almost identical to the absolute increase in gate fidelity. To be more specific, given that a Slepian-based trajectory achieves $x\%$ gate fidelity, the corresponding Chebyshev-based trajectory can on average achieves approximately $(x+0.0009)\%$ gate fidelity. We note that this improvement can become more prominent when the desired gate fidelity is higher (e.g., $x > 99.999$). This demonstrates that the Chebyshev-based trajectories can potentially enhance overall gate performance even when accounting for more than leakage error.

To determine scenarios where Chebyshev-based trajectories might significantly outperform other standard approaches, we find that in the region where the Slepian-based trajectory achieves a gate infidelity of $\sim 10^{-3}$ to $\sim 10^{-5}$, the Chebyshev-based trajectory appears to consistently decrease the infidelity by roughly $40.7\%$. However, in the infidelity range below $\sim 5\times10^{-6}$, we see mixed behavior and it becomes hard to conclude for sure which one is better. This indicates that the Chebyshev-based trajectory might outperform other standard approaches in relatively moderate infidelity ranges, but may not be as effective if the desired infidelity is very low. 

We note that there can be different operating points, and in our comparison, we further pick the one with the shortest pulse duration to be the best operating point because this aligns with our definition of a desired pulse in the first place. One could potentially pick any of these operating points as their operating points if a longer pulse duration is acceptable, and the comparison results will change accordingly. However, our approach of best operating point aims to avoid any bias in the evaluation.

\subsection{Hardware constraints} \label{sec:hardware bandwidth}

In this section, we study how hardware constraints, most notably, the sampling frequency and bandwidth of the arbitrary waveform generator (AWG), can affect the performance of the CZ gate. The hardware limitations are important, because the CZ gates we consider are based on fast-flux control, and it is necessary that current hardware be capable of implementing such fast pulses. Let $F_s$ denote the sampling frequency and $bw$ the bandwidth. Examples of state-of-the-art AWGs include QBLOX QCM with $F_s=1$ GSa/s and $bw=400$ MHz, Zurich Instrument SHFQC+ with specifications up to $F_s=2$ GSa/s and $bw=800$ MHz, QBLOX QCM with $F_s=2$ GSa/s and $bw=800$ MHz, Keysight M5300A with baseband sampling frequency $F_s=4.8$ GSa/s and $bw=2$ GHz, etc. Details on how we impose the hardware constraints in simulation are summarized in Appendix~\ref{app:simulating hardware limitations}.

\begin{table}[bh!]
\setlength{\tabcolsep}{10pt}
\centering
\begin{tabular}{| c | c c c c |} 
 \hline
 \rule{0pt}{1\normalbaselineskip} $F_s$ (GSa/s)  & 1 & 2 & 5 & 10 \\ [0.5ex]
 \hline
 \rule{0pt}{1\normalbaselineskip} $bw$ (GHz)  & 0.4 & 0.8 & 2 & 4 \\ [0.5ex]
 \hline
\end{tabular}
\caption{Sampling frequency and bandwidth for simulation.}
\label{table:hardware_parameters}
\end{table}

\begin{figure}[t]
    \centering  \includegraphics[width=0.5\textwidth]{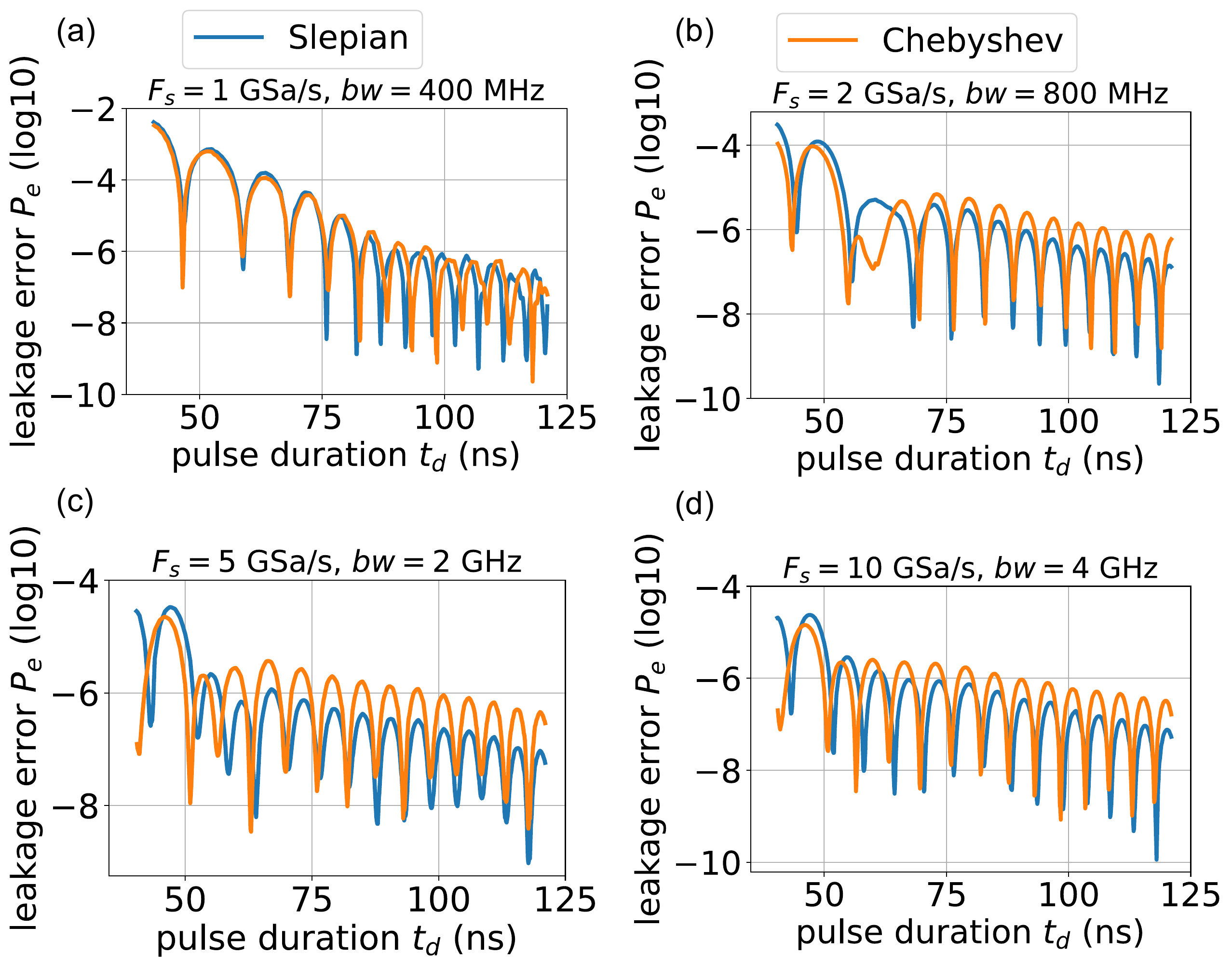}
    \caption{Comparison examples of the CZ gate performance using $\tilde g_{\textrm{sl2}}[n]$ and $\tilde g_{\textrm{ch2}}[n]$ with hardware parameters (a) $F_s=1$ GSa/s and $bw=400$ MHz, (b) $F_s=2$ GSa/s and $bw=800$ MHz, (c) $F_s=5$ GSa/s and $bw=2$ GHz, (d) $F_s=10$ GSa/s and $bw=4$ GHz.}
    \label{fig:harware_constraint}
\end{figure}


We utilize the same comparison example of $\tilde g_{\textrm{sl2}}[n]$ and $\tilde g_{\textrm{ch2}}[n]$ as in Section~\ref{sec:example1}. Fig.~\ref{fig:harware_constraint} presents the simulation results of the CZ gate using $\tilde g_{\textrm{sl2}}[n]$ and $\tilde g_{\textrm{ch2}}[n]$ with different practical hardware parameters. The hardware parameters are listed in Table~\ref{table:hardware_parameters}. When comparing Fig.~\ref{fig:harware_constraint} against Fig.~\ref{fig:p4}, it becomes evident that there is an overall increase in leakage error, regardless of the control trajectories. The difference between both trajectories also shrinks. As we enhance the hardware parameters, a better resemblance in the performance of the CZ gate to that without hardware constraints is observed. In Appendix~\ref{app: Additional simulation results}, we show an aggregate of the best operating points using $\tilde g_{\textrm{sl2}}[n]$ and $\tilde g_{\textrm{ch2}}[n]$ in Fig.~\ref{fig:harware_constraint} following the same argument as in Section~\ref{sec:example1}. Comparing the best operating points, we argue that the advantage of $\tilde g_{\textrm{ch2}}[n]$ over $\tilde g_{\textrm{sl2}}[n]$ can be mostly recovered with $F_s=5$ GSa/s and $bw=2$ GHz. Thus, the proposal in this paper is readily implementable using off-the-shelf state-of-the-art hardware.

\section{Conclusion and outlook} \label{sec: Conclusion and outlook}

In this work, we formulate the problem of baseband flux control design of an adiabatic CPHASE gate in superconducting circuits as a pulse-design problem, and further as a control-trajectory-design problem. Building upon knowledge from other contexts of pulse-design problems, we propose the Chebyshev-based trajectory as an alternative to the widely used Slepian-based trajectory. We then analytically show the advantage of the Chebyshev-based trajectory by using a two-level system abstraction. Furthermore, we compare the performance of the two types of trajectories by numerically simulating a CZ gate in two capacitively coupled transmon qubits. Our simulation results show that the Chebyshev-based trajectory can be designed to induce lower leakage error by roughly $6.0\%$ in certain cases, while maintaining similar or even smaller pulse duration. We note that in several cases the Slepian-based trajectory induces lower leakage error due to some unusual phenomenon not existing in analysis. On average, we find that the Chebyshev-based trajectory results in $23.3\%$ lower gate infidelity than its Slepian counterpart. 
From the perspective of quantum error correction, it is the reduction in gate infidelity—rather than the absolute increase in gate fidelity—that more directly impacts the efficacy of error correction protocols. The $23.3\%$ average reduction in gate infidelity of the Chebyshev-based trajectories is considerable, as physical error rates are exponentially leveraged in the suppression of logical error rates~\cite{google_2025}. This, in turn, can help reduce the required code distance for achieving a target logical error rate.
In addition, we study how practical hardware constraints including sampling frequency and bandwidth can influence the performance of the theoretically derived control pulses and affect the advantage of the Chebyshev-based trajectory over the Slepian-based trajectory in simulation. We find that the advantage can be mostly realized using state-of-the-art hardware.

The design of the flux pulse plays a crucial role in achieving high fidelity and fast speed for baseband flux-based gates in superconducting circuits. The weighted Chebyshev approximation, a versatile technique devised for tailoring pulses based on specific requirements and constraints, emerges as a valuable tool to design pulse shapes. Future work could explore extending this framework to other types of quantum gates, integrating these pulse design techniques into closed-loop calibration and optimal control pipelines, and studying the robustness of the gate performance against control noise in experimental deployment.

\begin{acknowledgments}
We gratefully acknowledge insightful conversations with Réouven Assouly, Youngkyu Sung and Junyoung An.
This material is based upon work supported by the U.S. Department of Energy, Office of Science, National Quantum Information Science Research Centers, Quantum Systems Accelerator. Additional support is acknowledged from the U.S. Department of Energy, Office of Science, National Quantum Information Science Research Centers, Co-design Center for Quantum Advantage (C2QA) under contract number DE-SC0012704. P.T.B. is exclusively supported by Mitsubishi Electric Research Laboratories (MERL). The views and conclusions contained herein are those of the authors and should not be interpreted as necessarily representing the official policies or endorsements, either expressed or implied, of the U.S. Government.

\end{acknowledgments}

\section*{Data Availability}

The data that support the findings of this article are openly available~\cite{ding2025data}.

\clearpage

\appendix

\section{The CPHASE gate} \label{app: CPHASE gate}
We focus on the adiabatic implementation of CPHASE gate in tunable transmon qubits and describe in detail a common implementation using baseband flux pulses.

The CPHASE gate is a two-qubit gate whose operation is represented by the unitary matrix
    \begin{equation} \label{eq:1}
    U_{\textrm{CPHASE}} = 
    \begin{bmatrix}
    1 & 0 & 0 & 0 \\
    0 & 1 & 0 & 0 \\
    0 & 0 & 1 & 0 \\
    0 & 0 & 0 & e^{i\phi}
    \end{bmatrix}.
    \end{equation}

   \begin{figure*}[tb]
   \begin{center}
   \includegraphics[width=0.97\textwidth]{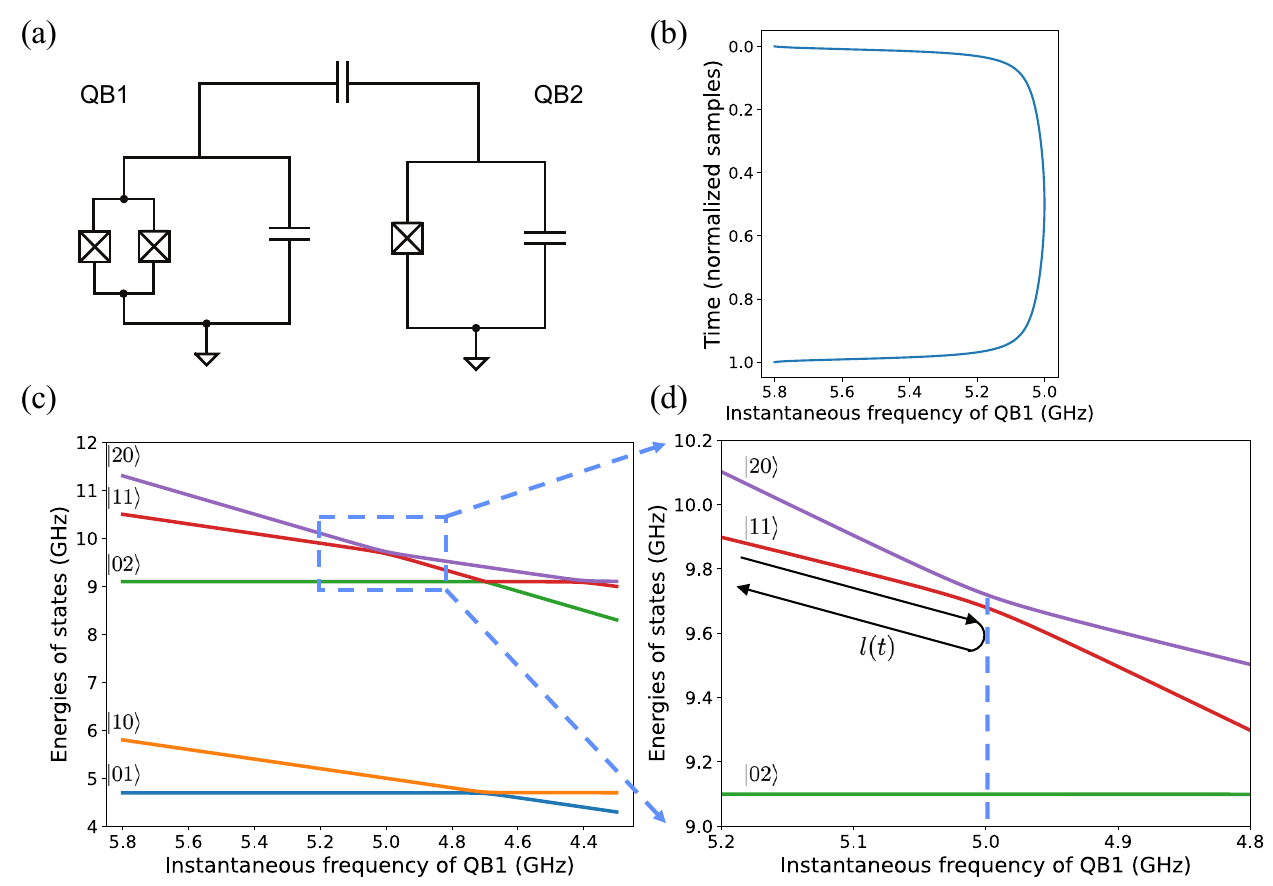}
   \caption{(a) Circuit diagram of a flux-tunable transmon capacitively coupled to a fixed-frequency transmon. The flux-tunable transmon is referred to as QB1 and the fixed-frequency transmon is referred to as QB2. (b) An example of a pulse to vary $\omega_1$ from $\omega_1 = 5.8$ GHz to $\omega_1 = 5.0$ GHz and then back to $\omega_1 = 5.8$ GHz. (c) Energy spectrum diagram of two coupled transmons as depicted in (a) as the frequency of QB1 is detuned by changing the local magnetic flux that threads its SQUID loop. The avoided crossing between the states $\ket{11}$ and $\ket{20}$ is used to implement the CPHASE gate. (d) A zoom-in plot around the avoided crossing of interest, where $l(t)$ represents a typical trajectory of the instantaneoues energy of state $\ket{11}$ during the process of the CPHASE gate. $l(t)$ can be described in terms of how the frequency of QB1 changes over time, as shown in (b).}
   \label{fig:p1}
   \end{center}
   \end{figure*}

The CPHASE gate adds a term $e^{i\phi}$ to the qubits only when both are in the excited state $\ket{11}$. To be more specific, if the original state of the qubits is $\ket{00}$, $\ket{01}$ or $\ket{10}$, the CPHASE gate effectively does nothing. If the original state of the qubits is $\ket{11}$, it will be transformed into $e^{i\phi}\ket{11}$ after the CPHASE gate operation.
    
One implementation of the CPHASE gate relies on the avoided crossing between states $\ket{11}$ and $\ket{20}$ that occurs when two transmon qubits are coupled to each other. Consider a system of two capacitively coupled qubits as depicted in Fig.~\ref{fig:p1}\panel{a}, where QB1 is a flux-tunable transmon qubit while QB2 is a fixed-frequency transmon qubit. The Hamiltonian - including states with two excitations in addition to the four computational states - can be written in the ${\ket{00},\ket{01},\ket{10},\ket{11},\ket{02},\ket{20}}$-basis as
\begin{equation} \label{eq:hamiltonian_cz}
    H = \begin{bmatrix}
    E_{00} & 0 & 0 & 0 & 0 & 0 \\
    0 & E_{01} & g & 0 & 0 & 0 \\
    0 & g & E_{10} & 0 & 0 & 0 \\
    0 & 0 & 0 & E_{11} & \sqrt{2}g & \sqrt{2}g \\
    0 & 0 & 0 & \sqrt{2}g & E_{02} & 0 \\
    0 & 0 & 0 & \sqrt{2}g & 0 & E_{20}
    \end{bmatrix},
\end{equation}
where $E_{ij}$ is the energy of state $\ket{ij}$ and $g$ is the coupling strength with a factor of $\sqrt{n}$ corresponding to the number of qubit excitations ($n=1,2$). Note that the frequency of QB1, and therefore the energies $E_{ij}$, depend on the external magnetic flux threading the SQUID loop. Fig.~\ref{fig:p1}\panel{c} shows an example of the energy spectrum of the system described by the Hamiltonian in Eq.~\ref{eq:hamiltonian_cz} as a function of the frequency detuning of QB1. We show energies of states $\ket{01},\ket{10},\ket{02},\ket{11},\ket{20}$.

The CPHASE gate is implemented by detuning the frequency of QB1 such that the instantaneoues energy of state $\ket{11}$ follows the trajectory $l(t)$ in Fig.~\ref{fig:p1}\panel{d}. To be more specific, we shift the frequency of QB1, thus in particular the energy of state $\ket{11}$, bringing it into resonance with state $\ket{20}$, which opens an avoided crossing due to the coupling. We then rewind the trajectory and return to the starting point. The trajectory $l(t)$ corresponds to the change of the instantaneous frequency of QB1 as time evolves as shown in Fig.~\ref{fig:p1}\panel{b}. In the adiabatic implementation of the process, we deliberately design $l(t)$ to detune the frequency of QB1 slowly enough so that there is only small leakage from $\ket{11}$ to $\ket{20}$ throughout the whole process. We note that due to the presence of the avoided crossing, the energy of state $\ket{11}$ is pushed lower than would be expected in an uncoupled system. This is the origin of an additional phase accumulation that only occurs for state $\ket{11}$, leading to the conditional phase accumulation. This process can be represented by a unitary matrix in the computational basis
\begin{equation} \label{eq:raw_cz}
    U_{raw} = \begin{bmatrix}
    1 & 0 & 0 & 0  \\
    0 & e^{i\phi_{01}} & 0 & 0 \\
    0 & 0 & e^{i\phi_{10}} & 0  \\
    0 & 0 & 0 & e^{i\phi_{11}}
    \end{bmatrix},
\end{equation}
where $\phi_{ij}$ is the accumulated phase
\begin{equation}
    \phi_{ij} = \int_0^{t_d} \omega_{ij}(t)\textrm{d}t \ ,
\end{equation}
with $t_d$ denoting the duration of the process.

The way we shift the frequency of QB1 is by changing the external magnetic flux $\Phi_{\textrm{ext}}$ threading the SQUID loop of QB1. The qubit frequency of QB1 $\omega_1$ as a function of $\Phi_{\textrm{ext}}$ is given by~\cite{Koch2007}
    \begin{equation} \label{eq:flux_frequency}
        \omega_1(\Phi_{\textrm{ext}}) = \frac{1}{\hbar}\bigg(\sqrt{8E_JE_C}\sqrt[\leftroot{-2}\uproot{10}4]{d^2+(1-d^2)\cos^2(\pi\frac{\Phi_{\textrm{ext}}}{\Phi_{0}})} -E_C\bigg) \ ,
    \end{equation}
where $E_J=E_{J1}+E_{J2}$ is the sum of the Josephson energies of the two junctions, namely $E_{J1}$, $E_{J2}$. $E_C$ is the charging energy. $\Phi_{0}$ is the superconducting flux quantum. $d$ is the junction symmetry parameter defined as
\begin{equation}
    d= \frac{|E_{J2}-E_{J1}|}{E_{J2}+E_{J1}} \ .
\end{equation}

In order to obtain the CPHASE gate as in Eq.~\ref{eq:1}, two single-qubit gates $R_z(-\phi_{01})$ and $R_z(-\phi_{10})$ need to be implemented to each qubit to cancel the phase accumulated by states $\ket{01}$ and $\ket{10}$. Therefore, the whole operation can be represented by
\begin{equation} \label{eq:raw_cz_after1qb}
    U_{\textrm{CPHASE}}' = \begin{bmatrix}
    1 & 0 & 0 & 0  \\
    0 & 1 & 0 & 0 \\
    0 & 0 & 1 & 0  \\
    0 & 0 & 0 & e^{i\phi'}
    \end{bmatrix},
\end{equation}
where $\phi'=\phi_{11}-\phi_{01}-\phi_{10}$. If there were no coupling between the two qubits, $\phi' = 0$. Because of the effect of the coupling between the two qubits, a nonzero phase will be acquired. The way the instantaneoues energy of state $\ket{11}$ is varied determines the value of $\phi'$. By choosing a suitable $l(t)$ as depicted in Fig.~\ref{fig:p1}\panel{d}, in principle we can always have $\phi'=\phi$ for any arbitrary desired phase $\phi$. When $\phi=\pi$, the operation is named a CZ gate.

There is an alternative way of implementing a CPHASE as opposed to the adiabatic method, which we refer to as a non-adiabatic implementation. Instead of gradually detuning the frequency of QB1, this approach involves making a sudden transition to the CPHASE operating point near the avoided crossing of $\ket{11}$ and $\ket{20}$. After a waiting period of time $t = \pi/\sqrt{2}g$, the state undergoes a single Larmor-type rotation from $\ket{11}$ to $\ket{20}$ and then back to $\ket{11}$. During this process, an overall conditional phase accumulation is obtained.

\section{Derivation of formula for leakage error} \label{app: Derived formula for leakage error and its approximation}
Two approaches to calculating the leakage error are presented based on the discussions in Ref.~\cite{martinis_fast_2014}. We first take advantage of the Bloch sphere representation and give an approximate but more intuitive solution from a geometric perspective. Then we go through a mathematical derivation and provide an analytical formula. We will comment on the efficacy of the formula by discussing the relationship of this formula to the more general Landau-Zener formulation~\cite{zener1932,landau1932}.

\subsection{Geometric approach} \label{sec:geo_approach}
\begin{figure}[tb]
     \centering
     \includegraphics[width=0.49\textwidth]{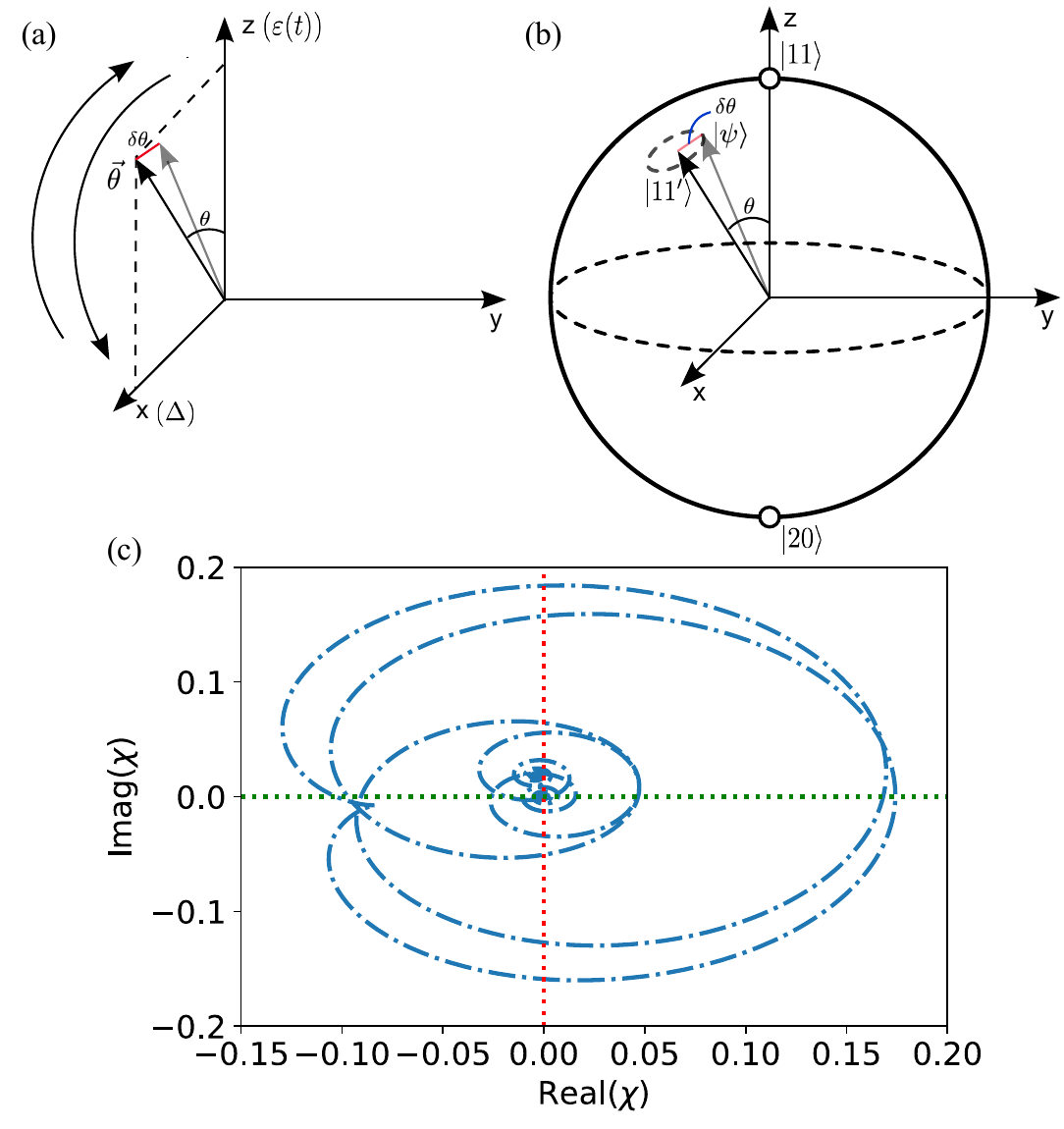}
    \caption{The geometric definition of $\theta(t)$ and the Bloch sphere picture of an infinitesimal step of evolution. (a) A control vector $\vec{\theta}$ representing the control variable $\theta(t)$ in a coordinate in terms of $\Delta$ and $\varepsilon(t)$. (b) An instantaneous basis vector $\ket{11'}$ parallel to the control vector $\vec{\theta}$. In an infinitesimal time $\delta t$, the state vector $\ket{\psi}$ deviates from the instantaneous basis vector $\ket{11'}$ by $-\delta\theta$ and precesses around it. (c) A bird's-eye plot supposing we stare at and stay in a moving frame with the control vector $\vec{\theta}$. The accumulated deviation $\chi$ of the state vector $\ket{\psi}$ from the control vector $\vec{\theta}$ as a function of time. The vertical axis Imag($\chi$) coincides with the longitude axis in the $x$-$z$ plane, while the horizontal axis Real($\chi$) denotes the latitude axis orthogonal to the $x$-$z$ plane. The origin represents the control vector $\vec{\theta}$.}
     \label{fig:appb1}
\end{figure}

Recall that in Eq.~\ref{eq:theta_def}, $\theta(t)$ is defined as $\theta(t) = \arctan (\Delta/\varepsilon(t))$. In Fig.~\ref{fig:appb1}\panel{a} we introduce a control vector $\vec{\theta}$ representing the control variable $\theta(t)$ in terms of $\Delta$ and $\varepsilon(t)$. Correspondingly, in Fig.~\ref{fig:appb1}\panel{b} we show an instantaneous basis vector $\ket{11'}$, which represents the ground state of the instantaneous Hamiltonian $H$ as $\theta(t)$ varies, in parallel to the control vector $\vec{\theta}$. As time progresses, we change our frame reference to coincide with the frame whose basis vectors are the eigenstates of the instantaneous Hamiltonian $H$.

We first show how the state evolves in an infinitesimal time $\delta t$. Suppose the angle between the initial state $\ket{\psi}$ at time $t_0$, represented by the gray Bloch vector in Fig.~\ref{fig:appb1}\panel{b}, and the $z$-axis is $\theta_0$. Ideally, $\ket{\psi}$ is aligned with the instantaneous ground vector $\ket{11'}$ at time $t_0$. After $\delta t$, a $\delta \theta$ change in the angle between instantaneous ground vector $\ket{11'}$ and the $z$-axis takes place. If we switch into the new reference frame, the state vector $\ket{\psi}$ deviates from the basis vector by $-\delta \theta$ and therefore starts to precess around the basis vector at frequency $\omega$, where $\omega$ refers to the eigenenergy difference of the instantaneous Hamiltonian. Therefore, during the infinitesimal time $\delta t$, the state vector $\ket{\psi}$ will pick up a deviation from the ground vector $\ket{11'}$ by $\delta \chi = -\delta \theta e^{-i\omega \delta t}$.

Next we consider a series of infinitesimal time $\delta t$'s. A simple approach is to move into the reference frame along with the control vector $\vec{\theta}$ and correspondingly the instantaneous ground vector $\ket{11'}$. Thus, the whole process can be viewed as the state vector $\ket{\psi}$ deviating from the basis vector by a series of $-\delta \theta_j$'s with an angle rotation $\phi_j=\sum_i \omega_i \delta t$, which is the accumulated phase up to the $j$-th $\delta t$. Since the angle rotation is orthogonal to the $-\delta \theta_j$ deviation, and both are sufficiently small in the adiabatic limit, we can accumulate them independently, i.e.,
\begin{align} \label{theta_sum}
    \chi =  \sum_j -\delta \theta_j e^{-i\phi_j}
\end{align}

Fig.~\ref{fig:appb1}\panel{c} is a bird’s-eye view plot supposing we stare at and stay in a moving frame with the control vector $\vec{\theta}$. The vertical axis Imag($\chi$) coincides with the longitude axis in the $x$-$z$ plane, while the horizontal axis Real($\chi$) denotes the latitude axis orthogonal to the $x$-$z$ plane. The origin represents the control vector $\vec{\theta}$. We plot the accumulated deviation $\chi$ as a function of time according to Eq.~\ref{theta_sum}, using a Slepian-based control trajectory.

Change the $\sum$ symbol into the $\int$ symbol and the $\delta$ symbol into the $\textrm{d}$ symbol, as in elementary calculus, and we have
\begin{align}
    \displaystyle \chi &= -\int \textrm{d} \theta e^{-i\int^t \omega(t') \textrm{d}t'} \\
    &= -\int \frac{\textrm{d}\theta}{\textrm{d}t} e^{-i\int^t \omega(t') \textrm{d}t'}\textrm{d}t \label{eq:approx_sum_theta} \ .
\end{align}

Therefore, the leakage error rate $P_e$ can be calculated as 1 minus the probability of measuring the state $\ket{\psi}$ in the instantaneous ground state $\ket{11'}$
\begin{equation} \label{eq:approx_leakage}
\begin{split}
    P_e&=1-\bigg(\cos \frac{\arcsin|\chi|}{2}\bigg)^2 \\ &=\bigg(\sin \frac{\arcsin|\chi|}{2}\bigg)^2 \\ &\approx |\chi|^2/4 \ .
\end{split}
\end{equation}
where the approximation holds valid when $\chi$ is sufficiently small. We plug the calculated $\chi$ as depicted in Fig.~\ref{fig:appb1}\panel{c} into Eq.~\ref{eq:approx_leakage} and find that the analytical leakage error matches well with the simulation result throughout the process.

Note that in this geometric derivation we assume that the changes at different infinitesimal time $\delta t$'s can be summed linearly. This approximation holds valid so long as the net overall change $\chi$ is small. In fact, in this research we concentrate on adiabatic control, and therefore, we are always interested in small leakage error rate $P_e$ incurred throughout the process. This small $P_e$ corresponds to the fact that $\chi$ should be small.

\subsection{Analytical approach} \label{sec:analytical_approach}
We continue to present an analytical approach to deriving the leakage error. Recall that in Section~\ref{sec:two_level}, the abstracted two-level system can be described by the Hamiltonian in Eq.~\ref{hamiltonian}. We have also defined a control variable $\theta(t) = \arctan (\Delta/\varepsilon(t))$. A pictorial representation of the control variable $\theta(t)$ in terms of $\Delta$ and $\varepsilon(t)$ is shown in Fig.~\ref{fig:appb1}\panel{a}.

Consider a state $\ket{\psi}$ given by $\displaystyle \ket{\psi}=\alpha_0 \ket{11}+\beta_0 \ket{20}$, where $\ket{11}$ and $\ket{20}$ denote the basis vectors of the $\sigma_z$-basis ($z$-axis). Suppose there exists another basis which rotates around the $y$-axis by an angle $\theta$ relative to the $\sigma_z$-basis ($z$-axis). In this new basis, the state $\ket{\psi}$ can be rewritten as 
\begin{equation}
    \ket{\psi} =\alpha\ket{11'}+\beta\ket{20'} \ ,
\end{equation}
where
\begin{subequations}
    \begin{align}
        \alpha &=\alpha_0 \cos \frac{\theta}{2}+\beta_0 \sin \frac{\theta}{2}  \ , \\
        \beta &=\beta_0 \cos \frac{\theta}{2}-\alpha_0 \sin \frac{\theta}{2} \ ,
    \end{align}
\end{subequations}
and $\ket{11'}$ and $\ket{20'}$ are the basis vectors of the new basis, which we now refer to as the $\theta$-rotated basis. Correspondingly, we refer to the Bloch sphere with the $\theta$-rotated basis as the $\theta$-rotated Bloch sphere.

In the $\theta$-rotated basis, we denote the eigenvalues of the basis states as $\pm \omega/2$. In a static (non-rotating) frame, the Bloch vector will precess around the $\theta$-rotated basis axis, which results in a phase induced time derivative
\begin{align}
    \dot{\alpha} &= -i\frac{\omega}{2}\alpha  \ , \\
    \dot{\beta} &= i\frac{\omega}{2}\beta  \ .
\end{align}

However, if we let $\theta(t)$ varies as a function of time, we will have an additional term in the time derivative of $\alpha$ and $\beta$ respectively
\begin{align}
    \dot{\alpha} &= -i\frac{\omega}{2}\alpha + (-\alpha_0 \sin\frac{\theta}{2}+\beta_0\cos \frac{\theta}{2})(\frac{1}{2} \dot{\theta}) \\ &= \frac{-i\omega\alpha+\beta \dot{\theta}}{2}  \ , \\
    \dot{\beta} &= i\frac{\omega}{2}\beta + (-\alpha_0 \cos \frac{\theta}{2}-\beta_0 \sin \frac{\theta}{2}) (\frac{1}{2} \dot{\theta}) \\ &= \frac{i\omega\beta-\alpha \dot{\theta}}{2}  \ ,
\end{align}
where we use the ``dot'' notation $\dot{x}$ as a shorthand to denote the time derivative of $x$. 

Now let us denote $\displaystyle \alpha = \cos \Theta/2$ and $\displaystyle \beta = e^{i\phi}\sin \Theta/2 $ where $\Theta$ and $\phi$ are the spherical coordinates on the $\theta$-rotated Bloch sphere. Here we have omitted an overall phase. Note that 
\begin{equation} \label{eq:alphabeta}
    \begin{split}
    \alpha^\ast \beta &= \cos \frac{\Theta}{2} \sin \frac{\Theta}{2} e^{i \phi} \\
    &=\frac{\sin\Theta e^{i\phi}}{2}  \ .
\end{split}
\end{equation}
It is interesting to see how $\alpha^\ast \beta$ evolves over time. We proceed by taking the time derivative of $\alpha^\ast \beta$
\begin{align} 
    \frac{\textrm{d}}{\textrm{d}t}(\alpha^\ast \beta) &= \dot{\alpha^\ast}\beta+\alpha^\ast\dot{\beta} \\
    &= \bigg[\frac{-i\omega\alpha+\beta \dot{\theta}}{2}\bigg]^\ast \beta + \alpha^\ast \frac{i\omega\beta-\alpha \dot{\theta}}{2} \\
    &= \frac{i\omega\alpha^\ast+\beta^\ast \dot{\theta}}{2} \beta + \alpha^\ast \frac{i\omega\beta-\alpha \dot{\theta}}{2} \\
    &= i\omega\alpha^\ast\beta + \frac{|\beta|^2-|\alpha|^2}{2} \dot{\theta}  \ . \label{eq:derivative_alphabeta}
\end{align}
Since
\begin{equation}
    \begin{split}
        (|\beta|^2-|\alpha|^2)^2 &= |\beta|^4 +|\alpha|^4 - 2|\beta|^2|\alpha|^2 \\
        &= |\beta|^4 +|\alpha|^4 + 2|\beta|^2|\alpha|^2 - 4|\alpha^\ast\beta|^2 \\
        &= 1-4|\alpha^\ast\beta|^2  \ ,
    \end{split} 
\end{equation}
we then have
\begin{equation} \label{eq:beta2_alpha2}
    |\beta|^2-|\alpha|^2 = \pm \sqrt{1-4|\alpha^\ast\beta|^2} \ .
\end{equation}
Substituting Eq.~\ref{eq:beta2_alpha2} into Eq.~\ref{eq:derivative_alphabeta}, we have
\begin{equation}
    \frac{\textrm{d}}{\textrm{d}t}(\alpha^\ast \beta) = i\omega\alpha^\ast\beta \pm \frac{\dot{\theta}}{2}\sqrt{1-4|\alpha^\ast\beta|^2}  \ .
\end{equation}
If we substitute $\alpha^\ast \beta = e^{i\phi} \sin\Theta /2$ as in Eq.~\ref{eq:alphabeta}, we will have
\begin{equation} \label{eq:dsin_dt}
    \frac{\textrm{d}}{\textrm{d}t}(\sin\Theta e^{i\phi}) = i\omega\sin\Theta e^{i\phi} \pm \dot{\theta}\cos\Theta  \ .
\end{equation}
Substituting $\phi = \phi' + \int^t \omega(t')\textrm{d}t'$, where $\phi'$ is some initial reference phase, in Eq.~\ref{eq:dsin_dt}, we have
\begin{align}
    \textrm{LHS} &= \frac{\textrm{d}}{\textrm{d}t}(\sin\Theta e^{i\phi'}) e^{i\int^t \omega(t')\textrm{d}t'} + i\omega \sin\Theta e^{i\phi}  \ ,\\
    \textrm{RHS} &= i\omega\sin\Theta e^{i\phi}  \pm \dot{\theta}\cos\Theta \ .
\end{align}
Therefore, we can derive
\begin{equation}
    \frac{\textrm{d}}{\textrm{d}t}(\sin\Theta e^{i\phi'}) e^{i\int^t \omega(t')\textrm{d}t'} =\pm \dot{\theta}\cos\Theta \ .
\end{equation}
which can be rewritten as
\begin{equation} \label{eq:d_sine}
    \textrm{d}(\sin\Theta  e^{i\phi'}) = \pm \dot{\theta}\cos\Theta e^{-i\int^t \omega(t')\textrm{d}t'} \textrm{d}t \ .
\end{equation}
If we integrate both sides of Eq.~\ref{eq:d_sine}, we will have
\begin{equation} \label{eq:overall_sine}
    \sin\Theta e^{i\phi'} = \pm \int \cos\Theta \frac{\textrm{d}\theta}{\textrm{d}t}e^{-i\int^t \omega(t')\textrm{d}t'} \textrm{d}t \ .
\end{equation}
Now we can write the leakage error $P_e$ throughout the whole dynamic process as
\begin{equation} \label{eq:analytical_pe}
\begin{split}
    P_e &= |\beta|^2 = \bigg|\sin \frac{\Theta}{2}\bigg|^2 \\ &\approx \bigg|\frac{\sin\Theta e^{i\phi'}}{2}\bigg|^2 \\ &= \frac{\displaystyle \bigg|\int \cos\Theta \frac{\textrm{d}\theta}{\textrm{d}t}e^{-i\int^t \omega(t')\textrm{d}t'} \textrm{d}t \bigg|^2}{4} \\
    &\approx \frac{\displaystyle \bigg|\int \frac{\textrm{d}\theta}{\textrm{d}t}e^{-i\int^t \omega(t')\textrm{d}t'} \textrm{d}t \bigg|^2}{4}  \ .
\end{split}
\end{equation}
Note that Eq.~\ref{eq:overall_sine} and Eq.~\ref{eq:approx_sum_theta} differ by a factor of $\cos\Theta$. Here, the term $\cos\Theta$ is due to the geometry of the $\theta$-rotated Bloch sphere relative to the original Bloch sphere whose basis vectors are $\ket{11}$ and $\ket{20}$. Since the whole process is performed adiabatically, the error is sufficiently small and hence $\Theta$ is sufficiently small, and therefore, the approximations in Eq.~\ref{eq:analytical_pe} are valid.

\subsection{Relationship to the Landau-Zener formula} \label{sec:lz_formula}
Consider a two-level system described by the Hamiltonian in Eq.~\ref{hamiltonian} and the energy diagram shown in Fig.~\ref{fig:p2}. Consider that the system is initially prepared in state $\ket{\psi_-}$ with $\varepsilon(t) \rightarrow -\infty$. Then $\varepsilon(t)$ increases in time and sweeps through the avoided crossing and eventually $\varepsilon(t) \rightarrow +\infty$. According to the Landau-Zener probability of transition~\cite{rubbmark_dynamical_1981}, we can derive the probability that the system will undergo a transition to $\ket{20}$ for the simple case where $\varepsilon(t) = \alpha t$ with $\alpha$ being a positive constant
\begin{equation} \label{eq:lz}
    P_{\textrm{LZ}} = e^{-\pi\Delta^2/ 2 \alpha} \ .
\end{equation}

In Eq.~\ref{eq:analytical_pe}, we do not assume that $\varepsilon(t) = \alpha t$ increases linearly with time. However, if we were to make this assumption, we could compute the transition probability $P_{\textrm{eLZ}}$ from Eq.~\ref{eq:analytical_pe} and compare it to $P_{\textrm{LZ}}$ in Eq.~\ref{eq:lz}. We further let that $\omega(t)=\Delta$ to simplify the comparison, i.e.,
\begin{equation} \label{eq:simply_pe_omega0}
    P_e = \frac{\displaystyle \bigg|\int \frac{\textrm{d}\theta}{\textrm{d}t}e^{-i\Delta t} \textrm{d}t \bigg|^2}{4} \ .
\end{equation}

First we compute the time derivative of $\theta(t)$
\begin{equation}
    \begin{split}
        \frac{\textrm{d}\theta}{\textrm{d}t} &= \frac{1}{1+(\Delta/\alpha t)^2} \times \frac{-\Delta}{\alpha t^2} \\
        &= \frac{-\Delta}{\alpha t^2+\Delta^2/\alpha} \\
        &= \frac{-\Delta/\alpha}{t^2+(\Delta/\alpha)^2} \ .
    \end{split}
\end{equation}
Then we compute the integral within $|\cdot|$ in Eq.~\ref{eq:simply_pe_omega0} and substitute $\Delta=\Delta/\hbar$ (let $\hbar=1$)
\begin{equation}
    \begin{split}
    \int \frac{\textrm{d}\theta}{\textrm{d}t}e^{-i\Delta t} \textrm{d}t &=\int \frac{-\Delta/\alpha}{t^2+(\Delta/\alpha)^2} e^{-i\Delta t} \textrm{d}t \\
    &=-\pi e^{-\Delta\Delta/\alpha} \\
    &=-\pi e^{-\Delta^2/\alpha} \ .
    \end{split}
\end{equation}
Therefore, we have
\begin{equation} \label{eq:error_lz}
    P_{\textrm{eLZ}} = \frac{\pi^2}{4}e^{-2\Delta^2/\alpha} \ .
\end{equation}
The relationship between $P_{\textrm{LZ}}$ in Eq.~\ref{eq:lz} and $P_{\textrm{eLZ}}$ in Eq.~\ref{eq:error_lz} can be written as
\begin{equation} \label{eq:relation_LZ_our}
    \log P_{\textrm{LZ}} = \frac{\pi}{4}(\log P_{\textrm{eLZ}} - C) \ ,
\end{equation}
where $C=\log (\pi^2/4)$ is some constant.

\section{Nonlinear time frame transformation} \label{app: Nonlinear time frame transformation}
We review a technique as proposed in Ref.~\cite{martinis_fast_2014}, which we term as nonlinear time-frame transformation. We note that in Ref.~\cite{QD_2023}, the authors did not consider a time frame transformation. 

Recall that in Eq.~\ref{eq:analytical_pe} we have landed on the leakage error with valid approximations
\begin{equation} \label{eq:pe_approx_repeated}
    P_e = \frac{\displaystyle \bigg|\int \frac{\textrm{d}\theta}{\textrm{d}t}e^{-i\int^t \omega(t')\textrm{d}t'} \textrm{d}t \bigg|^2}{4} \ .
\end{equation}
This formula for leakage error is quite complicated as there exists another integral within an integral and one of the integrals is itself an imaginary exponent. However, in the event that $\varepsilon(t)$ only changes slightly, i.e., $\displaystyle \max_t{|\varepsilon(t)|}-\min_t{|\varepsilon(t)|} \approx 0$, $\omega(t') \approx \omega_x$ with $\omega_x$ a constant. We can therefore make a further approximation
\begin{equation} \label{eq:pe_with_wx}
    P_e = \frac{\displaystyle \bigg|\int \frac{\textrm{d}\theta}{\textrm{d}t}e^{-i\omega_x t} \textrm{d}t \bigg|^2}{4} \ .
\end{equation}
The term inside $|\cdot|$ in Eq.~\ref{eq:pe_with_wx} is nothing but the Fourier transform of $\textrm{d}\theta/\textrm{d}t$ evaluated at $\omega_x$. This is great because we have now a very simple evaluation of the leakage error in terms of the control trajectory $\textrm{d}\theta/\textrm{d}t$ during the process. In order to generalize the simple form to an arbitrary $\varepsilon(t)$, a nonlinear time frame $\tau$ is introduced, where at any time $t$
\begin{equation} \label{eq:tau_constraint}
    \omega_\tau \textrm{d}\tau = \omega(t)\textrm{d}t \ ,
\end{equation}
with $\omega_\tau$ a constant. Clearly, $\tau=\tau(t)$ is some nonlinear function of $t$.
Plug Eq.~\ref{eq:tau_constraint} into Eq.~\ref{eq:pe_approx_repeated} and we have
\begin{equation}
\begin{split}
    P_e &= \frac{\displaystyle \bigg|\int \frac{\textrm{d}\tilde\theta}{\textrm{d}\tau}e^{-i\int^\tau \omega_\tau\textrm{d}\tau'} \textrm{d}\tau\bigg|^2}{4} \\ &= \frac{\displaystyle \bigg|\int \frac{\textrm{d}\tilde\theta}{\textrm{d}\tau}e^{-i\omega_\tau \tau} \textrm{d}\tau \bigg|^2}{4} \ .
\end{split}
\end{equation}
In this way, we achieve a simple form of the leakage error as in Eq.~\ref{eq:pe_with_wx}, and can design $\textrm{d}\tilde\theta/\textrm{d}\tau$ and hence $\tilde\theta(\tau)$ in the nonlinear time frame $\tau$ using approaches proposed in Section~\ref{sec: Problem formulation} and \ref{sec: Optimal solutions}.
In Eq.~\ref{eq:tau_constraint}, if we rearrange the terms by dividing both sides by $\omega(t)$, we have
\begin{equation}
    \textrm{d}t = \frac{\omega_\tau}{\omega(t)}\textrm{d}\tau = \frac{\omega_\tau}{\omega(\tau)}\textrm{d}\tau \ ,
\end{equation}
where we change the variable of $\omega(\cdot)$ from $t$ to $\tau$.
Integrate both sides and we have
\begin{equation}
    t(\tau) = \int_0^\tau \textrm{d}t = \int_0^\tau \frac{\omega_\tau}{\omega(\tau')}\textrm{d}\tau' \ .
\end{equation}
If we further set $\omega_\tau = \Delta$, we will have 
\begin{equation}
    \frac{\omega_\tau}{\omega(\tau')} = \frac{\Delta}{\omega(\tau')} = \sin\tilde\theta(\tau') \ ,
\end{equation}
where $\tilde\theta(\tau)$ is already known by design. Then we compute
\begin{equation} \label{eq:t_tau}
    t(\tau) = \int_0^\tau \sin\tilde\theta(\tau')\textrm{d}\tau' \ .
\end{equation}
Now that we have $\tilde\theta(\tau)$ and $t(\tau)$, we can numerically solve for $\theta(t)$ in the original time frame $t$.

\section{Validity of Eq.~\ref{eq:leakage_nonlinear_time_frame}} \label{app: validity}

We evaluate the validity of using Eq.~\ref{eq:leakage_nonlinear_time_frame} with the nonlinear time-frame transformation as discussed in Appendix~\ref{app: Nonlinear time frame transformation} for the leakage error. We proceed by comparing the analytically calculated leakage error $P_{e-\textrm{ana}}$ and the leakage error $P_{e-\textrm{sim}}$ by simulating a two-level system, using examples of the Slepian-based trajectory and the Chebyshev-based trajectory.

\begin{figure}[tb]
    \centering  \includegraphics[width=0.49\textwidth]{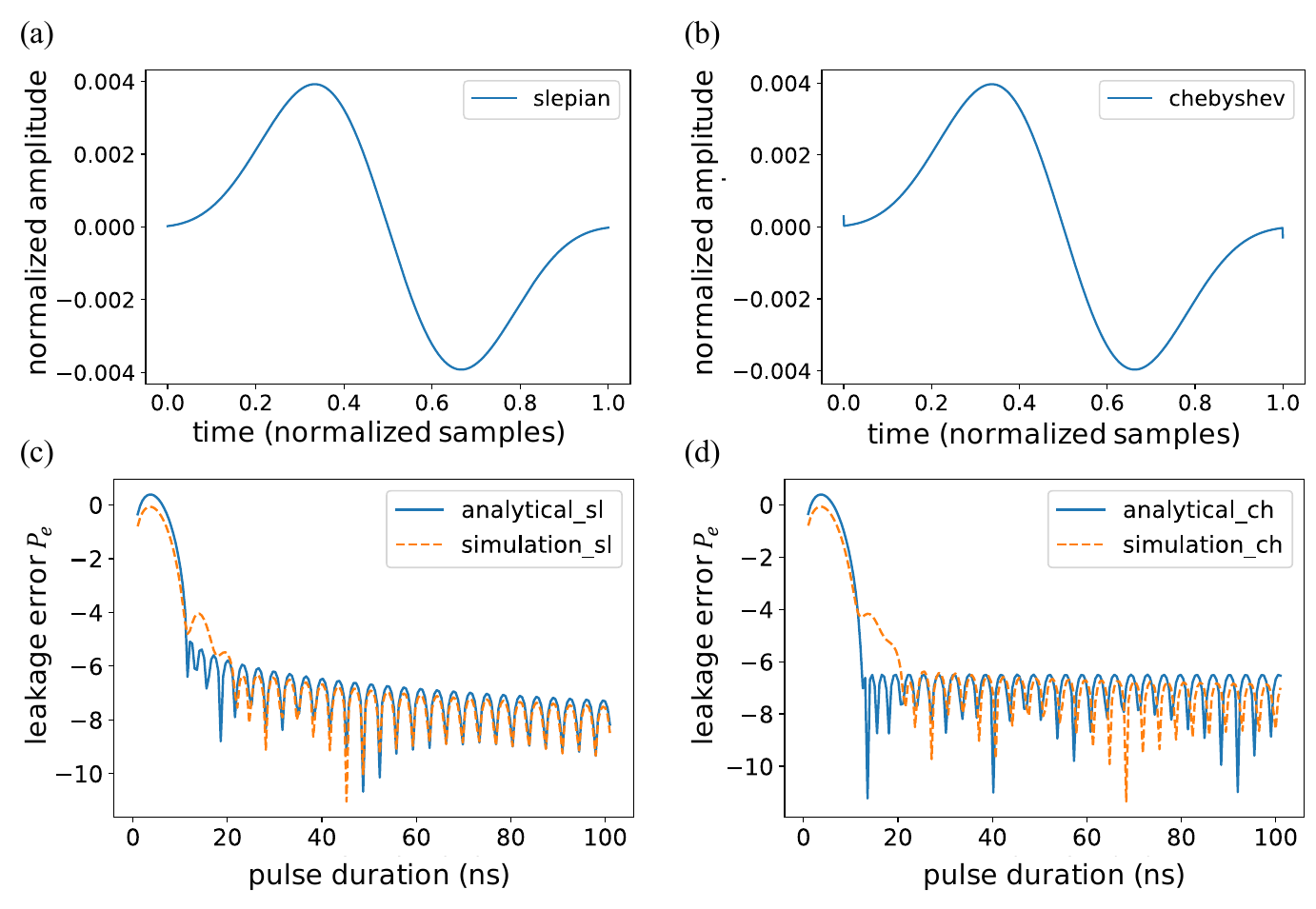}
    \caption{Comparison of analytically calculated leakage error $P_{e-\textrm{ana}}$ by Eq.~\ref{eq:leakage_nonlinear_time_frame} with the nonlinear time-frame transformation and leakage error $P_{e-\textrm{sim}}$ by simulating a two-level system. (a) Time-domain representation of an example of the Slepian-based trajectory. (b) Time-domain representation of an example of the Chebyshev-based trajectory. (c) $P_{e-\textrm{ana}}$ v.s. $P_{e-\textrm{sim}}$ as a function of pulse duration using (a). (d) $P_{e-\textrm{ana}}$ v.s. $P_{e-\textrm{sim}}$ as a function of pulse duration using (b).}
    \label{fig:app_validity}
\end{figure}

In Figs.~\ref{fig:app_validity}\panel{a}-\panel{b}, we show the time-domain representations of two control trajectories, namely an example of the Slepian-based trajectory and an example of the Chebyshev-based trajectory. We first calculate the analytical $P_{e-\textrm{ana}}$ using Eq.~\ref{eq:leakage_nonlinear_time_frame} considering the nonlinear time-frame transformation. The calculation is performed for a range of pulse duration. Then we perform a CZ gate type simulation (except that we do not consider the accumulation of a certain phase) using a two-level system and keep track of the leakage error throughout the process. The simulation is also performed for a range of pulse duration. In Figs.~\ref{fig:app_validity}\panel{c}-\panel{d}, we show the comparison of the analytically calculated leakage error $P_{e-\textrm{ana}}$ and simulated leakage error $P_{e-\textrm{sim}}$ for the two control trajectories in Figs.~\ref{fig:app_validity}\panel{a}-\panel{b} respectively. We observe that $P_{e-\textrm{sim}}$ (Slepian) manifests a feature of monotonously decreasing sidelobes as predicted by $P_{e-\textrm{ana}}$ (Slepian), while $P_{e-\textrm{sim}}$ (Chebyshev) characterizes relatively flat (slightly decreasing) sidelobes which should be exactly equiripple as predicted by $P_{e-\textrm{ana}}$ (Chebyshev). In addition, the sidelobes of $P_{e-\textrm{ana}}$ and $P_{e-\textrm{sim}}$ of both examples oscillate at a very close, if not exactly the same, frequency, which indicates the validity of Eq.~\ref{eq:leakage_nonlinear_time_frame} in the sidelobe region. However, the validity of Eq.~\ref{eq:leakage_nonlinear_time_frame} in the mainlobe region appears compromised. This is because we set $\Delta=50$ MHz, and when the pulse duration is less than approximately $1/\Delta=20$ ns, the whole process is essentially not in the adiabatic limit. This feature is not of too much concern since what we really care about is the sidelobe characteristic when designing and comparing the Slepian and the Chebyhsev trajectories so that we can push for a faster gate while keeping the process in the adiabatic regime. Also, we should note that the absolute value of $P_{e-\textrm{ana}}$ is not meaningful since it is calculated using a normalized control trajectory. We require control trajectories to be subject to the same normalization as discussed in Seciton~\ref{sec:problem_statement} to make sure the comparison is valid.

\section{Formula for accumulated phase} \label{sec:accumulated phase}

As discussed in Section~\ref{sec:physics_background}, the primary characteristic of the CPHASE gate is to accumulate some phase $\phi$. We derive a formula for the accumulated phase $\phi$ in the abstracted two-level system.

Recall the Hamiltonian defined in Eq.~\ref{hamiltonian} and the eigenenergies defined in Eq.~\ref{eq:eigen_states}. Now let $\Delta=0$ such that no coupling exists between the two levels. The eigenenergies are $E_{11}'=-\varepsilon(t)/2$ and $E_{20}'=\varepsilon(t)/2$, as depicted by the dashed lines in Fig.~\ref{fig:p2}. The difference between the eigenenergies of the ground state $\Delta E = E_{11}-E_{11}'$ with and without coupling is what determines the phase accumulation in the CPHASE gate. We can recast the eigenenergy difference in terms of $\theta(t)$
\begin{equation}
\begin{split}
    \Delta E &= E_{11}-E_{11}' = \frac{1}{2}(\varepsilon(t)-\sqrt{\varepsilon(t)^2+\Delta^2}) \\
    &= - \frac{\Delta}{2} \tan \frac{\theta(t)}{2} \ .
\end{split}
\end{equation}

The accumulated phase $\phi$ is the integral of the energy difference $\Delta E$ through the trajectory
\begin{equation}
    \phi= \int \Delta E \textrm{d}t = - \int \frac{\Delta}{2} \tan \frac{\theta(t)}{2} \textrm{d}t \ .
\end{equation}
    
By designing the shape of the control trajectory for $\theta(t)$, we can in principle apply an arbitrary CPHASE gate.

\section{The Slepian pulses and the Chebyshev pulses I and II} \label{app: Slepian and Chebyshev}
\subsection{The Slepian pulses}

The Slepian pulses, also known as discrete prolate spheroidal sequences (DPSSs), are a set of orthogonal pulses intended for the problem of maximal concentration in both the time domain and the frequency domain. Heisenberg’s uncertainty principle implies that pulses cannot be confined in both the time domain and the frequency domain. It is then natural to ask: how to optimally concentrate the energy in one domain if the pulse is strictly confined in the other domain. This problem, both in continuous time and in discrete time, was pursued and solved by Slepian, Landau, and Pollack~\cite{slepian_prolate_1961,landau_prolate_1961,landau_prolate_1962,slepian_prolate_1964,slepian_prolate_1978,slepian_comments_1983}. Here we briefly review the development and analysis of the Slepian pulses for the discrete-time case.

Consider a finite-length, discrete-time pulse $x[n], \ n=0,1,\dots,N-1$, which is specified to have finite energy, i.e., 
\begin{equation}
    E = \sum_{n=0}^{N-1}|x[n]|^2<\infty \ ,
\end{equation}
where $E$ denotes the energy of the pulse $x[n]$.

The discrete-time Fourier transform of the pulse $x[n]$ is given by
\begin{equation}
    X(e^{i\omega})=\sum_{n=0}^{N-1}x[n]e^{-i\omega n} \ .
\end{equation}

Let $0 < W < \pi$. The ratio $\lambda \in [0,1]$ that measures the percentage of the energy contained in the frequency band $[-W,W]$ over the total energy is defined as
\begin{equation} \label{eq:slepian_ratio}
    \lambda = \frac{\displaystyle \int_{-W}^{W}|X(e^{i\omega})|^2 d\omega}{\displaystyle \int_{- \pi}^{\pi}|X(e^{i\omega})|^2 d\omega} \ .
\end{equation}
The goal is to find the pulse $x[n]$ that maximizes $\lambda$ for all pulses $x[n],\ n=0,1,\dots,N-1$ of length $N$.

The Slepian pulses $\{v_n^{(k)}(N,W), \ k=0,1,\dots,N-1\}$ are the solutions to the optimization problem stated above~\cite{slepian_prolate_1978}, where $n=0,1,\dots,N-1$ is the index of the pulse, $k$ is the order of each pulse, and $N$ and $W$ are parameters referred to as the length and mainlobe width of the pulse, respectively. The Slepian pulses can be derived from the real solutions to the system of equations
\begin{equation} \label{eq:slepian_sequence}
    \sum_{m=0}^{N-1}\frac{\sin{2\pi W (n-m)}}{\pi (n-m)}v_m^{(k)}(N,W) = \lambda_k(N,W)v_n^{(k)}(N,W) \ ,
\end{equation}
for $n, m=0,1,\dots,N-1$ and $k=0,1,\dots,N-1$. When $n=m$, this simplifies to
\begin{equation}
    \frac{\sin{2\pi W (n-m)}}{\pi (n-m)}=2W \ .
\end{equation}
Eqs.~\ref{eq:slepian_sequence} can also be written in the matrix form
\begin{equation} \label{eq:slepian_matrix}
    {\bf A} {\bf v}^{(k)} = \lambda_k {\bf v}^{(k)} \ ,
\end{equation}
where 
\begin{subequations}
\begin{equation}
    {\bf A}_{n,m}=\frac{\sin{2\pi W (n-m)}}{\pi (n-m)} \ ,
\end{equation}
\begin{equation}
    {\bf v}^{(k)} = [v_0^{(k)}(N,W),v_1^{(k)}(N,W),\dots,v_{N-1}^{(k)}(N,W)]^T \ .
\end{equation}
\end{subequations}

Eq.~\ref{eq:slepian_matrix} is an eigenvalue problem, where $\lambda_k$'s are the $N$ distinct eigenvalues and ${\bf v}^{(k)}$'s are the corresponding eigenvectors. By convention the eigenvalues are ranked as $1>\lambda_0>\lambda_1>\dots>\lambda_{N-1}>0$. Therefore, the sequence $v_n^{(0)}(N,W)$ that corresponds to the largest eigenvalue $\lambda_0$ is referred to as the first Slepian pulse. Each successive Slepian pulse maximizes $\lambda$ while being orthogonal to the Slepian pulses preceding it.

\begin{figure}[tb]
    \centering  \includegraphics[width=0.5\textwidth]{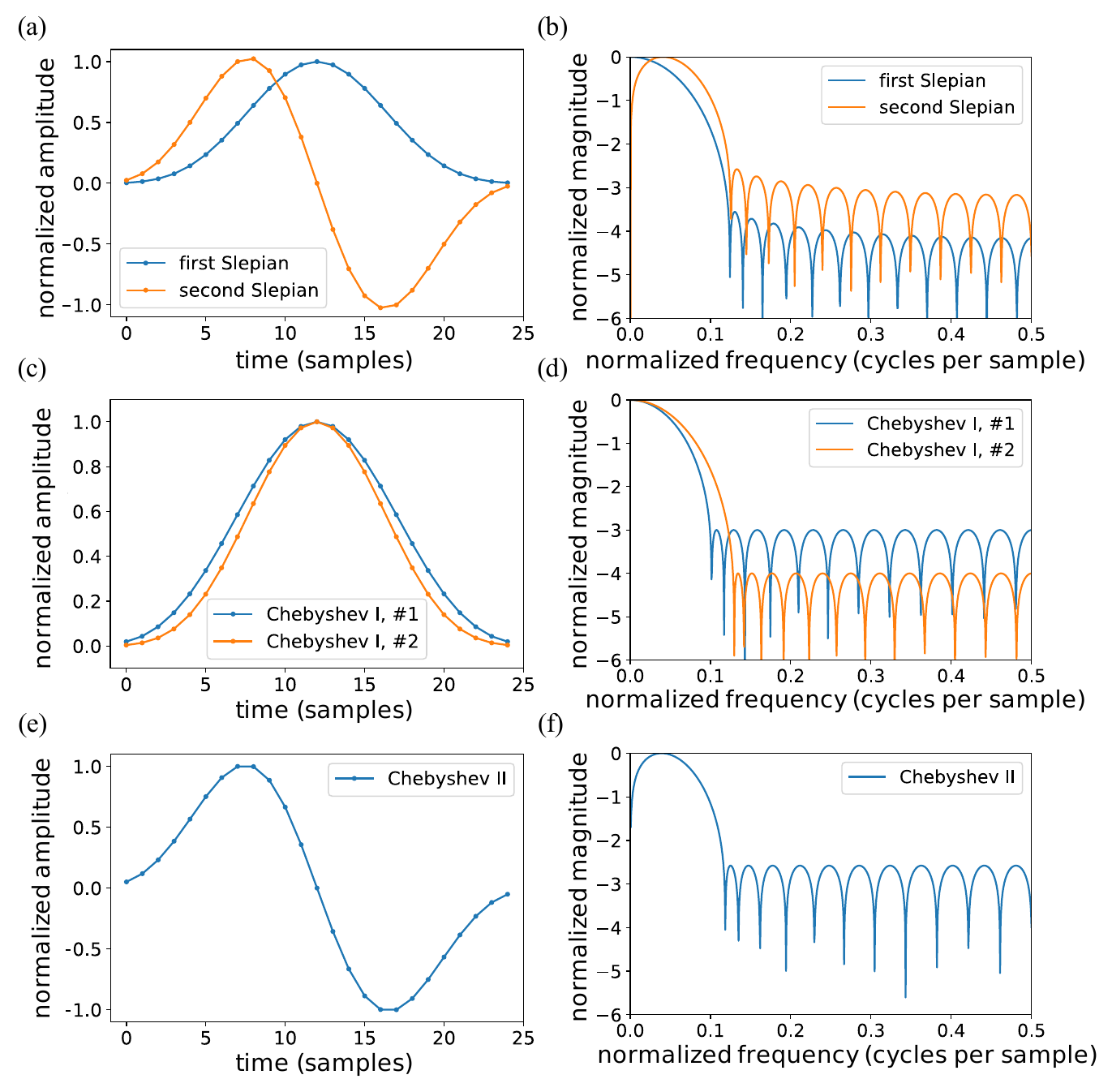}
    \caption{(a)(b) Time-domain and frequency-domain representations of the first and second Slepian pulses for $N=25$ and $NW=3$. (c)(d) Time-domain and frequency-domain representations of the Chebyshev pulses I for $N=25$ and sidelobe amplitudes specified to be $10^{-3}$ and $10^{-4}$. (e)(f) Time-domain and frequency-domain representations of a Case 3 pulse for $N=25$ using weighted Chebyshev approximation, an example of what we refer to as the Chebyshev pulses II.}
    \label{fig:p0}
\end{figure}

The time and frequency-domain representations of the first and second Slepian pulses for $N=25$ and $NW=3$ are shown in Figs.~\ref{fig:p0}\panel{a}-\panel{b}. The magnitude of the Fourier transform is normalized to be 1 at $\omega=0$ for the first Slepian pulse, while for the second Slepian pulse, it is normalized so that the peak magnitude of the mainlobe is 1. Compared to the rectangular pulse and raised cosine pulses, we find that the Slepian pulses have a relatively low sidelobe amplitude and small mainlobe width, which makes them a good candidate when a compromise between the sidelobe amplitude and mainlobe width is required.

To maintain consistent notation, we denote the first Slepian pulse ($k=0$) as 
\begin{equation}
    w_{\textrm{sl1}}^{NW}[n]=\begin{cases}
        v_n^{(0)}(N,W), & \ \ 0\leq n\leq N-1 \\
        0, & \ \ \textrm{otherwise}
        \end{cases} \ ,
\end{equation}
and the second Slepian pulse ($k=1$) as 
\begin{equation}
    w_{\textrm{sl2}}^{NW}[n]=\begin{cases}
        v_n^{(1)}(N,W), & \ \ 0\leq n\leq N-1 \\
        0, & \ \ \textrm{otherwise}
        \end{cases} \ ,
\end{equation}
where the superscript $NW$ may be omitted when it is given in context.

\subsection{The Chebyshev pulses} \label{sec:cheby_I}

Dolph formulated and solved the problem of finding a pulse that minimizes the mainlobe width given a specified sidelobe amplitude (or vice versa), in the context of antenna array design~\cite{dolph_current_1946}. The optimal solution to this problem is known as the Chebyshev pulse.

The Chebyshev pulse is based on the Chebyshev polynomials of the first kind defined as
    \begin{equation}
    T_n(x)=\begin{cases}
        \cos({n \arccos{(x)}}) & \ \ |x|\leq 1 \\
        \cosh({n \arccosh{(x)}}) & \ \ x\geq 1 \\
       (-1)^n\cosh({n \arccosh{(-x)}}) & \ \ x\leq -1 
    \end{cases} \ ,
    \end{equation}
where $n$ denotes the order of the Chebyshev polynomials. Plugging in the values $n=0$ and $n=1$, we have $T_0(x)=1$ and $T_1(x)=x$. Using the double angle trigonometric identity, i.e., $\cos{2\theta}=2\cos^2{\theta}-1$ or $\cosh{2\theta}=2\cosh^2{\theta}-1$, the following recurrence relation can be verified
\begin{equation}
    T_n(x)=2xT_{n-1}(x)-T_{n-2}(x), \ \ \ n \geq 2 \ .
\end{equation}
It can be further shown that $T_n(x)$ is an $n$th-order polynomial in $x$, i.e., $T_n(x)$ can be equivalently written as the ordinary polynomial
\begin{equation}
    T_n(x)=\sum_{k=0}^{n}b[k]x^k \ ,
\end{equation}
for some coefficients $b[k], \ k=0,1,\dots,n$. $T_n(x)$ is even or odd according to whether $n$ is even or odd. $T_n(x)$ oscillates between $-1$ and $1$ when $-1\leq x\leq 1$ and is monotonic when $x\geq1$ or $x\leq -1$.

The Chebyshev pulse $w_{\textrm{ch1}}[n]$ can be defined through its Fourier transform
\begin{equation} \label{eq:cheby_freq}
    W_{\textrm{ch1}}(e^{i\omega}) = e^{-i\omega\frac{N-1}{2}} \frac{T_{N-1}(x_0\cos{(\omega/2)})}{T_{N-1}(x_0)} \ ,
\end{equation}
where $N$ denotes the length of the pulse, and $x_0>1$ is a parameter related to the sidelobe amplitude of $W_{\textrm{ch1}}(e^{i\omega})$. Let $\omega_s$ be such that $x_0\cos{(\omega_s/2)}=1$. As $\omega$ increases from $0$ to $\omega_s$, the argument of the numerator in Eq.~\ref{eq:cheby_freq}, i.e., $x_0\cos{(\omega/2)}$, decreases from $x_0$ to $1$, and thus $W_{\textrm{ch1}}(e^{i\omega})$ decreases from $1$ to $\frac{1}{T_{N-1}(x_0)}:=r$. As $\omega$ increases from $\omega_s$ to $\pi$, $W_{\textrm{ch1}}(e^{i\omega})$ will oscillate between $-r$ and $r$.

Utilizing trigonometric identities and considering that $T_n(x)$ is an $n$th-order polynomial in $x$, it can be shown that Eq.~\ref{eq:cheby_freq} can further be written in a more structured form
\begin{equation}
    W_{\textrm{ch1}}(e^{i\omega}) = \sum_{n=0}^{N-1}w_{\textrm{ch1}}[n]e^{-i\omega n} \ ,
\end{equation}
where $w_{\textrm{ch1}}[n],\ n=0,1,\dots,N-1$ are the coefficients of the Chebyshev pulse. The Chebyshev pulse coefficients can also be evaluated from the inverse Fourier transform of Eq.~\ref{eq:cheby_freq}. The explicit analytical formula is given by
\begin{equation}
\begin{split}
    w_{\textrm{ch1}}[n]=\frac{1}{N}\bigg[&1+2r\sum_{k=0}^{N_s}(-1)^k T_{N-1}\Big(x_0 \cos{\frac{\pi k}{N}}\Big) \\ &\cos{\Big(\frac{2\pi k}{L}(n+\frac{1}{2})\Big)}\bigg], \ \ \ n=0,1,\dots,N-1 \ ,
\end{split}
\end{equation}
where $r=\frac{1}{T_{N-1}(x_0)}$ is as defined earlier, and
\begin{equation}
    N_s=\begin{cases}
        \frac{N-1}{2} & \ \ N  \text{ odd} \\
        \frac{N}{2}-1 & \ \ N \text{ even} 
    \end{cases} \ .
\end{equation}

The time-domain and frequency-domain representations of the Chebyshev pulses for $N=25$ and different specified sidelobe amplitudes ($10^{-3}$ for Chebyshev I, $\texttt{\#}1$ and $10^{-4}$ for Chebyshev I, $\texttt{\#}2$) are shown in Fig.~\ref{fig:p0}\panel{c}-\panel{d}. The magnitude of the Fourier transform is normalized to be 1 at $\omega=0$. One important characteristic of the Chebyshev pulse is the equiripple sidelobe amplitude for all sidelobes. From Fig.~\ref{fig:p0}\panel{d}, we can observe that as the sidelobe amplitude of the Chebyshev pulse is specified to be lower, its mainlobe width will be larger. In Appendix~\ref{app: Weighted Chebyshev approximation}, we will show that the Chebyshev pulse is a special case of the result of the weighted Chebyshev approximation. In the following sections, when it is necessary to discriminate between the Chebyshev pulse discussed in this section and the Chebyshev pulse II to be introduced in Section~\ref{sec:Chebyshev pulses II}, we will refer to the Chebyshev pulse as the Chebyshev pulse I to avoid confusion.

\subsection{The Chebyshev pulses II} \label{sec:Chebyshev pulses II}

We define an anti-symmetric counterpart of the Chebyshev pulse I $w_{\textrm{ch1}}[n]$ which we refer to as the Chebyshev pulse II $w_{\textrm{ch2}}^{\beta}[n]$, using the weighted Chebyshev approximation, a method for finding a polynomial that best approximates a given function in a weighted sense. Here, $\beta$ denotes the parameters that we feed into the weighted Chebyshev approximation problem. See Appendix~\ref{app: Weighted Chebyshev approximation} for more details. For simplicity, we will omit $\beta$. The Chebyshev pulse II shares the same characteristics of equiripple sidelobe amplitude and only one ripple in the passband as the Chebyshev pulse I. The time-domain and frequency-domain representations of an example of the Chebyshev pulse II for $N=25$ are shown in Figs.~\ref{fig:p0}\panel{e}-\panel{f}. The magnitude of the Fourier transform is normalized so that the peak magnitude of the mainlobe is 1. 

\begin{figure}[tb]
    \centering  \includegraphics[width=0.49\textwidth]{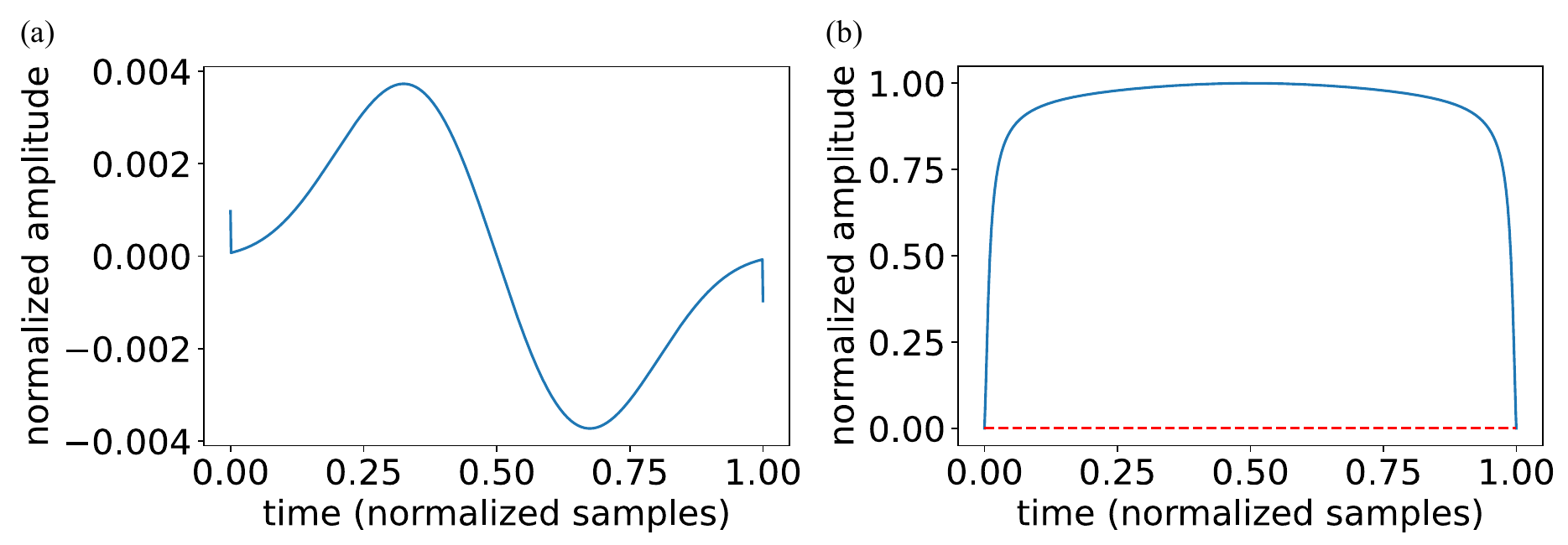}
    \caption{Time-domain and frequency-domain representations of a Case 1 pulse for $N=25$ using weighted Chebyshev approximation (WCA) to coincide with the Chebyshev pulse I for $N=25$ with sidelobe amplitude specified to be $10^{-3}$.}
    \label{fig:chebyimpulses}
\end{figure}

We note a special feature of the Chebyshev pulses I and II. The equiripple property in the frequency domain is enforced by the specifications of the Chebyshev pulses. Nonetheless, it carries the potential drawback of introducing ``impulses'' at the window endpoints. For example, in Figs.~\ref{fig:p0}\panel{c},~\ref{fig:p0}\panel{e}, both endpoints of each of the Chebyshev pulses I and II are not approaching zero. The same is true for the first and second Slepian pulses, i.e., their endpoints are not zero either. In other cases, such as what we show in Section~\ref{sec: Optimal solutions}, this feature can be more notable. For analysis purposes, it is an important feature because this is the primary reason for the equiripple property. However, as we transform the Chebyshev pulses II into the change of qubit frequency in simulation, the effect of these ``impulses'' will diminish due to interpolation and integration. For example, Figs.~\ref{fig:chebyimpulses}\panel{a}-\panel{b} show a Chebyshev-based trajectory and its corresponding $\varepsilon_{\textrm{ch2}}[n]$, which is the control pulse for the CPHASE gate. The ``impulses'' do not manifest in $ \varepsilon_{\textrm{ch2}}[n]$. This is good because we cannot implement sharp jumps in frequency changes of qubits.

\section{Weighted Chebyshev approximation} \label{app: Weighted Chebyshev approximation}
We review the basics of the weighted Chebyshev approximation (WCA) in the context of finite-length, discrete-time pulse design. 

Let $h[n],\ n=0,1,\dots,N-1$, be a real-valued finite-length, discrete-time pulse of length $N$ defined over the discrete-time interval $0 \leq n \leq N-1$. The Fourier transform of $h[n]$ is
\begin{equation}
    H(e^{i\omega})=\sum_{n=0}^{N-1}h[n]e^{-i\omega n} \ .
\end{equation}
$H(e^{i\omega})$ can also be written in terms of its amplitude and phase
\begin{equation}
    H(e^{i\omega}) = A(\omega)e^{i\phi(\omega)} \ ,
\end{equation}
where $A(\omega)$ and $\phi(\omega)$ are real-valued functions of $\omega$. 

We further require that $h[n]$ be symmetric or anti-symmetric. Here, when $h[n]$ is referred to as being symmetric, it means
\begin{equation}
    h[n]=h[N-1-n], \ \ \ n=0,1,\dots,N-1 \ .
\end{equation}
Similarly, when $h[n]$ is referred to as being anti-symmetric, it means
\begin{equation}
    h[n]=-h[N-1-n], \ \ \ n=0,1,\dots,N-1 \ .
\end{equation}

Depending on the value of $N$ being odd or even and $h[n]$ being symmetric or anti-symmetric, there exist four cases of pulses $h[n]$. With the symmetry constraints, it can be shown that $\phi(\omega)$ can be written in the form of $\phi(\omega) = C+B\omega$, which is a linear function of $\omega$, where $C$ and $B=-\frac{N-1}{2}$ are real-valued. Therefore, the Fourier transform of the four cases of pulses can be written in the form
\begin{equation}
    H(e^{i\omega}) = A(\omega) e^{iC}e^{iB\omega} \ .
\end{equation}

Values of $C$ and forms of $A(\omega)$ are given in Table~\ref{table:four_linear_phase}.

\begin{table}[ht!]
\centering
\begin{tabular}{|| c | c | c ||} 
 \hline
  & $C$ & $A(\omega)$  \\ [0.5ex] 
 \hline
 \makecell{Case 1: \\ $N$ odd,  $h[n]$ symmetric} & 0 & $\displaystyle \sum_{n=0}^{\frac{N-1}{2}}a[n]\cos{(\omega n)}$  \\ [0.5ex] \hline
 \makecell{Case 2: \\ $N$ even, $h[n]$ symmetric} & 0 & $\displaystyle \sum_{n=0}^{\frac{N}{2}}b[n]\cos{(\omega (n-1/2))}$  \\ [0.5ex] \hline
 \makecell{Case 3: \\ $N$ odd, $h[n]$ anti-symmetric} & 1 & $\displaystyle \sum_{n=0}^{\frac{N-1}{2}}c[n]\sin{(\omega n)}$  \\ [0.5ex] \hline
 \makecell{Case 4: \\ $N$ even, $h[n]$ anti-symmetric} & 1 & $\displaystyle \sum_{n=0}^{\frac{N}{2}}d[n]\sin{(\omega (n-1/2))}$ \\[0.5ex] \hline
\end{tabular}
\caption{Values of $C$ and forms of $A(\omega)$ for the four cases of pulses. Here, $a[n], b[n], c[n], d[n]$ are coefficients that can be determined given $h[n]$.}
\label{table:four_linear_phase}
\end{table}

Note that the forms of $A(\omega)$ are either a sum of cosines or sines, with the argument being either $\omega n$ or $\omega (n-1/2)$. Utilizing basic trigonometric identities, the forms of $A(\omega)$ for all four cases can be rewritten in the form $A(\omega)=Q(\omega)P(\omega)$, where $Q(\omega)$ is specific to each case and $P(\omega)$ is always a sum of cosines. Forms of $Q(\omega)$ and $P(\omega)$ are given in Table~\ref{table:four_linear_phase_q_p}.

\begin{table}[ht!]
\centering
\begin{tabular}{|| c | c | c ||} 
 \hline
  & $Q(\omega)$ & $P(\omega)$  \\ [0.5ex] 
 \hline
 \makecell{Case 1: \\ $N$ odd,  $h[n]$ symmetric} & 1 & $\displaystyle \sum_{n=0}^{\frac{N-1}{2}}\bar{a}[n]\cos{(\omega n)}$  \\ [0.5ex] \hline
 \makecell{Case 2: \\ $N$ even, $h[n]$ symmetric} & $\cos{(\omega/2)}$ & $\displaystyle \sum_{n=0}^{\frac{N}{2}-1}\bar{b}[n]\cos{(\omega n)}$ \\ [0.5ex] \hline
 \makecell{Case 3: \\ $N$ odd, $h[n]$ anti-symmetric} & $\sin{(\omega)}$ & $\displaystyle \sum_{n=0}^{\frac{N-3}{2}}\bar{c}[n]\cos{(\omega n)}$  \\ [0.5ex] \hline
 \makecell{Case 4: \\ $N$ even, $h[n]$ anti-symmetric} & $\sin{(\omega/2)}$ & $\displaystyle \sum_{n=0}^{\frac{N}{2}-1}\bar{d}[n]\cos{(\omega n)}$ \\[0.5ex] \hline
\end{tabular}
\caption{Forms of $Q(\omega)$ and $P(\omega)$ for the four cases of pulses. Here, $\bar{a}[n], \bar{b}[n], \bar{c}[n], \bar{d}[n]$ are coefficients that can be determined given $h[n]$. For Case 1, we have $\bar{a}[n]=a[n]$.}
\label{table:four_linear_phase_q_p}
\end{table}

Having established the notations, the Chebyshev approximation problem may be stated as follows. Given a disjoint union of frequency bands of interest $\mathcal{F} \subset [0,\pi]$, a desired function $D(\omega)$ defined and continuous on $\mathcal{F}$, a positive weighting function $W(\omega)$ defined and continuous on $\mathcal{F}$, and a desired choice of one of the four cases of $h[n]$, the minimum of the following quantity
\begin{equation} \label{eq:minimax_opt}
    \displaystyle ||E(\omega)||:=\max_{\omega \in \mathcal{F}} W(\omega)|D(\omega)-A(\omega)| \ ,
\end{equation}
and the corresponding $h[n]$ are desired. Here, $E(\omega):=W(\omega)|D(\omega)-A(\omega)|$ is referred to as the weighted approximation error and the optimization problem is a minimax problem of $E(\omega)$.

Considering we have the form $A(\omega)=Q(\omega)P(\omega)$, we can rewrite the weighted approximation error as
\begin{align}
    E(\omega)&=W(\omega)|D(\omega)-A(\omega)| \\
    &=W(\omega)|D(\omega)-Q(\omega)P(\omega)| \\
    &=W(\omega)Q(\omega)\bigg|\frac{D(\omega)}{Q(\omega)}-P(\omega)\bigg|  \ . \label{eq:devide_by_q}
\end{align}
Note that Eq.~\ref{eq:devide_by_q} is valid except possibly at $\omega = 0$ or $\pi$. To avoid those scenarios where $Q(\omega)=0$, it suffices to restrict that $\mathcal{F} \subset [0,\pi)$ for Case 2 problems, $\mathcal{F} \subset (0,\pi)$ for Case 3 problems, and $\mathcal{F} \subset (0,\pi]$ for Case 4 problems.

Let $\hat{W}(\omega)=W(\omega)Q(\omega)$ and $\hat{D}(\omega)=D(\omega)/Q(\omega)$, and we have
\begin{equation} \label{eq:new_error_form}
    E(\omega)=\hat{W}(\omega)|\hat{D}(\omega)-P(\omega)| \ .
\end{equation}

With the form of weighted approximation error in Eq.~\ref{eq:new_error_form}, one algorithmic solution to the above mentioned problem makes use of the alternation theorem, the Remez exchange algorithm, and/or the Parks-McClellan algorithm~\cite{parks_chebyshev_1972,parks_program_1972,mcclellan_unified_1973}. Other solutions make use of linear programming with additional constraints~\cite{Rabiner1972,Rabiner1975}. We refer the readers to the included references for more details.

\begin{figure}[tb]
    \centering  \includegraphics[width=0.49\textwidth]{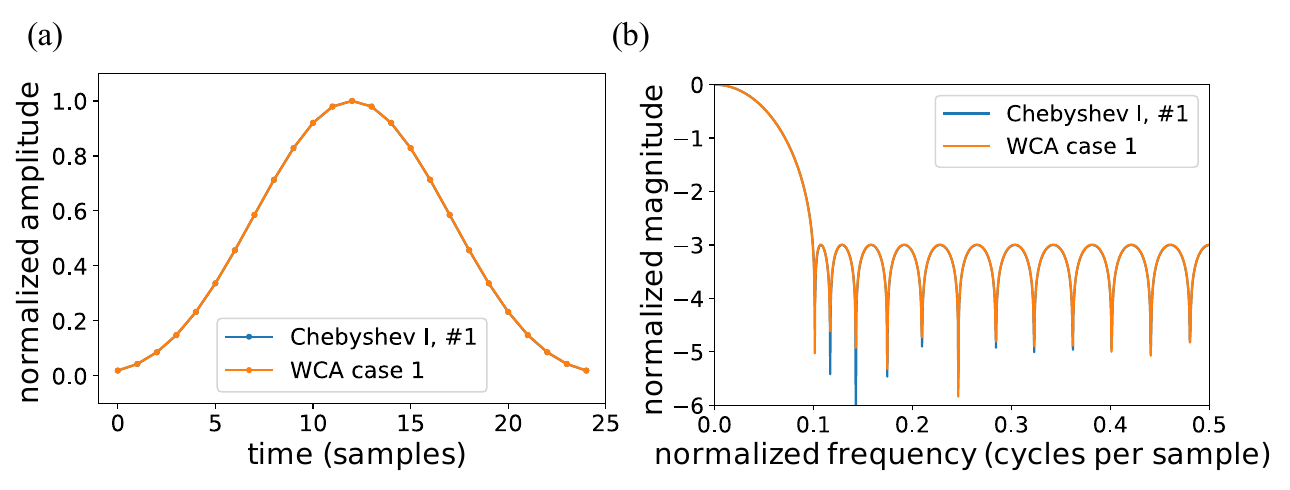}
    \caption{Time-domain and frequency-domain representations of a Case 1 pulse for $N=25$ using weighted Chebyshev approximation (WCA) to coincide with the Chebyshev pulse I for $N=25$ with sidelobe amplitude specified to be $10^{-3}$.}
    \label{fig:appa1}
\end{figure}

The solution for designing pulses in the minimax sense as in Eq.~\ref{eq:minimax_opt} is often in a numerical form without explicit analytical form. However, with a special set of $\mathcal{F} \subset [0,\pi]$, $D(\omega)$ and $W(\omega)$, and with $h[n]$ specified to be Case 1 or 2, the solution coincides with the Chebyshev pulse I discussed in Section~\ref{sec:cheby_I}. In other words, the Chebyshev pulse I is a special case in the weighted Chebyshev approximation problem. We refer the readers to Chapter 3 of Ref.~\cite{rabiner_theory_1975} for more details. Figs.~\ref{fig:appa1}\panel{a}-\panel{b} show the time-domain and frequency-domain representations of an example of using the weighted Chebyshev approximation (WCA) to design a pulse, which coincides with the Chebyshev pulse I with sidelobe amplitude specified to be $10^{-3}$ in Fig.~\ref{fig:p0}\panel{c}. Note that there is only one ripple in the passband, which is otherwise referred to as the mainlobe. This is also one of the reasons why we name the Chebyshev pulses II, since they are an anti-symmetric counterpart of the Chebyshev pulses I, and both can be considered as a special result of the weighted Chebyshev approximation problem.

If we provide an appropriate set of $\mathcal{F} \subset [0,\pi]$, $D(\omega)$ and $W(\omega)$, but specify $h[n]$ to be Case 3 or 4, the optimal solution to the weighted Chebyshev approximation problem will be an anti-symmetric counterpart of the Chebyshev pulse I $w_{\textrm{ch1}}[n]$, which we refer to as the Chebyshev pulse II $w_{\textrm{ch2}}^{\beta}[n]$, where $\beta = [\mathcal{F}, D(\omega), W(\omega)]$. Note that it is not necessarily true that the solution given by the weighted Chebyshev approximation will always be an instance of the Chebyshev pulse II, for any set of $\mathcal{F} \subset [0,\pi]$, $D(\omega)$ and $W(\omega)$. In order to find an appropriate Chebyshev pulse II $w_{\textrm{ch2}}[n]$, the parameters need to be properly chosen.
In Ref.~\cite{Ding2024}, the authors demonstrate the design process of $w_{\textrm{ch2}}[n]$ through an illustrative example.

\section{Phase accumulation and leakage error of the example in Section~\ref{sec:example1}} \label{app: Phase accumulation and leakage error of example}
\begin{figure*}[tb]
   \centering
   \includegraphics[width=0.9\textwidth]{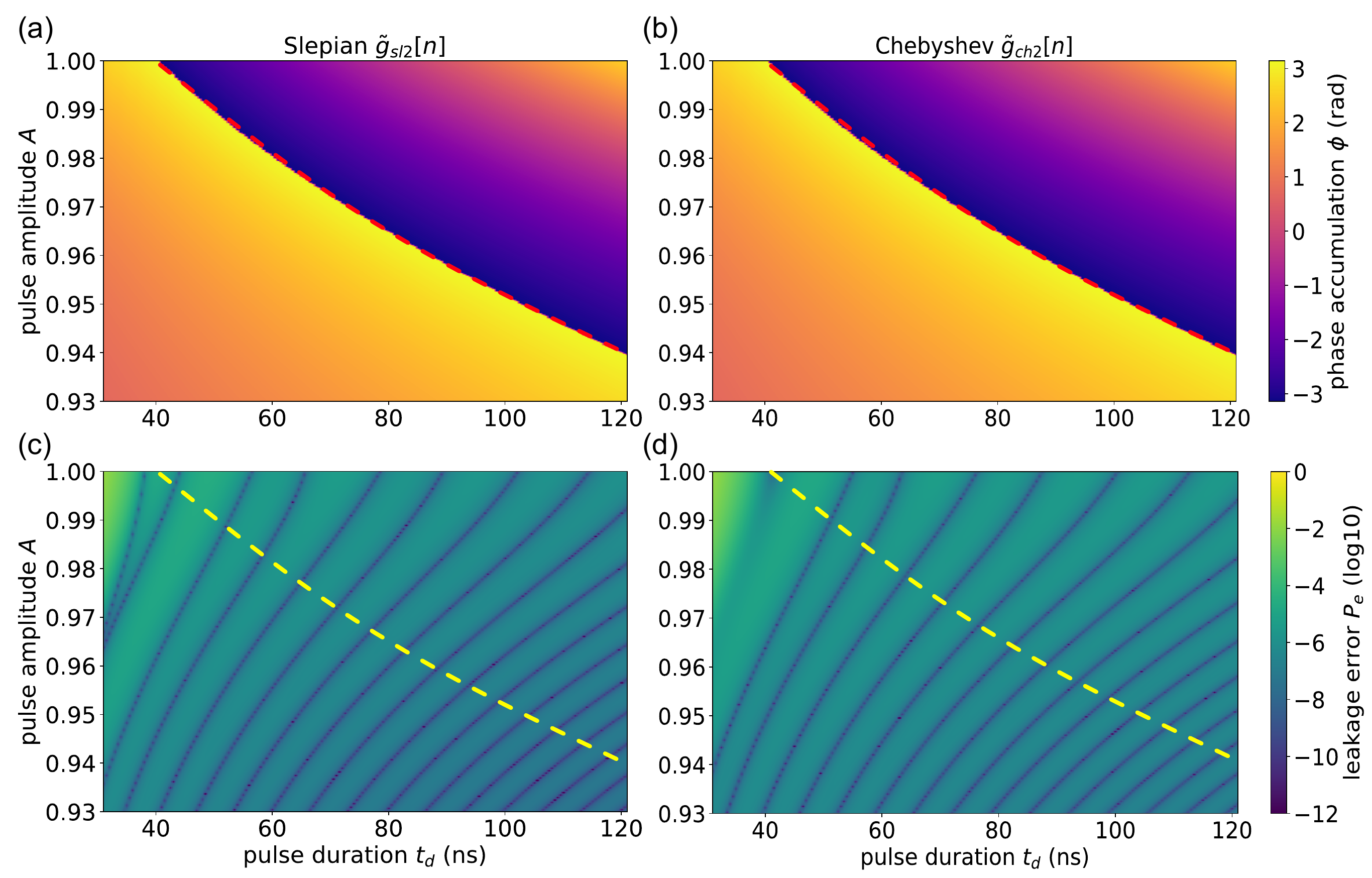}
   \caption{Phase accumulation and leakage error of the example in Section~\ref{sec:example1}. (a)(c) The phase accumulation and leakage error for a range of control pulse duration $t_d$ and amplitude $A$ using $\tilde g_{\textrm{sl2}}[n]$ as shown in Fig.~\ref{fig:p3}\panel{a}. The red dashed curve describes all the amplitude and duration pairs that result in the phase accumulation $\phi=\pi$. (b)(d) The phase accumulation and leakage error for a range of control pulse duration $t_d$ and amplitude $A$ using $\tilde g_{\textrm{ch2}}[n]$ as shown in Fig.~\ref{fig:p3}\panel{a}. The yellow dashed curve describes the corresponding leakage error data points to those amplitude and duration pairs.}
   \label{fig:phase_leakage_sim_example}
\end{figure*}

Figures~\ref{fig:phase_leakage_sim_example}\panel{a},~\ref{fig:phase_leakage_sim_example}\panel{c} present the phase accumulation and leakage error for a range of control pulse duration $t_d$ and amplitude $A$ using the $\tilde g_{\textrm{sl2}}[n]$ shown in Fig.~\ref{fig:p3}\panel{a}. As a comparison, Figs.~\ref{fig:phase_leakage_sim_example}\panel{b},~\ref{fig:phase_leakage_sim_example}\panel{d} present the phase accumulation and leakage error for a range of control pulse duration $t_d$ and amplitude $A$ using the same $\tilde g_{\textrm{ch2}}[n]$ as shown in Fig.~\ref{fig:p3}\panel{a}. Following the procedure in Section~\ref{sec:setting_procedure}, we first find all the amplitude and duration pairs that result in the phase accumulation $\phi=\pi$, described by the red dashed curve in Figs.~\ref{fig:phase_leakage_sim_example}\panel{a}-\panel{b}. Then we determine the corresponding leakage error data points described by the yellow dashed curve in Figs.~\ref{fig:phase_leakage_sim_example}\panel{c}-\panel{d}.

\section{Additional simulation results} \label{app: Additional simulation results}
\begin{figure*}[bt]
    \centering  \includegraphics[width=0.823\textwidth]{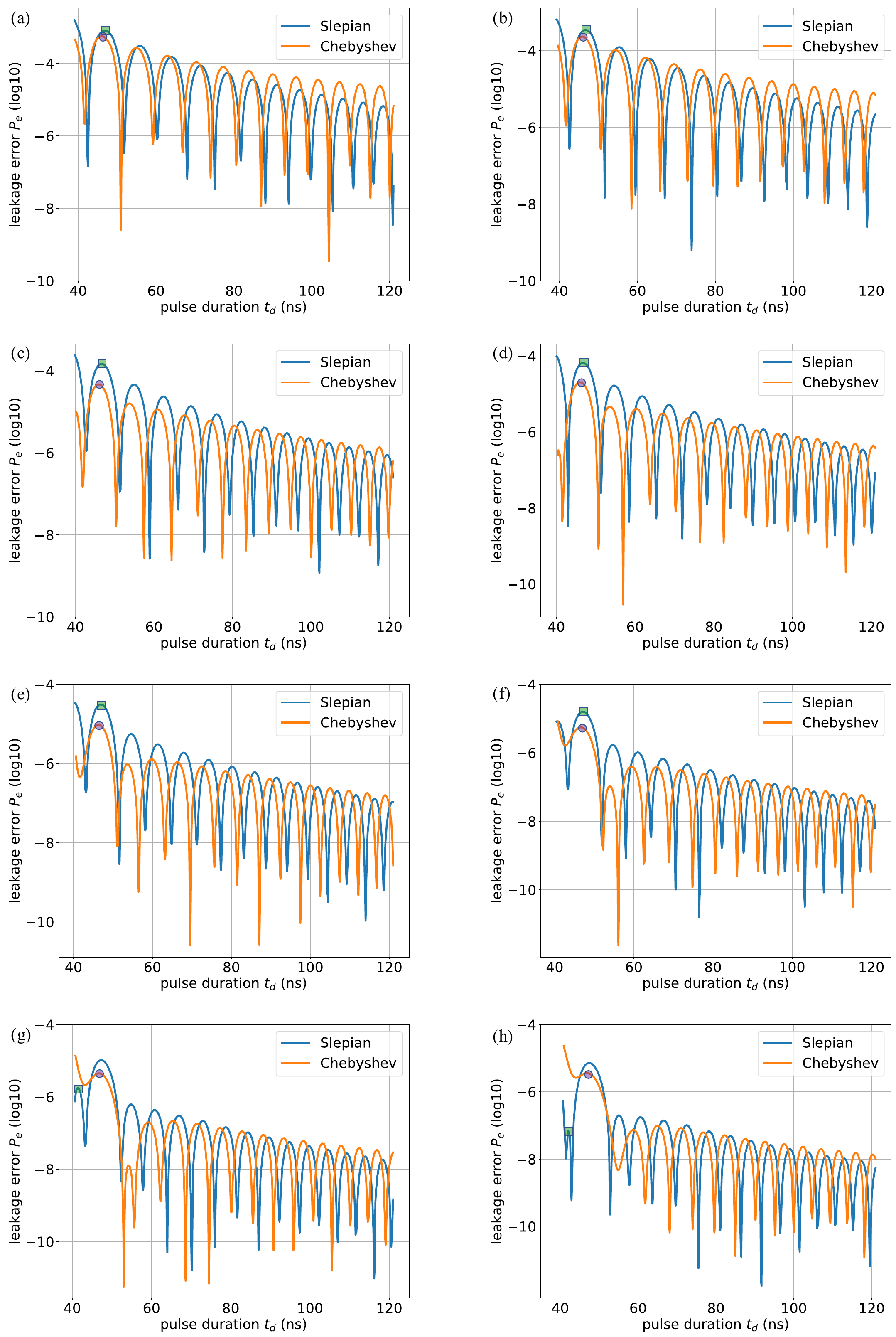}
    \caption{Additional simulation results comparing different pairs of the Slepian-based trajectories and Chebyshev-based trajectories. The best operating points with shortest pulse duration are indicated by green squares and purple dots. (a) through (h) are arranged in a descending order of leakage error of best operating points.}
    \label{fig:aggregate_sim}
\end{figure*}

We present additional simulation comparison results between different pairs of benchmark Slepian-based trajectories $\tilde g_{\textrm{sl2}}[n]$ and Chebyshev-based trajectories $\tilde g_{\textrm{ch2}}[n]$ in Fig.~\ref{fig:aggregate_sim}. We follow the simulation and analysis procedure as discussed in Section~\ref{sec: Simulation results}. In all these examples, we can see that $\tilde g_{\textrm{ch2}}[n]$ pushes the leakage lower in the range of smaller pulse duration while sacrificing a higher leakage error in the range of larger pulse duration. The best operating points with shortest pulse duration are indicated by green squares and purple dots. Note that in Figs.~\ref{fig:aggregate_sim}\panel{g}-\panel{h}, we observe an unusually small lobe appearing before the first main leakage error lobe. Table~\ref{table:aggregate} shows an aggregate of comparisons of best operating points following the same argument as in Section~\ref{sec:example1}. 

We also present additional simulation comparison results under certain hardware constraints. Table~\ref{table:aggregate_hardware} shows an aggregation of the best operating points using $\tilde g_{\textrm{sl2}}[n]$ and $\tilde g_{\textrm{ch2}}[n]$ in Fig.~\ref{fig:harware_constraint} following the same argument as in Section~\ref{sec:example1}.

\begin{table*}[h]
\centering
\setlength{\tabcolsep}{10pt}
\def\arraystretch{1.5}
\begin{tabular}{|c|c|c|c|c|c|c|}
\hline
\multirow{2}{*}{Index} & \multicolumn{3}{c|}{Slepian} & \multicolumn{3}{c|}{Chebyshev}\\
\cline{2-7}
 & $t_d$ (ns) & $\log(P_e)$ & $1-F_g$ & $t_d$ (ns) & $\log(P_e)$ & $1-F_g$ \\
\hline
(a) & 46.8 & $-3.09$ & $1.8\times10^{-4}$ & 46.0 & $-3.13$ & $1.5\times10^{-4}$ \\
(b) & 46.5 & $-3.46$ & $8.4\times10^{-5}$ & 46.0 & $-3.49$ & $6.8\times10^{-5}$ \\
(c) & 46.6 & $-3.83$ & $3.7\times10^{-5}$ & 46.0 & $-3.85$ & $1.8\times10^{-5}$ \\
(d) & 46.7 & $-4.18$ & $1.7\times10^{-5}$ & 46.0 & $-4.22$ & $9.5\times10^{-6}$ \\
(e) & 47.0 & $-4.51$ & $7.3\times10^{-6}$ & 46.5 & $-4.52$ & $4.0\times10^{-6}$ \\
(f) & 47.0 & $-4.80$ & $4.2\times10^{-6}$ & 46.5 & $-4.82$ & $2.9\times10^{-6}$ \\ 
(g) & 41.5 & $-5.75$ & $6.8\times10^{-6}$ & 46.7 & $-5.01$ & $2.2\times10^{-6}$ \\
(h) & 42.0 & $-7.16$ & $1.2\times10^{-6}$ & 47.0 & $-5.15$ & $1.7\times10^{-6}$ \\
\hline
\end{tabular}
\caption{An aggregate of comparisons between various pairs of benchmark second Slepian-based trajectories and Chebyshev-based trajectories, designed with different leakage error thresholds. The index corresponds to that in Fig.~\ref{fig:aggregate_sim}.}
\label{table:aggregate}
\end{table*}

\begin{table*}[bth]
\centering
\def\arraystretch{1.5}
\setlength{\tabcolsep}{10pt}
\begin{tabular}{|c|c|c|c|c|c|c|}
\hline
\multirow{2}{*}{Index} & \multicolumn{3}{c|}{Slepian} & \multicolumn{3}{c|}{Chebyshev} \\
\cline{2-7}
 & $t_d$ (ns) & $log(P_e)$ & $1-F_g$ & $t_d$ (ns) & $log(P_e)$ & $1-F_g$ \\
\hline
(a) & 51.9 & $-3.14$ & $1.5\times10^{-4}$ & 51.5 & $-3.15$ & $1.4\times10^{-4}$ \\
(b) & 48.5 & $-3.91$ & $2.5\times10^{-5}$ & 47.6 & $-3.96$ & $1.8\times10^{-5}$ \\
(c) & 47.0 & $-4.47$ & $7.9\times10^{-6}$ & 46.0 & $-4.55$ & $9.5\times10^{-6}$ \\
(d) & 47.0 & $-4.62$ & $5.9\times10^{-6}$ & 46.0 & $-4.69$ & $7.5\times10^{-6}$ \\
\hline
\end{tabular}
\caption{Best operating points using $\tilde g_{\textrm{sl2}}[n]$ and $\tilde g_{\textrm{ch2}}[n]$ under certain hardware constraints in Fig.~\ref{fig:harware_constraint}. The index corresponds to that in Fig.~\ref{fig:harware_constraint}.}
\label{table:aggregate_hardware}
\end{table*}

\section{Average gate fidelity} \label{app: Average gate fidelity}
We compute the average gate fidelity $F_g$ following Refs.~\cite{Christopher2018,sung_realization_2021,Nielsen2002}. We first compute the average process fidelity $F_p$ by numerically simulating quantum process tomography. We prepare 16 input states $\{\ket{0}, \ket{1}, \ket{+}, \ket{-}\}\otimes\{\ket{0}, \ket{1}, \ket{+}, \ket{-}\}$ and construct the process matrix (chi matrix representation) $\chi$ from the output states. We then compute the average process fidelity $F_p = \text{Tr}(\chi_{\text{ideal}}\chi)$ by comparing it to the ideal process matrix $\chi_{\text{ideal}}$.
To quantify leakage error in average gate fidelity, we partition the Hilbert space $\mathcal{H}$ into two disjoint subspaces $\mathcal{H}=\mathcal{H}_1\oplus\mathcal{H}_2$. Here, $\mathcal{H}_1$ is the $d_1$-dimensional subspace whose bases are computational states, while $\mathcal{H}_2$ is the $d_2$-dimensional subspace where leakage to additional noncomputational states occurs. The state leakage term $L_1$ is defined as $L_1 = 1-\text{Tr}(\mathbb{1}_1\rho)$, where $\mathbb{1}_1$ denotes the projector onto $\mathcal{H}_1$ and $\rho$ denotes the state. $L_1$ is mainly determined by $P_e$ in our case. The average gate fidelity $F_g$ can then be computed through the relationship between $F_g$ and $F_p$ introduced in~\cite{Christopher2018},
\begin{equation}
    F_g = \frac{d_1 F_p +1-L_1}{d_1+1}.
\end{equation}
We define gate infidelity to be $1-F_g$.

\section{Details on simulating hardware limitations} \label{app:simulating hardware limitations}
In Eq.~\ref{eq:design_pipeline} we present the design pipeline from $\tilde g(\tau)$ to $\omega_{1}(t)$ and finally to $\Phi_{\text{ext}}(t)$. In experiments, $\Phi_{\text{ext}}(t)$ is induced by passing an electric current through a flux line, which is connected to an antenna positioned in proximity to the target qubit and linked inductively to its SQUID loop. The current is generated at room temperature, employing either an active current source or a voltage source that applies a voltage across a series resistance. In either case, $\Phi_{\text{ext}}(t)$ can be modeled as a linear function of physical control parameter $P(t)$, i.e., $\Phi_{\text{ext}}(t)=kP(t)$. $P(t)$ can be the output current of the current source or the output voltage of the voltage source. 

\begin{figure*}[bt]
    \centering  \includegraphics[width=0.99\textwidth]{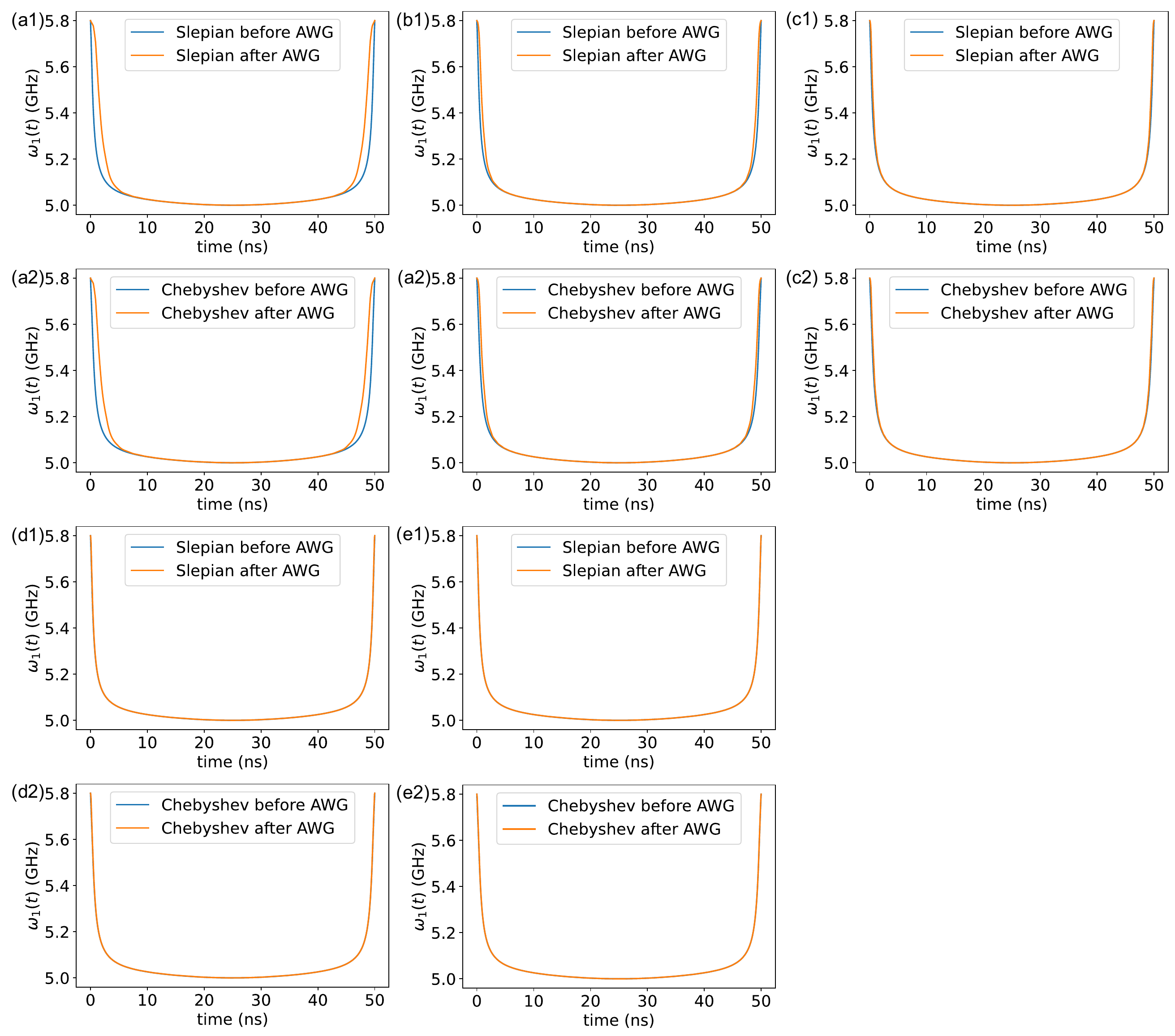}
    \caption{$\omega_{1}(t)$ and $\hat{\omega}_{1}(t)$ before and after imposing the hardware limitation for $t_d=50$ ns based on $\tilde g_{\textrm{sl2}}[n]$ and $\tilde g_{\textrm{ch2}}[n]$ as in Section~\ref{sec:example1}. (a1)(a2) $F_s=0.5$ GSa/s and $bw=200$ MHz, (b1)(b2) $F_s=1$ GSa/s and $bw=400$ MHz, (c1)(c2) $F_s=2$ GSa/s and $bw=800$ MHz, (d1)(d2) $F_s=5$ GSa/s and $bw=2$ GHz, (e1)(e2) $F_s=10$ GSa/s and $bw=4$ GHz.}
    \label{fig:pulse_hardware_comparison}
\end{figure*}

\begin{figure*}[bt]
    \centering  \includegraphics[width=0.88\textwidth]{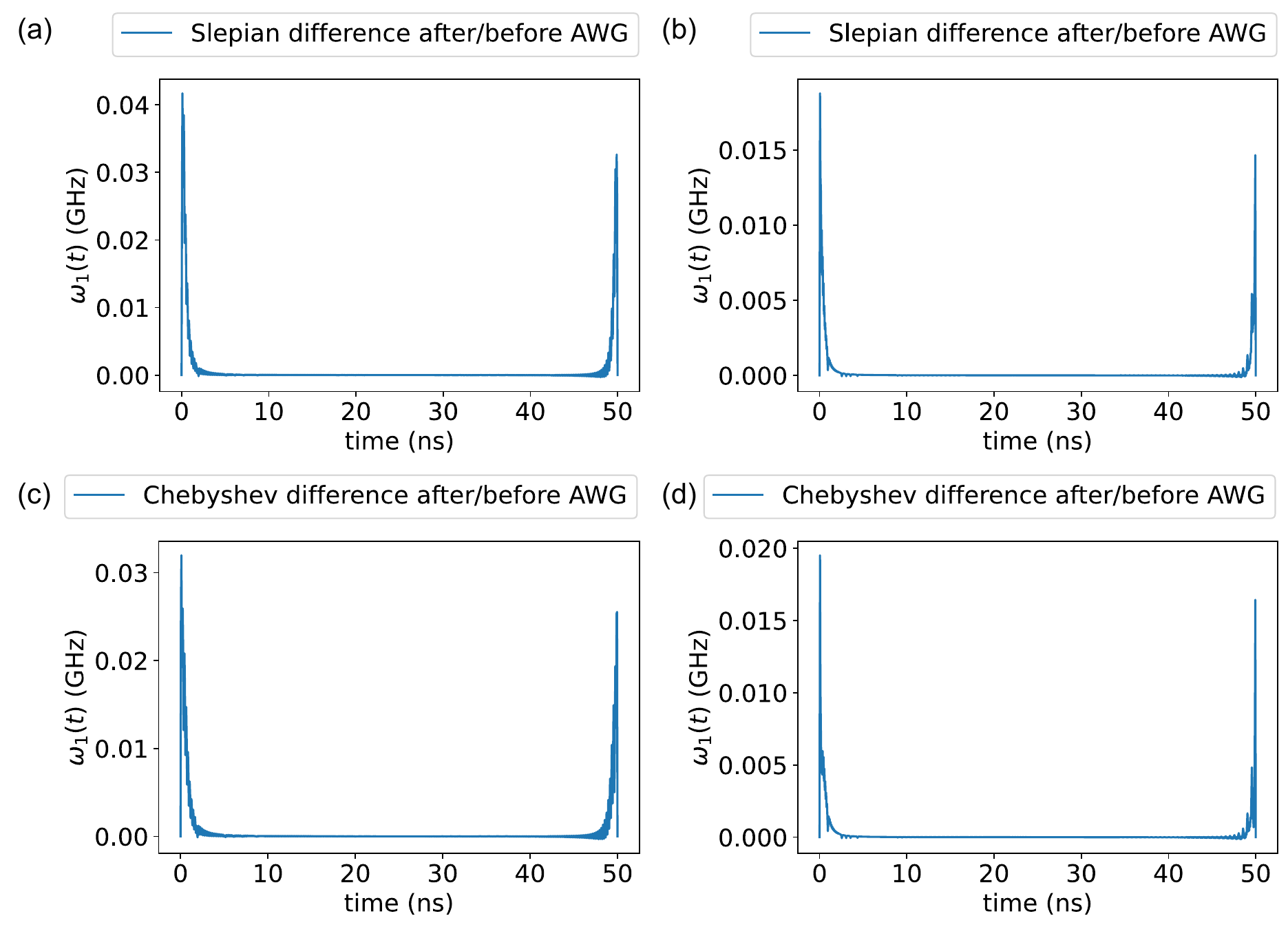}
    \caption{Difference between $\hat{\omega}_{1}(t)$ and $\omega_{1}(t)$  after and before imposing the hardware limitation. \panel{a}\panel{b}\panel{c}\panel{d} correspond to Fig.~\ref{fig:pulse_hardware_comparison}\panel{d1}\panel{e1}\panel{d2}\panel{e2} respectively. The difference shrinks as we improve the hardware parameters.}
    \label{fig:pulse_hardware_diff}
\end{figure*}

We consider the sampling frequency $F_s$ and bandwidth $bw$ as the hardware limitations. For a designed $\Phi_{\text{ext}}(t)$, we have a corresponding $P(t)$. We first sample $P(t)$ with the sampling frequency $F_s$ and interpolate the samples with zero holdings to simulate the function of the digital-to-analog converter (DAC). We then put the interpolated pulse into a lowpass filter with bandwidth $bw$ and obtain $\hat{P}(t)$. In our simulation, we use a first-order Butterworth filter. In order to convert  $\hat{P}(t)$ to $\hat{\omega}_{1}(t)$, we numerically compute the inverse function of Eq.~\ref{eq:flux_frequency} to obtain $f_1: \Phi_{\text{ext}} \rightarrow \omega_{1}$ and feed $\hat{\Phi}_{\text{ext}}(t)=k\hat{P}(t)$ as the input. Finally we follow the same simulation procedure as discussed in Section~\ref{sec:setting_procedure} with $\hat{\omega}_{1}(t)$. Fig.~\ref{fig:pulse_hardware_comparison} shows the comparison of $\omega_{1}(t)$ and $\hat{\omega}_{1}(t)$ before and after imposing the hardware limitation for $t_d=50$ ns based on $\tilde g_{\textrm{sl2}}[n]$ and $\tilde g_{\textrm{ch2}}[n]$ as in Section~\ref{sec:example1}, with different $F_s$ and $bw$. The difference between the two pulses in Figs.~\ref{fig:pulse_hardware_comparison}\panel{d1},~\ref{fig:pulse_hardware_comparison}\panel{e1},~\ref{fig:pulse_hardware_comparison}\panel{d2},~\ref{fig:pulse_hardware_comparison}\panel{e2} are shown in Figs.~\ref{fig:pulse_hardware_diff}\panel{a},~\ref{fig:pulse_hardware_diff}\panel{b},~\ref{fig:pulse_hardware_diff}\panel{c},~\ref{fig:pulse_hardware_diff}\panel{d}, respectively. As the sampling frequency and bandwidth of the hardware enhance, the distinction between the control pulses prior to and following the AWG diminishes.

\clearpage

\bibliography{cheby_ref}

\begin{thebibliography}{69}%
\makeatletter
\providecommand \@ifxundefined [1]{%
 \@ifx{#1\undefined}
}%
\providecommand \@ifnum [1]{%
 \ifnum #1\expandafter \@firstoftwo
 \else \expandafter \@secondoftwo
 \fi
}%
\providecommand \@ifx [1]{%
 \ifx #1\expandafter \@firstoftwo
 \else \expandafter \@secondoftwo
 \fi
}%
\providecommand \natexlab [1]{#1}%
\providecommand \enquote  [1]{``#1''}%
\providecommand \bibnamefont  [1]{#1}%
\providecommand \bibfnamefont [1]{#1}%
\providecommand \citenamefont [1]{#1}%
\providecommand \href@noop [0]{\@secondoftwo}%
\providecommand \href [0]{\begingroup \@sanitize@url \@href}%
\providecommand \@href[1]{\@@startlink{#1}\@@href}%
\providecommand \@@href[1]{\endgroup#1\@@endlink}%
\providecommand \@sanitize@url [0]{\catcode `\\12\catcode `\$12\catcode
  `\&12\catcode `\#12\catcode `\^12\catcode `\_12\catcode `\%12\relax}%
\providecommand \@@startlink[1]{}%
\providecommand \@@endlink[0]{}%
\providecommand \url  [0]{\begingroup\@sanitize@url \@url }%
\providecommand \@url [1]{\endgroup\@href {#1}{\urlprefix }}%
\providecommand \urlprefix  [0]{URL }%
\providecommand \Eprint [0]{\href }%
\providecommand \doibase [0]{https://doi.org/}%
\providecommand \selectlanguage [0]{\@gobble}%
\providecommand \bibinfo  [0]{\@secondoftwo}%
\providecommand \bibfield  [0]{\@secondoftwo}%
\providecommand \translation [1]{[#1]}%
\providecommand \BibitemOpen [0]{}%
\providecommand \bibitemStop [0]{}%
\providecommand \bibitemNoStop [0]{.\EOS\space}%
\providecommand \EOS [0]{\spacefactor3000\relax}%
\providecommand \BibitemShut  [1]{\csname bibitem#1\endcsname}%
\let\auto@bib@innerbib\@empty
\bibitem [{\citenamefont {Zajac}\ \emph {et~al.}(2018)\citenamefont {Zajac},
  \citenamefont {Sigillito}, \citenamefont {Russ}, \citenamefont {Borjans},
  \citenamefont {Taylor}, \citenamefont {Burkard},\ and\ \citenamefont
  {Petta}}]{Zajac2018}%
  \BibitemOpen
  \bibfield  {author} {\bibinfo {author} {\bibfnamefont {D.~M.}\ \bibnamefont
  {Zajac}}, \bibinfo {author} {\bibfnamefont {A.~J.}\ \bibnamefont
  {Sigillito}}, \bibinfo {author} {\bibfnamefont {M.}~\bibnamefont {Russ}},
  \bibinfo {author} {\bibfnamefont {F.}~\bibnamefont {Borjans}}, \bibinfo
  {author} {\bibfnamefont {J.~M.}\ \bibnamefont {Taylor}}, \bibinfo {author}
  {\bibfnamefont {G.}~\bibnamefont {Burkard}},\ and\ \bibinfo {author}
  {\bibfnamefont {J.~R.}\ \bibnamefont {Petta}},\ }\bibfield  {title} {\bibinfo
  {title} {Resonantly driven cnot gate for electron spins},\ }\href
  {https://doi.org/10.1126/science.aao5965} {\bibfield  {journal} {\bibinfo
  {journal} {Science}\ }\textbf {\bibinfo {volume} {359}},\ \bibinfo {pages}
  {439} (\bibinfo {year} {2018})}\BibitemShut {NoStop}%
\bibitem [{\citenamefont {Wright}\ \emph {et~al.}(2019)\citenamefont {Wright},
  \citenamefont {Beck}, \citenamefont {Debnath}, \citenamefont {Amini},
  \citenamefont {Nam}, \citenamefont {Grzesiak}, \citenamefont {Chen},
  \citenamefont {Pisenti}, \citenamefont {Chmielewski}, \citenamefont
  {Collins}, \citenamefont {Hudek}, \citenamefont {Mizrahi}, \citenamefont
  {Wong-Campos}, \citenamefont {Allen}, \citenamefont {Apisdorf}, \citenamefont
  {Solomon}, \citenamefont {Williams}, \citenamefont {Ducore}, \citenamefont
  {Blinov}, \citenamefont {Kreikemeier}, \citenamefont {Chaplin}, \citenamefont
  {Keesan}, \citenamefont {Monroe},\ and\ \citenamefont
  {Kim}}]{Wright2019BenchmarkingA1}%
  \BibitemOpen
  \bibfield  {author} {\bibinfo {author} {\bibfnamefont {K.}~\bibnamefont
  {Wright}}, \bibinfo {author} {\bibfnamefont {K.~M.}\ \bibnamefont {Beck}},
  \bibinfo {author} {\bibfnamefont {S.}~\bibnamefont {Debnath}}, \bibinfo
  {author} {\bibfnamefont {J.~M.}\ \bibnamefont {Amini}}, \bibinfo {author}
  {\bibfnamefont {Y.~S.}\ \bibnamefont {Nam}}, \bibinfo {author} {\bibfnamefont
  {N.}~\bibnamefont {Grzesiak}}, \bibinfo {author} {\bibfnamefont {J.-S.}\
  \bibnamefont {Chen}}, \bibinfo {author} {\bibfnamefont {N.~C.}\ \bibnamefont
  {Pisenti}}, \bibinfo {author} {\bibfnamefont {M.}~\bibnamefont
  {Chmielewski}}, \bibinfo {author} {\bibfnamefont {C.}~\bibnamefont
  {Collins}}, \bibinfo {author} {\bibfnamefont {K.~M.}\ \bibnamefont {Hudek}},
  \bibinfo {author} {\bibfnamefont {J.}~\bibnamefont {Mizrahi}}, \bibinfo
  {author} {\bibfnamefont {J.~D.}\ \bibnamefont {Wong-Campos}}, \bibinfo
  {author} {\bibfnamefont {S.}~\bibnamefont {Allen}}, \bibinfo {author}
  {\bibfnamefont {J.}~\bibnamefont {Apisdorf}}, \bibinfo {author}
  {\bibfnamefont {P.}~\bibnamefont {Solomon}}, \bibinfo {author} {\bibfnamefont
  {M.}~\bibnamefont {Williams}}, \bibinfo {author} {\bibfnamefont {A.~M.}\
  \bibnamefont {Ducore}}, \bibinfo {author} {\bibfnamefont {A.}~\bibnamefont
  {Blinov}}, \bibinfo {author} {\bibfnamefont {S.~M.}\ \bibnamefont
  {Kreikemeier}}, \bibinfo {author} {\bibfnamefont {V.}~\bibnamefont
  {Chaplin}}, \bibinfo {author} {\bibfnamefont {M.~J.}\ \bibnamefont {Keesan}},
  \bibinfo {author} {\bibfnamefont {C.}~\bibnamefont {Monroe}},\ and\ \bibinfo
  {author} {\bibfnamefont {J.}~\bibnamefont {Kim}},\ }\bibfield  {title}
  {\bibinfo {title} {Benchmarking an 11-qubit quantum computer},\ }\href
  {https://api.semanticscholar.org/CorpusID:84846167} {\bibfield  {journal}
  {\bibinfo  {journal} {Nature Communications}\ }\textbf {\bibinfo {volume}
  {10}} (\bibinfo {year} {2019})}\BibitemShut {NoStop}%
\bibitem [{\citenamefont {He}\ \emph {et~al.}(2019)\citenamefont {He},
  \citenamefont {Gorman}, \citenamefont {Keith}, \citenamefont {Kranz},
  \citenamefont {Keizer},\ and\ \citenamefont {Simmons}}]{He2019ATG}%
  \BibitemOpen
  \bibfield  {author} {\bibinfo {author} {\bibfnamefont {Y.}~\bibnamefont
  {He}}, \bibinfo {author} {\bibfnamefont {S.~K.}\ \bibnamefont {Gorman}},
  \bibinfo {author} {\bibfnamefont {D.}~\bibnamefont {Keith}}, \bibinfo
  {author} {\bibfnamefont {L.}~\bibnamefont {Kranz}}, \bibinfo {author}
  {\bibfnamefont {J.~G.}\ \bibnamefont {Keizer}},\ and\ \bibinfo {author}
  {\bibfnamefont {M.~Y.}\ \bibnamefont {Simmons}},\ }\bibfield  {title}
  {\bibinfo {title} {A two-qubit gate between phosphorus donor electrons in
  silicon},\ }\href {https://api.semanticscholar.org/CorpusID:197542823}
  {\bibfield  {journal} {\bibinfo  {journal} {Nature}\ }\textbf {\bibinfo
  {volume} {571}},\ \bibinfo {pages} {371 } (\bibinfo {year}
  {2019})}\BibitemShut {NoStop}%
\bibitem [{\citenamefont {Xu}\ \emph {et~al.}(2020)\citenamefont {Xu},
  \citenamefont {Chu}, \citenamefont {Yuan}, \citenamefont {Qiu}, \citenamefont
  {Zhou}, \citenamefont {Zhang}, \citenamefont {Tan}, \citenamefont {Yu},
  \citenamefont {Liu}, \citenamefont {Li}, \citenamefont {Yan},\ and\
  \citenamefont {Yu}}]{PhysRevLett.125.240503}%
  \BibitemOpen
  \bibfield  {author} {\bibinfo {author} {\bibfnamefont {Y.}~\bibnamefont
  {Xu}}, \bibinfo {author} {\bibfnamefont {J.}~\bibnamefont {Chu}}, \bibinfo
  {author} {\bibfnamefont {J.}~\bibnamefont {Yuan}}, \bibinfo {author}
  {\bibfnamefont {J.}~\bibnamefont {Qiu}}, \bibinfo {author} {\bibfnamefont
  {Y.}~\bibnamefont {Zhou}}, \bibinfo {author} {\bibfnamefont {L.}~\bibnamefont
  {Zhang}}, \bibinfo {author} {\bibfnamefont {X.}~\bibnamefont {Tan}}, \bibinfo
  {author} {\bibfnamefont {Y.}~\bibnamefont {Yu}}, \bibinfo {author}
  {\bibfnamefont {S.}~\bibnamefont {Liu}}, \bibinfo {author} {\bibfnamefont
  {J.}~\bibnamefont {Li}}, \bibinfo {author} {\bibfnamefont {F.}~\bibnamefont
  {Yan}},\ and\ \bibinfo {author} {\bibfnamefont {D.}~\bibnamefont {Yu}},\
  }\bibfield  {title} {\bibinfo {title} {High-fidelity, high-scalability
  two-qubit gate scheme for superconducting qubits},\ }\href
  {https://doi.org/10.1103/PhysRevLett.125.240503} {\bibfield  {journal}
  {\bibinfo  {journal} {Phys. Rev. Lett.}\ }\textbf {\bibinfo {volume} {125}},\
  \bibinfo {pages} {240503} (\bibinfo {year} {2020})}\BibitemShut {NoStop}%
\bibitem [{\citenamefont {Krantz}\ \emph {et~al.}(2019)\citenamefont {Krantz},
  \citenamefont {Kjaergaard}, \citenamefont {Yan}, \citenamefont {Orlando},
  \citenamefont {Gustavsson},\ and\ \citenamefont
  {Oliver}}]{krantz_quantum_2019}%
  \BibitemOpen
  \bibfield  {author} {\bibinfo {author} {\bibfnamefont {P.}~\bibnamefont
  {Krantz}}, \bibinfo {author} {\bibfnamefont {M.}~\bibnamefont {Kjaergaard}},
  \bibinfo {author} {\bibfnamefont {F.}~\bibnamefont {Yan}}, \bibinfo {author}
  {\bibfnamefont {T.}~\bibnamefont {Orlando}}, \bibinfo {author} {\bibfnamefont
  {S.}~\bibnamefont {Gustavsson}},\ and\ \bibinfo {author} {\bibfnamefont
  {W.}~\bibnamefont {Oliver}},\ }\bibfield  {title} {\bibinfo {title} {A
  quantum engineer's guide to superconducting qubits},\ }\href
  {https://doi.org/10.1063/1.5089550} {\bibfield  {journal} {\bibinfo
  {journal} {Applied Physics Reviews}\ }\textbf {\bibinfo {volume} {6}},\
  \bibinfo {pages} {021318} (\bibinfo {year} {2019})}\BibitemShut {NoStop}%
\bibitem [{\citenamefont {Arute}\ \emph {et~al.}(2019)\citenamefont {Arute},
  \citenamefont {Arya}, \citenamefont {Babbush}, \citenamefont {Bacon},
  \citenamefont {Bardin}, \citenamefont {Barends}, \citenamefont {Biswas},
  \citenamefont {Boixo}, \citenamefont {Brandao}, \citenamefont {Buell},
  \citenamefont {Burkett}, \citenamefont {Chen}, \citenamefont {Chen},
  \citenamefont {Chiaro}, \citenamefont {Collins}, \citenamefont {Courtney},
  \citenamefont {Dunsworth}, \citenamefont {Farhi}, \citenamefont {Foxen},
  \citenamefont {Fowler}, \citenamefont {Gidney}, \citenamefont {Giustina},
  \citenamefont {Graff}, \citenamefont {Guerin}, \citenamefont {Habegger},
  \citenamefont {Harrigan}, \citenamefont {Hartmann}, \citenamefont {Ho},
  \citenamefont {Hoffmann}, \citenamefont {Huang}, \citenamefont {Humble},
  \citenamefont {Isakov}, \citenamefont {Jeffrey}, \citenamefont {Jiang},
  \citenamefont {Kafri}, \citenamefont {Kechedzhi}, \citenamefont {Kelly},
  \citenamefont {Klimov}, \citenamefont {Knysh}, \citenamefont {Korotkov},
  \citenamefont {Kostritsa}, \citenamefont {Landhuis}, \citenamefont
  {Lindmark}, \citenamefont {Lucero}, \citenamefont {Lyakh}, \citenamefont
  {Mandrà}, \citenamefont {McClean}, \citenamefont {McEwen}, \citenamefont
  {Megrant}, \citenamefont {Mi}, \citenamefont {Michielsen}, \citenamefont
  {Mohseni}, \citenamefont {Mutus}, \citenamefont {Naaman}, \citenamefont
  {Neeley}, \citenamefont {Neill}, \citenamefont {Niu}, \citenamefont {Ostby},
  \citenamefont {Petukhov}, \citenamefont {Platt}, \citenamefont {Quintana},
  \citenamefont {Rieffel}, \citenamefont {Roushan}, \citenamefont {Rubin},
  \citenamefont {Sank}, \citenamefont {Satzinger}, \citenamefont {Smelyanskiy},
  \citenamefont {Sung}, \citenamefont {Trevithick}, \citenamefont
  {Vainsencher}, \citenamefont {Villalonga}, \citenamefont {White},
  \citenamefont {Yao}, \citenamefont {Yeh}, \citenamefont {Zalcman},
  \citenamefont {Neven},\ and\ \citenamefont {Martinis}}]{arute_quantum_2019}%
  \BibitemOpen
  \bibfield  {author} {\bibinfo {author} {\bibfnamefont {F.}~\bibnamefont
  {Arute}}, \bibinfo {author} {\bibfnamefont {K.}~\bibnamefont {Arya}},
  \bibinfo {author} {\bibfnamefont {R.}~\bibnamefont {Babbush}}, \bibinfo
  {author} {\bibfnamefont {D.}~\bibnamefont {Bacon}}, \bibinfo {author}
  {\bibfnamefont {J.~C.}\ \bibnamefont {Bardin}}, \bibinfo {author}
  {\bibfnamefont {R.}~\bibnamefont {Barends}}, \bibinfo {author} {\bibfnamefont
  {R.}~\bibnamefont {Biswas}}, \bibinfo {author} {\bibfnamefont
  {S.}~\bibnamefont {Boixo}}, \bibinfo {author} {\bibfnamefont {F.~G. S.~L.}\
  \bibnamefont {Brandao}}, \bibinfo {author} {\bibfnamefont {D.~A.}\
  \bibnamefont {Buell}}, \bibinfo {author} {\bibfnamefont {B.}~\bibnamefont
  {Burkett}}, \bibinfo {author} {\bibfnamefont {Y.}~\bibnamefont {Chen}},
  \bibinfo {author} {\bibfnamefont {Z.}~\bibnamefont {Chen}}, \bibinfo {author}
  {\bibfnamefont {B.}~\bibnamefont {Chiaro}}, \bibinfo {author} {\bibfnamefont
  {R.}~\bibnamefont {Collins}}, \bibinfo {author} {\bibfnamefont
  {W.}~\bibnamefont {Courtney}}, \bibinfo {author} {\bibfnamefont
  {A.}~\bibnamefont {Dunsworth}}, \bibinfo {author} {\bibfnamefont
  {E.}~\bibnamefont {Farhi}}, \bibinfo {author} {\bibfnamefont
  {B.}~\bibnamefont {Foxen}}, \bibinfo {author} {\bibfnamefont
  {A.}~\bibnamefont {Fowler}}, \bibinfo {author} {\bibfnamefont
  {C.}~\bibnamefont {Gidney}}, \bibinfo {author} {\bibfnamefont
  {M.}~\bibnamefont {Giustina}}, \bibinfo {author} {\bibfnamefont
  {R.}~\bibnamefont {Graff}}, \bibinfo {author} {\bibfnamefont
  {K.}~\bibnamefont {Guerin}}, \bibinfo {author} {\bibfnamefont
  {S.}~\bibnamefont {Habegger}}, \bibinfo {author} {\bibfnamefont {M.~P.}\
  \bibnamefont {Harrigan}}, \bibinfo {author} {\bibfnamefont {M.~J.}\
  \bibnamefont {Hartmann}}, \bibinfo {author} {\bibfnamefont {A.}~\bibnamefont
  {Ho}}, \bibinfo {author} {\bibfnamefont {M.}~\bibnamefont {Hoffmann}},
  \bibinfo {author} {\bibfnamefont {T.}~\bibnamefont {Huang}}, \bibinfo
  {author} {\bibfnamefont {T.~S.}\ \bibnamefont {Humble}}, \bibinfo {author}
  {\bibfnamefont {S.~V.}\ \bibnamefont {Isakov}}, \bibinfo {author}
  {\bibfnamefont {E.}~\bibnamefont {Jeffrey}}, \bibinfo {author} {\bibfnamefont
  {Z.}~\bibnamefont {Jiang}}, \bibinfo {author} {\bibfnamefont
  {D.}~\bibnamefont {Kafri}}, \bibinfo {author} {\bibfnamefont
  {K.}~\bibnamefont {Kechedzhi}}, \bibinfo {author} {\bibfnamefont
  {J.}~\bibnamefont {Kelly}}, \bibinfo {author} {\bibfnamefont {P.~V.}\
  \bibnamefont {Klimov}}, \bibinfo {author} {\bibfnamefont {S.}~\bibnamefont
  {Knysh}}, \bibinfo {author} {\bibfnamefont {A.}~\bibnamefont {Korotkov}},
  \bibinfo {author} {\bibfnamefont {F.}~\bibnamefont {Kostritsa}}, \bibinfo
  {author} {\bibfnamefont {D.}~\bibnamefont {Landhuis}}, \bibinfo {author}
  {\bibfnamefont {M.}~\bibnamefont {Lindmark}}, \bibinfo {author}
  {\bibfnamefont {E.}~\bibnamefont {Lucero}}, \bibinfo {author} {\bibfnamefont
  {D.}~\bibnamefont {Lyakh}}, \bibinfo {author} {\bibfnamefont
  {S.}~\bibnamefont {Mandrà}}, \bibinfo {author} {\bibfnamefont {J.~R.}\
  \bibnamefont {McClean}}, \bibinfo {author} {\bibfnamefont {M.}~\bibnamefont
  {McEwen}}, \bibinfo {author} {\bibfnamefont {A.}~\bibnamefont {Megrant}},
  \bibinfo {author} {\bibfnamefont {X.}~\bibnamefont {Mi}}, \bibinfo {author}
  {\bibfnamefont {K.}~\bibnamefont {Michielsen}}, \bibinfo {author}
  {\bibfnamefont {M.}~\bibnamefont {Mohseni}}, \bibinfo {author} {\bibfnamefont
  {J.}~\bibnamefont {Mutus}}, \bibinfo {author} {\bibfnamefont
  {O.}~\bibnamefont {Naaman}}, \bibinfo {author} {\bibfnamefont
  {M.}~\bibnamefont {Neeley}}, \bibinfo {author} {\bibfnamefont
  {C.}~\bibnamefont {Neill}}, \bibinfo {author} {\bibfnamefont {M.~Y.}\
  \bibnamefont {Niu}}, \bibinfo {author} {\bibfnamefont {E.}~\bibnamefont
  {Ostby}}, \bibinfo {author} {\bibfnamefont {A.}~\bibnamefont {Petukhov}},
  \bibinfo {author} {\bibfnamefont {J.~C.}\ \bibnamefont {Platt}}, \bibinfo
  {author} {\bibfnamefont {C.}~\bibnamefont {Quintana}}, \bibinfo {author}
  {\bibfnamefont {E.~G.}\ \bibnamefont {Rieffel}}, \bibinfo {author}
  {\bibfnamefont {P.}~\bibnamefont {Roushan}}, \bibinfo {author} {\bibfnamefont
  {N.~C.}\ \bibnamefont {Rubin}}, \bibinfo {author} {\bibfnamefont
  {D.}~\bibnamefont {Sank}}, \bibinfo {author} {\bibfnamefont {K.~J.}\
  \bibnamefont {Satzinger}}, \bibinfo {author} {\bibfnamefont {V.}~\bibnamefont
  {Smelyanskiy}}, \bibinfo {author} {\bibfnamefont {K.~J.}\ \bibnamefont
  {Sung}}, \bibinfo {author} {\bibfnamefont {M.~D.}\ \bibnamefont
  {Trevithick}}, \bibinfo {author} {\bibfnamefont {A.}~\bibnamefont
  {Vainsencher}}, \bibinfo {author} {\bibfnamefont {B.}~\bibnamefont
  {Villalonga}}, \bibinfo {author} {\bibfnamefont {T.}~\bibnamefont {White}},
  \bibinfo {author} {\bibfnamefont {Z.~J.}\ \bibnamefont {Yao}}, \bibinfo
  {author} {\bibfnamefont {P.}~\bibnamefont {Yeh}}, \bibinfo {author}
  {\bibfnamefont {A.}~\bibnamefont {Zalcman}}, \bibinfo {author} {\bibfnamefont
  {H.}~\bibnamefont {Neven}},\ and\ \bibinfo {author} {\bibfnamefont {J.~M.}\
  \bibnamefont {Martinis}},\ }\bibfield  {title} {\bibinfo {title} {Quantum
  supremacy using a programmable superconducting processor},\ }\href
  {https://doi.org/10.1038/s41586-019-1666-5} {\bibfield  {journal} {\bibinfo
  {journal} {Nature}\ }\textbf {\bibinfo {volume} {574}},\ \bibinfo {pages}
  {505} (\bibinfo {year} {2019})}\BibitemShut {NoStop}%
\bibitem [{\citenamefont {Kjaergaard}\ \emph {et~al.}(2020)\citenamefont
  {Kjaergaard}, \citenamefont {Schwartz}, \citenamefont {Braum\"{u}ller},
  \citenamefont {Krantz}, \citenamefont {Wang}, \citenamefont {Gustavsson},\
  and\ \citenamefont {Oliver}}]{doi:10.1146/annurev-conmatphys-031119-050605}%
  \BibitemOpen
  \bibfield  {author} {\bibinfo {author} {\bibfnamefont {M.}~\bibnamefont
  {Kjaergaard}}, \bibinfo {author} {\bibfnamefont {M.~E.}\ \bibnamefont
  {Schwartz}}, \bibinfo {author} {\bibfnamefont {J.}~\bibnamefont
  {Braum\"{u}ller}}, \bibinfo {author} {\bibfnamefont {P.}~\bibnamefont
  {Krantz}}, \bibinfo {author} {\bibfnamefont {J.~I.-J.}\ \bibnamefont {Wang}},
  \bibinfo {author} {\bibfnamefont {S.}~\bibnamefont {Gustavsson}},\ and\
  \bibinfo {author} {\bibfnamefont {W.~D.}\ \bibnamefont {Oliver}},\ }\bibfield
   {title} {\bibinfo {title} {Superconducting qubits: Current state of play},\
  }\href {https://doi.org/10.1146/annurev-conmatphys-031119-050605} {\bibfield
  {journal} {\bibinfo  {journal} {Annual Review of Condensed Matter Physics}\
  }\textbf {\bibinfo {volume} {11}},\ \bibinfo {pages} {369} (\bibinfo {year}
  {2020})}\BibitemShut {NoStop}%
\bibitem [{\citenamefont {Kim}\ \emph {et~al.}(2023)\citenamefont {Kim},
  \citenamefont {Eddins}, \citenamefont {Anand}, \citenamefont {Wei},
  \citenamefont {Berg}, \citenamefont {Rosenblatt}, \citenamefont {Nayfeh},
  \citenamefont {Wu}, \citenamefont {Zaletel}, \citenamefont {Temme},\ and\
  \citenamefont {Kandala}}]{Youngseok_2023}%
  \BibitemOpen
  \bibfield  {author} {\bibinfo {author} {\bibfnamefont {Y.}~\bibnamefont
  {Kim}}, \bibinfo {author} {\bibfnamefont {A.}~\bibnamefont {Eddins}},
  \bibinfo {author} {\bibfnamefont {S.}~\bibnamefont {Anand}}, \bibinfo
  {author} {\bibfnamefont {K.}~\bibnamefont {Wei}}, \bibinfo {author}
  {\bibfnamefont {E.}~\bibnamefont {Berg}}, \bibinfo {author} {\bibfnamefont
  {S.}~\bibnamefont {Rosenblatt}}, \bibinfo {author} {\bibfnamefont
  {H.}~\bibnamefont {Nayfeh}}, \bibinfo {author} {\bibfnamefont
  {Y.}~\bibnamefont {Wu}}, \bibinfo {author} {\bibfnamefont {M.}~\bibnamefont
  {Zaletel}}, \bibinfo {author} {\bibfnamefont {K.}~\bibnamefont {Temme}},\
  and\ \bibinfo {author} {\bibfnamefont {A.}~\bibnamefont {Kandala}},\
  }\bibfield  {title} {\bibinfo {title} {Evidence for the utility of quantum
  computing before fault tolerance},\ }\href
  {https://doi.org/10.1038/s41586-023-06096-3} {\bibfield  {journal} {\bibinfo
  {journal} {Nature}\ }\textbf {\bibinfo {volume} {618}},\ \bibinfo {pages}
  {500} (\bibinfo {year} {2023})}\BibitemShut {NoStop}%
\bibitem [{\citenamefont {Ristè}\ \emph {et~al.}(2014)\citenamefont {Ristè},
  \citenamefont {Poletto}, \citenamefont {Huang}, \citenamefont {Bruno},
  \citenamefont {Vesterinen}, \citenamefont {Saira},\ and\ \citenamefont
  {DiCarlo}}]{Riste_2014}%
  \BibitemOpen
  \bibfield  {author} {\bibinfo {author} {\bibfnamefont {D.}~\bibnamefont
  {Ristè}}, \bibinfo {author} {\bibfnamefont {S.}~\bibnamefont {Poletto}},
  \bibinfo {author} {\bibfnamefont {M.-Z.}\ \bibnamefont {Huang}}, \bibinfo
  {author} {\bibfnamefont {A.}~\bibnamefont {Bruno}}, \bibinfo {author}
  {\bibfnamefont {V.}~\bibnamefont {Vesterinen}}, \bibinfo {author}
  {\bibfnamefont {O.}~\bibnamefont {Saira}},\ and\ \bibinfo {author}
  {\bibfnamefont {L.}~\bibnamefont {DiCarlo}},\ }\bibfield  {title} {\bibinfo
  {title} {Detecting bit-flip errors in a logical qubit using stabilizer
  measurements},\ }\href {https://doi.org/10.1038/ncomms7983} {\bibfield
  {journal} {\bibinfo  {journal} {Nature communications}\ }\textbf {\bibinfo
  {volume} {6}} (\bibinfo {year} {2014})}\BibitemShut {NoStop}%
\bibitem [{\citenamefont {Takita}\ \emph {et~al.}(2016)\citenamefont {Takita},
  \citenamefont {C\'orcoles}, \citenamefont {Magesan}, \citenamefont {Abdo},
  \citenamefont {Brink}, \citenamefont {Cross}, \citenamefont {Chow},\ and\
  \citenamefont {Gambetta}}]{Maika2016}%
  \BibitemOpen
  \bibfield  {author} {\bibinfo {author} {\bibfnamefont {M.}~\bibnamefont
  {Takita}}, \bibinfo {author} {\bibfnamefont {A.~D.}\ \bibnamefont
  {C\'orcoles}}, \bibinfo {author} {\bibfnamefont {E.}~\bibnamefont {Magesan}},
  \bibinfo {author} {\bibfnamefont {B.}~\bibnamefont {Abdo}}, \bibinfo {author}
  {\bibfnamefont {M.}~\bibnamefont {Brink}}, \bibinfo {author} {\bibfnamefont
  {A.}~\bibnamefont {Cross}}, \bibinfo {author} {\bibfnamefont {J.~M.}\
  \bibnamefont {Chow}},\ and\ \bibinfo {author} {\bibfnamefont {J.~M.}\
  \bibnamefont {Gambetta}},\ }\bibfield  {title} {\bibinfo {title}
  {Demonstration of weight-four parity measurements in the surface code
  architecture},\ }\href {https://doi.org/10.1103/PhysRevLett.117.210505}
  {\bibfield  {journal} {\bibinfo  {journal} {Phys. Rev. Lett.}\ }\textbf
  {\bibinfo {volume} {117}},\ \bibinfo {pages} {210505} (\bibinfo {year}
  {2016})}\BibitemShut {NoStop}%
\bibitem [{\citenamefont {Andersen}\ \emph {et~al.}(2019)\citenamefont
  {Andersen}, \citenamefont {Remm}, \citenamefont {Lazar}, \citenamefont
  {Krinner}, \citenamefont {Heinsoo}, \citenamefont {Besse}, \citenamefont
  {Gabureac}, \citenamefont {Wallraff},\ and\ \citenamefont
  {Eichler}}]{Christian_2019}%
  \BibitemOpen
  \bibfield  {author} {\bibinfo {author} {\bibfnamefont {C.}~\bibnamefont
  {Andersen}}, \bibinfo {author} {\bibfnamefont {A.}~\bibnamefont {Remm}},
  \bibinfo {author} {\bibfnamefont {S.}~\bibnamefont {Lazar}}, \bibinfo
  {author} {\bibfnamefont {S.}~\bibnamefont {Krinner}}, \bibinfo {author}
  {\bibfnamefont {J.}~\bibnamefont {Heinsoo}}, \bibinfo {author} {\bibfnamefont
  {J.-C.}\ \bibnamefont {Besse}}, \bibinfo {author} {\bibfnamefont
  {M.}~\bibnamefont {Gabureac}}, \bibinfo {author} {\bibfnamefont
  {A.}~\bibnamefont {Wallraff}},\ and\ \bibinfo {author} {\bibfnamefont
  {C.}~\bibnamefont {Eichler}},\ }\bibfield  {title} {\bibinfo {title}
  {Entanglement stabilization using ancilla-based parity detection and
  real-time feedback in superconducting circuits},\ }\href
  {https://doi.org/10.1038/s41534-019-0185-4} {\bibfield  {journal} {\bibinfo
  {journal} {npj Quantum Information}\ }\textbf {\bibinfo {volume} {5}}
  (\bibinfo {year} {2019})}\BibitemShut {NoStop}%
\bibitem [{\citenamefont {Acharya}\ \emph {et~al.}(2023)\citenamefont
  {Acharya}, \citenamefont {Aleiner}, \citenamefont {Allen}, \citenamefont
  {Andersen}, \citenamefont {Ansmann}, \citenamefont {Arute}, \citenamefont
  {Arya}, \citenamefont {Asfaw}, \citenamefont {Atalaya}, \citenamefont
  {Babbush}, \citenamefont {Bacon}, \citenamefont {Bardin}, \citenamefont
  {Basso}, \citenamefont {Bengtsson}, \citenamefont {Boixo}, \citenamefont
  {Bortoli}, \citenamefont {Bourassa}, \citenamefont {Bovaird}, \citenamefont
  {Brill},\ and\ \citenamefont {Zhu}}]{Rajeev_2023}%
  \BibitemOpen
  \bibfield  {author} {\bibinfo {author} {\bibfnamefont {R.}~\bibnamefont
  {Acharya}}, \bibinfo {author} {\bibfnamefont {I.}~\bibnamefont {Aleiner}},
  \bibinfo {author} {\bibfnamefont {R.}~\bibnamefont {Allen}}, \bibinfo
  {author} {\bibfnamefont {T.}~\bibnamefont {Andersen}}, \bibinfo {author}
  {\bibfnamefont {M.}~\bibnamefont {Ansmann}}, \bibinfo {author} {\bibfnamefont
  {F.}~\bibnamefont {Arute}}, \bibinfo {author} {\bibfnamefont
  {K.}~\bibnamefont {Arya}}, \bibinfo {author} {\bibfnamefont {A.}~\bibnamefont
  {Asfaw}}, \bibinfo {author} {\bibfnamefont {J.}~\bibnamefont {Atalaya}},
  \bibinfo {author} {\bibfnamefont {R.}~\bibnamefont {Babbush}}, \bibinfo
  {author} {\bibfnamefont {D.}~\bibnamefont {Bacon}}, \bibinfo {author}
  {\bibfnamefont {J.}~\bibnamefont {Bardin}}, \bibinfo {author} {\bibfnamefont
  {J.}~\bibnamefont {Basso}}, \bibinfo {author} {\bibfnamefont
  {A.}~\bibnamefont {Bengtsson}}, \bibinfo {author} {\bibfnamefont
  {S.}~\bibnamefont {Boixo}}, \bibinfo {author} {\bibfnamefont
  {G.}~\bibnamefont {Bortoli}}, \bibinfo {author} {\bibfnamefont
  {A.}~\bibnamefont {Bourassa}}, \bibinfo {author} {\bibfnamefont
  {J.}~\bibnamefont {Bovaird}}, \bibinfo {author} {\bibfnamefont
  {L.}~\bibnamefont {Brill}},\ and\ \bibinfo {author} {\bibfnamefont
  {N.}~\bibnamefont {Zhu}},\ }\bibfield  {title} {\bibinfo {title} {Suppressing
  quantum errors by scaling a surface code logical qubit},\ }\href
  {https://doi.org/10.1038/s41586-022-05434-1} {\bibfield  {journal} {\bibinfo
  {journal} {Nature}\ }\textbf {\bibinfo {volume} {614}},\ \bibinfo {pages}
  {676} (\bibinfo {year} {2023})}\BibitemShut {NoStop}%
\bibitem [{\citenamefont {Nakamura}\ \emph {et~al.}(1999)\citenamefont
  {Nakamura}, \citenamefont {Pashkin},\ and\ \citenamefont
  {Tsai}}]{Nakamura1999}%
  \BibitemOpen
  \bibfield  {author} {\bibinfo {author} {\bibfnamefont {Y.}~\bibnamefont
  {Nakamura}}, \bibinfo {author} {\bibfnamefont {Y.~A.}\ \bibnamefont
  {Pashkin}},\ and\ \bibinfo {author} {\bibfnamefont {J.~S.}\ \bibnamefont
  {Tsai}},\ }\bibfield  {title} {\bibinfo {title} {Coherent control of
  macroscopic quantum states in a single-cooper-pair box},\ }\href
  {https://www.nature.com/articles/19718} {\bibfield  {journal} {\bibinfo
  {journal} {Nature}\ }\textbf {\bibinfo {volume} {398}},\ \bibinfo {pages}
  {786} (\bibinfo {year} {1999})}\BibitemShut {NoStop}%
\bibitem [{\citenamefont {Ioffe}\ \emph {et~al.}(1999)\citenamefont {Ioffe},
  \citenamefont {Geshkenbein}, \citenamefont {Feigel'man}, \citenamefont
  {Fauch{\`e}re},\ and\ \citenamefont {Blatter}}]{Ioffe1999}%
  \BibitemOpen
  \bibfield  {author} {\bibinfo {author} {\bibfnamefont {L.~B.}\ \bibnamefont
  {Ioffe}}, \bibinfo {author} {\bibfnamefont {V.~B.}\ \bibnamefont
  {Geshkenbein}}, \bibinfo {author} {\bibfnamefont {M.~V.}\ \bibnamefont
  {Feigel'man}}, \bibinfo {author} {\bibfnamefont {A.~L.}\ \bibnamefont
  {Fauch{\`e}re}},\ and\ \bibinfo {author} {\bibfnamefont {G.}~\bibnamefont
  {Blatter}},\ }\bibfield  {title} {\bibinfo {title} {Environmentally decoupled
  sds -wave josephson junctions for quantum computing},\ }\href
  {https://www.nature.com/articles/19718} {\bibfield  {journal} {\bibinfo
  {journal} {Nature}\ }\textbf {\bibinfo {volume} {398}},\ \bibinfo {pages}
  {679} (\bibinfo {year} {1999})}\BibitemShut {NoStop}%
\bibitem [{\citenamefont {Mooij}\ \emph {et~al.}(1999)\citenamefont {Mooij},
  \citenamefont {Orlando}, \citenamefont {Levitov}, \citenamefont {Tian},
  \citenamefont {Wal},\ and\ \citenamefont {Lloyd}}]{Mooij1999}%
  \BibitemOpen
  \bibfield  {author} {\bibinfo {author} {\bibfnamefont {J.~E.}\ \bibnamefont
  {Mooij}}, \bibinfo {author} {\bibfnamefont {T.~P.}\ \bibnamefont {Orlando}},
  \bibinfo {author} {\bibfnamefont {L.}~\bibnamefont {Levitov}}, \bibinfo
  {author} {\bibfnamefont {L.}~\bibnamefont {Tian}}, \bibinfo {author}
  {\bibfnamefont {C.~H.}\ \bibnamefont {Wal}},\ and\ \bibinfo {author}
  {\bibfnamefont {S.}~\bibnamefont {Lloyd}},\ }\bibfield  {title} {\bibinfo
  {title} {Josephson persistent-current qubit},\ }\href
  {https://www.science.org/doi/10.1126/science.285.5430.1036} {\bibfield
  {journal} {\bibinfo  {journal} {Science}\ }\textbf {\bibinfo {volume} {285
  5430}},\ \bibinfo {pages} {1036} (\bibinfo {year} {1999})}\BibitemShut
  {NoStop}%
\bibitem [{\citenamefont {Koch}\ \emph {et~al.}(2007)\citenamefont {Koch},
  \citenamefont {Yu}, \citenamefont {Gambetta}, \citenamefont {Houck},
  \citenamefont {Schuster}, \citenamefont {Majer}, \citenamefont {Blais},
  \citenamefont {Devoret}, \citenamefont {Girvin},\ and\ \citenamefont
  {Schoelkopf}}]{Koch2007}%
  \BibitemOpen
  \bibfield  {author} {\bibinfo {author} {\bibfnamefont {J.}~\bibnamefont
  {Koch}}, \bibinfo {author} {\bibfnamefont {T.~M.}\ \bibnamefont {Yu}},
  \bibinfo {author} {\bibfnamefont {J.}~\bibnamefont {Gambetta}}, \bibinfo
  {author} {\bibfnamefont {A.~A.}\ \bibnamefont {Houck}}, \bibinfo {author}
  {\bibfnamefont {D.~I.}\ \bibnamefont {Schuster}}, \bibinfo {author}
  {\bibfnamefont {J.}~\bibnamefont {Majer}}, \bibinfo {author} {\bibfnamefont
  {A.}~\bibnamefont {Blais}}, \bibinfo {author} {\bibfnamefont {M.~H.}\
  \bibnamefont {Devoret}}, \bibinfo {author} {\bibfnamefont {S.~M.}\
  \bibnamefont {Girvin}},\ and\ \bibinfo {author} {\bibfnamefont {R.~J.}\
  \bibnamefont {Schoelkopf}},\ }\bibfield  {title} {\bibinfo {title}
  {Charge-insensitive qubit design derived from the cooper pair box},\ }\href
  {https://doi.org/10.1103/PhysRevA.76.042319} {\bibfield  {journal} {\bibinfo
  {journal} {Phys. Rev. A}\ }\textbf {\bibinfo {volume} {76}},\ \bibinfo
  {pages} {042319} (\bibinfo {year} {2007})}\BibitemShut {NoStop}%
\bibitem [{\citenamefont {Manucharyan}\ \emph {et~al.}(2009)\citenamefont
  {Manucharyan}, \citenamefont {Koch}, \citenamefont {Glazman},\ and\
  \citenamefont {Devoret}}]{Manucharyan2009}%
  \BibitemOpen
  \bibfield  {author} {\bibinfo {author} {\bibfnamefont {V.~E.}\ \bibnamefont
  {Manucharyan}}, \bibinfo {author} {\bibfnamefont {J.}~\bibnamefont {Koch}},
  \bibinfo {author} {\bibfnamefont {L.~I.}\ \bibnamefont {Glazman}},\ and\
  \bibinfo {author} {\bibfnamefont {M.~H.}\ \bibnamefont {Devoret}},\
  }\bibfield  {title} {\bibinfo {title} {Fluxonium: Single cooper-pair circuit
  free of charge offsets},\ }\href
  {https://www.nature.com/articles/ncomms12964} {\bibfield  {journal} {\bibinfo
   {journal} {Science}\ }\textbf {\bibinfo {volume} {326}},\ \bibinfo {pages}
  {113 } (\bibinfo {year} {2009})}\BibitemShut {NoStop}%
\bibitem [{\citenamefont {Yan}\ \emph {et~al.}(2016)\citenamefont {Yan},
  \citenamefont {Gustavsson}, \citenamefont {Kamal}, \citenamefont {Birenbaum},
  \citenamefont {Sears}, \citenamefont {Hover}, \citenamefont {Gudmundsen},
  \citenamefont {Rosenberg}, \citenamefont {Samach}, \citenamefont {Weber},
  \citenamefont {Yoder}, \citenamefont {Orlando}, \citenamefont {Clarke},
  \citenamefont {Kerman},\ and\ \citenamefont {Oliver}}]{Yan2016}%
  \BibitemOpen
  \bibfield  {author} {\bibinfo {author} {\bibfnamefont {F.}~\bibnamefont
  {Yan}}, \bibinfo {author} {\bibfnamefont {S.}~\bibnamefont {Gustavsson}},
  \bibinfo {author} {\bibfnamefont {A.}~\bibnamefont {Kamal}}, \bibinfo
  {author} {\bibfnamefont {J.}~\bibnamefont {Birenbaum}}, \bibinfo {author}
  {\bibfnamefont {A.~P.}\ \bibnamefont {Sears}}, \bibinfo {author}
  {\bibfnamefont {D.~J.}\ \bibnamefont {Hover}}, \bibinfo {author}
  {\bibfnamefont {T.~J.}\ \bibnamefont {Gudmundsen}}, \bibinfo {author}
  {\bibfnamefont {D.}~\bibnamefont {Rosenberg}}, \bibinfo {author}
  {\bibfnamefont {G.~O.}\ \bibnamefont {Samach}}, \bibinfo {author}
  {\bibfnamefont {S.~J.}\ \bibnamefont {Weber}}, \bibinfo {author}
  {\bibfnamefont {J.~L.}\ \bibnamefont {Yoder}}, \bibinfo {author}
  {\bibfnamefont {T.~P.}\ \bibnamefont {Orlando}}, \bibinfo {author}
  {\bibfnamefont {J.}~\bibnamefont {Clarke}}, \bibinfo {author} {\bibfnamefont
  {A.~J.}\ \bibnamefont {Kerman}},\ and\ \bibinfo {author} {\bibfnamefont
  {W.~D.}\ \bibnamefont {Oliver}},\ }\bibfield  {title} {\bibinfo {title} {The
  flux qubit revisited to enhance coherence and reproducibility},\ }\href
  {https://api.semanticscholar.org/CorpusID:17306643} {\bibfield  {journal}
  {\bibinfo  {journal} {Nature Communications}\ }\textbf {\bibinfo {volume}
  {7}} (\bibinfo {year} {2016})}\BibitemShut {NoStop}%
\bibitem [{\citenamefont {Gyenis}\ \emph {et~al.}(2021)\citenamefont {Gyenis},
  \citenamefont {Mundada}, \citenamefont {Di~Paolo}, \citenamefont {Hazard},
  \citenamefont {You}, \citenamefont {Schuster}, \citenamefont {Koch},
  \citenamefont {Blais},\ and\ \citenamefont {Houck}}]{Gyenis2021}%
  \BibitemOpen
  \bibfield  {author} {\bibinfo {author} {\bibfnamefont {A.}~\bibnamefont
  {Gyenis}}, \bibinfo {author} {\bibfnamefont {P.~S.}\ \bibnamefont {Mundada}},
  \bibinfo {author} {\bibfnamefont {A.}~\bibnamefont {Di~Paolo}}, \bibinfo
  {author} {\bibfnamefont {T.~M.}\ \bibnamefont {Hazard}}, \bibinfo {author}
  {\bibfnamefont {X.}~\bibnamefont {You}}, \bibinfo {author} {\bibfnamefont
  {D.~I.}\ \bibnamefont {Schuster}}, \bibinfo {author} {\bibfnamefont
  {J.}~\bibnamefont {Koch}}, \bibinfo {author} {\bibfnamefont {A.}~\bibnamefont
  {Blais}},\ and\ \bibinfo {author} {\bibfnamefont {A.~A.}\ \bibnamefont
  {Houck}},\ }\bibfield  {title} {\bibinfo {title} {Experimental realization of
  a protected superconducting circuit derived from the $0$--$\ensuremath{\pi}$
  qubit},\ }\href {https://doi.org/10.1103/PRXQuantum.2.010339} {\bibfield
  {journal} {\bibinfo  {journal} {PRX Quantum}\ }\textbf {\bibinfo {volume}
  {2}},\ \bibinfo {pages} {010339} (\bibinfo {year} {2021})}\BibitemShut
  {NoStop}%
\bibitem [{\citenamefont {Majer}\ \emph {et~al.}(2007)\citenamefont {Majer},
  \citenamefont {Chow}, \citenamefont {Gambetta}, \citenamefont {Koch},
  \citenamefont {Johnson}, \citenamefont {Schreier}, \citenamefont {Frunzio},
  \citenamefont {Schuster}, \citenamefont {Houck}, \citenamefont {Wallraff},
  \citenamefont {Blais}, \citenamefont {Devoret}, \citenamefont {Girvin},\ and\
  \citenamefont {Schoelkopf}}]{Majer2007}%
  \BibitemOpen
  \bibfield  {author} {\bibinfo {author} {\bibfnamefont {J.}~\bibnamefont
  {Majer}}, \bibinfo {author} {\bibfnamefont {J.}~\bibnamefont {Chow}},
  \bibinfo {author} {\bibfnamefont {J.}~\bibnamefont {Gambetta}}, \bibinfo
  {author} {\bibfnamefont {J.}~\bibnamefont {Koch}}, \bibinfo {author}
  {\bibfnamefont {B.}~\bibnamefont {Johnson}}, \bibinfo {author} {\bibfnamefont
  {J.}~\bibnamefont {Schreier}}, \bibinfo {author} {\bibfnamefont
  {L.}~\bibnamefont {Frunzio}}, \bibinfo {author} {\bibfnamefont
  {D.}~\bibnamefont {Schuster}}, \bibinfo {author} {\bibfnamefont
  {A.}~\bibnamefont {Houck}}, \bibinfo {author} {\bibfnamefont
  {A.}~\bibnamefont {Wallraff}}, \bibinfo {author} {\bibfnamefont
  {A.}~\bibnamefont {Blais}}, \bibinfo {author} {\bibfnamefont
  {M.}~\bibnamefont {Devoret}}, \bibinfo {author} {\bibfnamefont
  {S.}~\bibnamefont {Girvin}},\ and\ \bibinfo {author} {\bibfnamefont
  {R.}~\bibnamefont {Schoelkopf}},\ }\bibfield  {title} {\bibinfo {title}
  {Coupling superconducting qubits via a cavity bus},\ }\href
  {https://doi.org/10.1038/nature06184} {\bibfield  {journal} {\bibinfo
  {journal} {Nature}\ }\textbf {\bibinfo {volume} {449}},\ \bibinfo {pages}
  {443} (\bibinfo {year} {2007})}\BibitemShut {NoStop}%
\bibitem [{\citenamefont {Rigetti}\ and\ \citenamefont
  {Devoret}(2010)}]{Rigetti2010}%
  \BibitemOpen
  \bibfield  {author} {\bibinfo {author} {\bibfnamefont {C.}~\bibnamefont
  {Rigetti}}\ and\ \bibinfo {author} {\bibfnamefont {M.}~\bibnamefont
  {Devoret}},\ }\bibfield  {title} {\bibinfo {title} {Fully microwave-tunable
  universal gates in superconducting qubits with linear couplings and fixed
  transition frequencies},\ }\href {https://doi.org/10.1103/PhysRevB.81.134507}
  {\bibfield  {journal} {\bibinfo  {journal} {Phys. Rev. B}\ }\textbf {\bibinfo
  {volume} {81}},\ \bibinfo {pages} {134507} (\bibinfo {year}
  {2010})}\BibitemShut {NoStop}%
\bibitem [{\citenamefont {Poletto}\ \emph {et~al.}(2012)\citenamefont
  {Poletto}, \citenamefont {Gambetta}, \citenamefont {Merkel}, \citenamefont
  {Smolin}, \citenamefont {Chow}, \citenamefont {C\'orcoles}, \citenamefont
  {Keefe}, \citenamefont {Rothwell}, \citenamefont {Rozen}, \citenamefont
  {Abraham}, \citenamefont {Rigetti},\ and\ \citenamefont
  {Steffen}}]{Poletto2012}%
  \BibitemOpen
  \bibfield  {author} {\bibinfo {author} {\bibfnamefont {S.}~\bibnamefont
  {Poletto}}, \bibinfo {author} {\bibfnamefont {J.~M.}\ \bibnamefont
  {Gambetta}}, \bibinfo {author} {\bibfnamefont {S.~T.}\ \bibnamefont
  {Merkel}}, \bibinfo {author} {\bibfnamefont {J.~A.}\ \bibnamefont {Smolin}},
  \bibinfo {author} {\bibfnamefont {J.~M.}\ \bibnamefont {Chow}}, \bibinfo
  {author} {\bibfnamefont {A.~D.}\ \bibnamefont {C\'orcoles}}, \bibinfo
  {author} {\bibfnamefont {G.~A.}\ \bibnamefont {Keefe}}, \bibinfo {author}
  {\bibfnamefont {M.~B.}\ \bibnamefont {Rothwell}}, \bibinfo {author}
  {\bibfnamefont {J.~R.}\ \bibnamefont {Rozen}}, \bibinfo {author}
  {\bibfnamefont {D.~W.}\ \bibnamefont {Abraham}}, \bibinfo {author}
  {\bibfnamefont {C.}~\bibnamefont {Rigetti}},\ and\ \bibinfo {author}
  {\bibfnamefont {M.}~\bibnamefont {Steffen}},\ }\bibfield  {title} {\bibinfo
  {title} {Entanglement of two superconducting qubits in a waveguide cavity via
  monochromatic two-photon excitation},\ }\href
  {https://doi.org/10.1103/PhysRevLett.109.240505} {\bibfield  {journal}
  {\bibinfo  {journal} {Phys. Rev. Lett.}\ }\textbf {\bibinfo {volume} {109}},\
  \bibinfo {pages} {240505} (\bibinfo {year} {2012})}\BibitemShut {NoStop}%
\bibitem [{\citenamefont {Chow}\ \emph {et~al.}(2013)\citenamefont {Chow},
  \citenamefont {Gambetta}, \citenamefont {Cross}, \citenamefont {Merkel},
  \citenamefont {Rigetti},\ and\ \citenamefont {Steffen}}]{Chow_2013}%
  \BibitemOpen
  \bibfield  {author} {\bibinfo {author} {\bibfnamefont {J.~M.}\ \bibnamefont
  {Chow}}, \bibinfo {author} {\bibfnamefont {J.~M.}\ \bibnamefont {Gambetta}},
  \bibinfo {author} {\bibfnamefont {A.~W.}\ \bibnamefont {Cross}}, \bibinfo
  {author} {\bibfnamefont {S.~T.}\ \bibnamefont {Merkel}}, \bibinfo {author}
  {\bibfnamefont {C.}~\bibnamefont {Rigetti}},\ and\ \bibinfo {author}
  {\bibfnamefont {M.}~\bibnamefont {Steffen}},\ }\bibfield  {title} {\bibinfo
  {title} {Microwave-activated conditional-phase gate for superconducting
  qubits},\ }\href {https://doi.org/10.1088/1367-2630/15/11/115012} {\bibfield
  {journal} {\bibinfo  {journal} {New Journal of Physics}\ }\textbf {\bibinfo
  {volume} {15}},\ \bibinfo {pages} {115012} (\bibinfo {year}
  {2013})}\BibitemShut {NoStop}%
\bibitem [{\citenamefont {Martinis}\ and\ \citenamefont
  {Geller}(2014)}]{martinis_fast_2014}%
  \BibitemOpen
  \bibfield  {author} {\bibinfo {author} {\bibfnamefont {J.~M.}\ \bibnamefont
  {Martinis}}\ and\ \bibinfo {author} {\bibfnamefont {M.~R.}\ \bibnamefont
  {Geller}},\ }\bibfield  {title} {\bibinfo {title} {Fast adiabatic qubit gates
  using only $\sigma_z$ control},\ }\href
  {https://doi.org/10.1103/PhysRevA.90.022307} {\bibfield  {journal} {\bibinfo
  {journal} {Physical Review A}\ }\textbf {\bibinfo {volume} {90}},\ \bibinfo
  {pages} {022307} (\bibinfo {year} {2014})}\BibitemShut {NoStop}%
\bibitem [{\citenamefont {Kelly}\ \emph {et~al.}(2014)\citenamefont {Kelly},
  \citenamefont {Barends}, \citenamefont {Campbell}, \citenamefont {Chen},
  \citenamefont {Chen}, \citenamefont {Chiaro}, \citenamefont {Dunsworth},
  \citenamefont {Fowler}, \citenamefont {Hoi}, \citenamefont {Jeffrey},
  \citenamefont {Megrant}, \citenamefont {Mutus}, \citenamefont {Neill},
  \citenamefont {O'Malley}, \citenamefont {Quintana}, \citenamefont {Roushan},
  \citenamefont {Sank}, \citenamefont {Vainsencher}, \citenamefont {Wenner},
  \citenamefont {White}, \citenamefont {Cleland},\ and\ \citenamefont
  {Martinis}}]{kelly_prl_2014}%
  \BibitemOpen
  \bibfield  {author} {\bibinfo {author} {\bibfnamefont {J.}~\bibnamefont
  {Kelly}}, \bibinfo {author} {\bibfnamefont {R.}~\bibnamefont {Barends}},
  \bibinfo {author} {\bibfnamefont {B.}~\bibnamefont {Campbell}}, \bibinfo
  {author} {\bibfnamefont {Y.}~\bibnamefont {Chen}}, \bibinfo {author}
  {\bibfnamefont {Z.}~\bibnamefont {Chen}}, \bibinfo {author} {\bibfnamefont
  {B.}~\bibnamefont {Chiaro}}, \bibinfo {author} {\bibfnamefont
  {A.}~\bibnamefont {Dunsworth}}, \bibinfo {author} {\bibfnamefont {A.~G.}\
  \bibnamefont {Fowler}}, \bibinfo {author} {\bibfnamefont {I.-C.}\
  \bibnamefont {Hoi}}, \bibinfo {author} {\bibfnamefont {E.}~\bibnamefont
  {Jeffrey}}, \bibinfo {author} {\bibfnamefont {A.}~\bibnamefont {Megrant}},
  \bibinfo {author} {\bibfnamefont {J.}~\bibnamefont {Mutus}}, \bibinfo
  {author} {\bibfnamefont {C.}~\bibnamefont {Neill}}, \bibinfo {author}
  {\bibfnamefont {P.~J.~J.}\ \bibnamefont {O'Malley}}, \bibinfo {author}
  {\bibfnamefont {C.}~\bibnamefont {Quintana}}, \bibinfo {author}
  {\bibfnamefont {P.}~\bibnamefont {Roushan}}, \bibinfo {author} {\bibfnamefont
  {D.}~\bibnamefont {Sank}}, \bibinfo {author} {\bibfnamefont {A.}~\bibnamefont
  {Vainsencher}}, \bibinfo {author} {\bibfnamefont {J.}~\bibnamefont {Wenner}},
  \bibinfo {author} {\bibfnamefont {T.~C.}\ \bibnamefont {White}}, \bibinfo
  {author} {\bibfnamefont {A.~N.}\ \bibnamefont {Cleland}},\ and\ \bibinfo
  {author} {\bibfnamefont {J.~M.}\ \bibnamefont {Martinis}},\ }\bibfield
  {title} {\bibinfo {title} {Optimal quantum control using randomized
  benchmarking},\ }\href {https://doi.org/10.1103/PhysRevLett.112.240504}
  {\bibfield  {journal} {\bibinfo  {journal} {Phys. Rev. Lett.}\ }\textbf
  {\bibinfo {volume} {112}},\ \bibinfo {pages} {240504} (\bibinfo {year}
  {2014})}\BibitemShut {NoStop}%
\bibitem [{\citenamefont {Yan}\ \emph {et~al.}(2018)\citenamefont {Yan},
  \citenamefont {Krantz}, \citenamefont {Sung}, \citenamefont {Kjaergaard},
  \citenamefont {Campbell}, \citenamefont {Orlando}, \citenamefont
  {Gustavsson},\ and\ \citenamefont {Oliver}}]{fei_tunable_2018}%
  \BibitemOpen
  \bibfield  {author} {\bibinfo {author} {\bibfnamefont {F.}~\bibnamefont
  {Yan}}, \bibinfo {author} {\bibfnamefont {P.}~\bibnamefont {Krantz}},
  \bibinfo {author} {\bibfnamefont {Y.}~\bibnamefont {Sung}}, \bibinfo {author}
  {\bibfnamefont {M.}~\bibnamefont {Kjaergaard}}, \bibinfo {author}
  {\bibfnamefont {D.~L.}\ \bibnamefont {Campbell}}, \bibinfo {author}
  {\bibfnamefont {T.~P.}\ \bibnamefont {Orlando}}, \bibinfo {author}
  {\bibfnamefont {S.}~\bibnamefont {Gustavsson}},\ and\ \bibinfo {author}
  {\bibfnamefont {W.~D.}\ \bibnamefont {Oliver}},\ }\bibfield  {title}
  {\bibinfo {title} {Tunable coupling scheme for implementing high-fidelity
  two-qubit gates},\ }\href {https://doi.org/10.1103/PhysRevApplied.10.054062}
  {\bibfield  {journal} {\bibinfo  {journal} {Phys. Rev. Appl.}\ }\textbf
  {\bibinfo {volume} {10}},\ \bibinfo {pages} {054062} (\bibinfo {year}
  {2018})}\BibitemShut {NoStop}%
\bibitem [{\citenamefont {Sung}\ \emph {et~al.}(2021)\citenamefont {Sung},
  \citenamefont {Braumüller}, \citenamefont {Vepsäläinen}, \citenamefont
  {Kannan}, \citenamefont {Kjaergaard}, \citenamefont {Greene}, \citenamefont
  {Samach}, \citenamefont {McNally}, \citenamefont {Kim}, \citenamefont
  {Melville}, \citenamefont {Niedzielski}, \citenamefont {Schwartz},
  \citenamefont {Yoder}, \citenamefont {Orlando}, \citenamefont {Gustavsson},\
  and\ \citenamefont {Oliver}}]{sung_realization_2021}%
  \BibitemOpen
  \bibfield  {author} {\bibinfo {author} {\bibfnamefont {Y.}~\bibnamefont
  {Sung}}, \bibinfo {author} {\bibfnamefont {J.}~\bibnamefont {Braumüller}},
  \bibinfo {author} {\bibfnamefont {A.}~\bibnamefont {Vepsäläinen}}, \bibinfo
  {author} {\bibfnamefont {B.}~\bibnamefont {Kannan}}, \bibinfo {author}
  {\bibfnamefont {M.}~\bibnamefont {Kjaergaard}}, \bibinfo {author}
  {\bibfnamefont {A.}~\bibnamefont {Greene}}, \bibinfo {author} {\bibfnamefont
  {G.~O.}\ \bibnamefont {Samach}}, \bibinfo {author} {\bibfnamefont
  {C.}~\bibnamefont {McNally}}, \bibinfo {author} {\bibfnamefont
  {D.}~\bibnamefont {Kim}}, \bibinfo {author} {\bibfnamefont {A.}~\bibnamefont
  {Melville}}, \bibinfo {author} {\bibfnamefont {B.~M.}\ \bibnamefont
  {Niedzielski}}, \bibinfo {author} {\bibfnamefont {M.~E.}\ \bibnamefont
  {Schwartz}}, \bibinfo {author} {\bibfnamefont {J.~L.}\ \bibnamefont {Yoder}},
  \bibinfo {author} {\bibfnamefont {T.~P.}\ \bibnamefont {Orlando}}, \bibinfo
  {author} {\bibfnamefont {S.}~\bibnamefont {Gustavsson}},\ and\ \bibinfo
  {author} {\bibfnamefont {W.~D.}\ \bibnamefont {Oliver}},\ }\bibfield  {title}
  {\bibinfo {title} {Realization of high-fidelity {CZ} and {ZZ}-free {iSWAP}
  gates with a tunable coupler},\ }\href
  {https://doi.org/10.1103/PhysRevX.11.021058} {\bibfield  {journal} {\bibinfo
  {journal} {Physical Review X}\ }\textbf {\bibinfo {volume} {11}},\ \bibinfo
  {pages} {021058} (\bibinfo {year} {2021})},\ \bibinfo {note} {arXiv:
  2011.01261}\BibitemShut {NoStop}%
\bibitem [{\citenamefont {Zhang}\ \emph {et~al.}(2021)\citenamefont {Zhang},
  \citenamefont {Chakram}, \citenamefont {Roy}, \citenamefont {Earnest},
  \citenamefont {Lu}, \citenamefont {Huang}, \citenamefont {Weiss},
  \citenamefont {Koch},\ and\ \citenamefont {Schuster}}]{Zhang2021}%
  \BibitemOpen
  \bibfield  {author} {\bibinfo {author} {\bibfnamefont {H.}~\bibnamefont
  {Zhang}}, \bibinfo {author} {\bibfnamefont {S.}~\bibnamefont {Chakram}},
  \bibinfo {author} {\bibfnamefont {T.}~\bibnamefont {Roy}}, \bibinfo {author}
  {\bibfnamefont {N.}~\bibnamefont {Earnest}}, \bibinfo {author} {\bibfnamefont
  {Y.}~\bibnamefont {Lu}}, \bibinfo {author} {\bibfnamefont {Z.}~\bibnamefont
  {Huang}}, \bibinfo {author} {\bibfnamefont {D.~K.}\ \bibnamefont {Weiss}},
  \bibinfo {author} {\bibfnamefont {J.}~\bibnamefont {Koch}},\ and\ \bibinfo
  {author} {\bibfnamefont {D.~I.}\ \bibnamefont {Schuster}},\ }\bibfield
  {title} {\bibinfo {title} {Universal fast-flux control of a coherent,
  low-frequency qubit},\ }\href {https://doi.org/10.1103/PhysRevX.11.011010}
  {\bibfield  {journal} {\bibinfo  {journal} {Phys. Rev. X}\ }\textbf {\bibinfo
  {volume} {11}},\ \bibinfo {pages} {011010} (\bibinfo {year}
  {2021})}\BibitemShut {NoStop}%
\bibitem [{\citenamefont {Ding}\ \emph {et~al.}(2023)\citenamefont {Ding},
  \citenamefont {Hays}, \citenamefont {Sung}, \citenamefont {Kannan},
  \citenamefont {An}, \citenamefont {Di~Paolo}, \citenamefont {Karamlou},
  \citenamefont {Hazard}, \citenamefont {Azar}, \citenamefont {Kim},
  \citenamefont {Niedzielski}, \citenamefont {Melville}, \citenamefont
  {Schwartz}, \citenamefont {Yoder}, \citenamefont {Orlando}, \citenamefont
  {Gustavsson}, \citenamefont {Grover}, \citenamefont {Serniak},\ and\
  \citenamefont {Oliver}}]{Ding2023}%
  \BibitemOpen
  \bibfield  {author} {\bibinfo {author} {\bibfnamefont {L.}~\bibnamefont
  {Ding}}, \bibinfo {author} {\bibfnamefont {M.}~\bibnamefont {Hays}}, \bibinfo
  {author} {\bibfnamefont {Y.}~\bibnamefont {Sung}}, \bibinfo {author}
  {\bibfnamefont {B.}~\bibnamefont {Kannan}}, \bibinfo {author} {\bibfnamefont
  {J.}~\bibnamefont {An}}, \bibinfo {author} {\bibfnamefont {A.}~\bibnamefont
  {Di~Paolo}}, \bibinfo {author} {\bibfnamefont {A.~H.}\ \bibnamefont
  {Karamlou}}, \bibinfo {author} {\bibfnamefont {T.~M.}\ \bibnamefont
  {Hazard}}, \bibinfo {author} {\bibfnamefont {K.}~\bibnamefont {Azar}},
  \bibinfo {author} {\bibfnamefont {D.~K.}\ \bibnamefont {Kim}}, \bibinfo
  {author} {\bibfnamefont {B.~M.}\ \bibnamefont {Niedzielski}}, \bibinfo
  {author} {\bibfnamefont {A.}~\bibnamefont {Melville}}, \bibinfo {author}
  {\bibfnamefont {M.~E.}\ \bibnamefont {Schwartz}}, \bibinfo {author}
  {\bibfnamefont {J.~L.}\ \bibnamefont {Yoder}}, \bibinfo {author}
  {\bibfnamefont {T.~P.}\ \bibnamefont {Orlando}}, \bibinfo {author}
  {\bibfnamefont {S.}~\bibnamefont {Gustavsson}}, \bibinfo {author}
  {\bibfnamefont {J.~A.}\ \bibnamefont {Grover}}, \bibinfo {author}
  {\bibfnamefont {K.}~\bibnamefont {Serniak}},\ and\ \bibinfo {author}
  {\bibfnamefont {W.~D.}\ \bibnamefont {Oliver}},\ }\bibfield  {title}
  {\bibinfo {title} {High-fidelity, frequency-flexible two-qubit fluxonium
  gates with a transmon coupler},\ }\href
  {https://doi.org/10.1103/PhysRevX.13.031035} {\bibfield  {journal} {\bibinfo
  {journal} {Phys. Rev. X}\ }\textbf {\bibinfo {volume} {13}},\ \bibinfo
  {pages} {031035} (\bibinfo {year} {2023})}\BibitemShut {NoStop}%
\bibitem [{\citenamefont {Chow}\ \emph {et~al.}(2012)\citenamefont {Chow},
  \citenamefont {Gambetta}, \citenamefont {C\'orcoles}, \citenamefont {Merkel},
  \citenamefont {Smolin}, \citenamefont {Rigetti}, \citenamefont {Poletto},
  \citenamefont {Keefe}, \citenamefont {Rothwell}, \citenamefont {Rozen},
  \citenamefont {Ketchen},\ and\ \citenamefont {Steffen}}]{Chow2012}%
  \BibitemOpen
  \bibfield  {author} {\bibinfo {author} {\bibfnamefont {J.~M.}\ \bibnamefont
  {Chow}}, \bibinfo {author} {\bibfnamefont {J.~M.}\ \bibnamefont {Gambetta}},
  \bibinfo {author} {\bibfnamefont {A.~D.}\ \bibnamefont {C\'orcoles}},
  \bibinfo {author} {\bibfnamefont {S.~T.}\ \bibnamefont {Merkel}}, \bibinfo
  {author} {\bibfnamefont {J.~A.}\ \bibnamefont {Smolin}}, \bibinfo {author}
  {\bibfnamefont {C.}~\bibnamefont {Rigetti}}, \bibinfo {author} {\bibfnamefont
  {S.}~\bibnamefont {Poletto}}, \bibinfo {author} {\bibfnamefont {G.~A.}\
  \bibnamefont {Keefe}}, \bibinfo {author} {\bibfnamefont {M.~B.}\ \bibnamefont
  {Rothwell}}, \bibinfo {author} {\bibfnamefont {J.~R.}\ \bibnamefont {Rozen}},
  \bibinfo {author} {\bibfnamefont {M.~B.}\ \bibnamefont {Ketchen}},\ and\
  \bibinfo {author} {\bibfnamefont {M.}~\bibnamefont {Steffen}},\ }\bibfield
  {title} {\bibinfo {title} {Universal quantum gate set approaching
  fault-tolerant thresholds with superconducting qubits},\ }\href
  {https://doi.org/10.1103/PhysRevLett.109.060501} {\bibfield  {journal}
  {\bibinfo  {journal} {Phys. Rev. Lett.}\ }\textbf {\bibinfo {volume} {109}},\
  \bibinfo {pages} {060501} (\bibinfo {year} {2012})}\BibitemShut {NoStop}%
\bibitem [{\citenamefont {Cross}\ and\ \citenamefont
  {Gambetta}(2015)}]{Gambetta2015}%
  \BibitemOpen
  \bibfield  {author} {\bibinfo {author} {\bibfnamefont {A.~W.}\ \bibnamefont
  {Cross}}\ and\ \bibinfo {author} {\bibfnamefont {J.~M.}\ \bibnamefont
  {Gambetta}},\ }\bibfield  {title} {\bibinfo {title} {Optimized pulse shapes
  for a resonator-induced phase gate},\ }\href
  {https://doi.org/10.1103/PhysRevA.91.032325} {\bibfield  {journal} {\bibinfo
  {journal} {Phys. Rev. A}\ }\textbf {\bibinfo {volume} {91}},\ \bibinfo
  {pages} {032325} (\bibinfo {year} {2015})}\BibitemShut {NoStop}%
\bibitem [{\citenamefont {Paik}\ \emph {et~al.}(2016)\citenamefont {Paik},
  \citenamefont {Mezzacapo}, \citenamefont {Sandberg}, \citenamefont {McClure},
  \citenamefont {Abdo}, \citenamefont {C\'orcoles}, \citenamefont {Dial},
  \citenamefont {Bogorin}, \citenamefont {Plourde}, \citenamefont {Steffen},
  \citenamefont {Cross}, \citenamefont {Gambetta},\ and\ \citenamefont
  {Chow}}]{Paik2016}%
  \BibitemOpen
  \bibfield  {author} {\bibinfo {author} {\bibfnamefont {H.}~\bibnamefont
  {Paik}}, \bibinfo {author} {\bibfnamefont {A.}~\bibnamefont {Mezzacapo}},
  \bibinfo {author} {\bibfnamefont {M.}~\bibnamefont {Sandberg}}, \bibinfo
  {author} {\bibfnamefont {D.~T.}\ \bibnamefont {McClure}}, \bibinfo {author}
  {\bibfnamefont {B.}~\bibnamefont {Abdo}}, \bibinfo {author} {\bibfnamefont
  {A.~D.}\ \bibnamefont {C\'orcoles}}, \bibinfo {author} {\bibfnamefont
  {O.}~\bibnamefont {Dial}}, \bibinfo {author} {\bibfnamefont {D.~F.}\
  \bibnamefont {Bogorin}}, \bibinfo {author} {\bibfnamefont {B.~L.~T.}\
  \bibnamefont {Plourde}}, \bibinfo {author} {\bibfnamefont {M.}~\bibnamefont
  {Steffen}}, \bibinfo {author} {\bibfnamefont {A.~W.}\ \bibnamefont {Cross}},
  \bibinfo {author} {\bibfnamefont {J.~M.}\ \bibnamefont {Gambetta}},\ and\
  \bibinfo {author} {\bibfnamefont {J.~M.}\ \bibnamefont {Chow}},\ }\bibfield
  {title} {\bibinfo {title} {Experimental demonstration of a resonator-induced
  phase gate in a multiqubit circuit-qed system},\ }\href
  {https://doi.org/10.1103/PhysRevLett.117.250502} {\bibfield  {journal}
  {\bibinfo  {journal} {Phys. Rev. Lett.}\ }\textbf {\bibinfo {volume} {117}},\
  \bibinfo {pages} {250502} (\bibinfo {year} {2016})}\BibitemShut {NoStop}%
\bibitem [{\citenamefont {Krinner}\ \emph {et~al.}(2020)\citenamefont
  {Krinner}, \citenamefont {Kurpiers}, \citenamefont {Royer}, \citenamefont
  {Magnard}, \citenamefont {Tsitsilin}, \citenamefont {Besse}, \citenamefont
  {Remm}, \citenamefont {Blais},\ and\ \citenamefont {Wallraff}}]{Krinner2020}%
  \BibitemOpen
  \bibfield  {author} {\bibinfo {author} {\bibfnamefont {S.}~\bibnamefont
  {Krinner}}, \bibinfo {author} {\bibfnamefont {P.}~\bibnamefont {Kurpiers}},
  \bibinfo {author} {\bibfnamefont {B.}~\bibnamefont {Royer}}, \bibinfo
  {author} {\bibfnamefont {P.}~\bibnamefont {Magnard}}, \bibinfo {author}
  {\bibfnamefont {I.}~\bibnamefont {Tsitsilin}}, \bibinfo {author}
  {\bibfnamefont {J.-C.}\ \bibnamefont {Besse}}, \bibinfo {author}
  {\bibfnamefont {A.}~\bibnamefont {Remm}}, \bibinfo {author} {\bibfnamefont
  {A.}~\bibnamefont {Blais}},\ and\ \bibinfo {author} {\bibfnamefont
  {A.}~\bibnamefont {Wallraff}},\ }\bibfield  {title} {\bibinfo {title}
  {Demonstration of an all-microwave controlled-phase gate between far-detuned
  qubits},\ }\href {https://doi.org/10.1103/PhysRevApplied.14.044039}
  {\bibfield  {journal} {\bibinfo  {journal} {Phys. Rev. Appl.}\ }\textbf
  {\bibinfo {volume} {14}},\ \bibinfo {pages} {044039} (\bibinfo {year}
  {2020})}\BibitemShut {NoStop}%
\bibitem [{\citenamefont {Mitchell}\ \emph {et~al.}(2021)\citenamefont
  {Mitchell}, \citenamefont {Naik}, \citenamefont {Morvan}, \citenamefont
  {Hashim}, \citenamefont {Kreikebaum}, \citenamefont {Marinelli},
  \citenamefont {Lavrijsen}, \citenamefont {Nowrouzi}, \citenamefont
  {Santiago},\ and\ \citenamefont {Siddiqi}}]{Mitchell2021}%
  \BibitemOpen
  \bibfield  {author} {\bibinfo {author} {\bibfnamefont {B.~K.}\ \bibnamefont
  {Mitchell}}, \bibinfo {author} {\bibfnamefont {R.~K.}\ \bibnamefont {Naik}},
  \bibinfo {author} {\bibfnamefont {A.}~\bibnamefont {Morvan}}, \bibinfo
  {author} {\bibfnamefont {A.}~\bibnamefont {Hashim}}, \bibinfo {author}
  {\bibfnamefont {J.~M.}\ \bibnamefont {Kreikebaum}}, \bibinfo {author}
  {\bibfnamefont {B.}~\bibnamefont {Marinelli}}, \bibinfo {author}
  {\bibfnamefont {W.}~\bibnamefont {Lavrijsen}}, \bibinfo {author}
  {\bibfnamefont {K.}~\bibnamefont {Nowrouzi}}, \bibinfo {author}
  {\bibfnamefont {D.~I.}\ \bibnamefont {Santiago}},\ and\ \bibinfo {author}
  {\bibfnamefont {I.}~\bibnamefont {Siddiqi}},\ }\bibfield  {title} {\bibinfo
  {title} {Hardware-efficient microwave-activated tunable coupling between
  superconducting qubits},\ }\href
  {https://doi.org/10.1103/PhysRevLett.127.200502} {\bibfield  {journal}
  {\bibinfo  {journal} {Phys. Rev. Lett.}\ }\textbf {\bibinfo {volume} {127}},\
  \bibinfo {pages} {200502} (\bibinfo {year} {2021})}\BibitemShut {NoStop}%
\bibitem [{\citenamefont {Kandala}\ \emph {et~al.}(2021)\citenamefont
  {Kandala}, \citenamefont {Wei}, \citenamefont {Srinivasan}, \citenamefont
  {Magesan}, \citenamefont {Carnevale}, \citenamefont {Keefe}, \citenamefont
  {Klaus}, \citenamefont {Dial},\ and\ \citenamefont {McKay}}]{Kandala2021}%
  \BibitemOpen
  \bibfield  {author} {\bibinfo {author} {\bibfnamefont {A.}~\bibnamefont
  {Kandala}}, \bibinfo {author} {\bibfnamefont {K.~X.}\ \bibnamefont {Wei}},
  \bibinfo {author} {\bibfnamefont {S.}~\bibnamefont {Srinivasan}}, \bibinfo
  {author} {\bibfnamefont {E.}~\bibnamefont {Magesan}}, \bibinfo {author}
  {\bibfnamefont {S.}~\bibnamefont {Carnevale}}, \bibinfo {author}
  {\bibfnamefont {G.~A.}\ \bibnamefont {Keefe}}, \bibinfo {author}
  {\bibfnamefont {D.}~\bibnamefont {Klaus}}, \bibinfo {author} {\bibfnamefont
  {O.}~\bibnamefont {Dial}},\ and\ \bibinfo {author} {\bibfnamefont {D.~C.}\
  \bibnamefont {McKay}},\ }\bibfield  {title} {\bibinfo {title} {Demonstration
  of a high-fidelity cnot gate for fixed-frequency transmons with engineered
  $zz$ suppression},\ }\href {https://doi.org/10.1103/PhysRevLett.127.130501}
  {\bibfield  {journal} {\bibinfo  {journal} {Phys. Rev. Lett.}\ }\textbf
  {\bibinfo {volume} {127}},\ \bibinfo {pages} {130501} (\bibinfo {year}
  {2021})}\BibitemShut {NoStop}%
\bibitem [{\citenamefont {Neeley}\ \emph {et~al.}(2010)\citenamefont {Neeley},
  \citenamefont {Bialczak}, \citenamefont {Lenander}, \citenamefont {Lucero},
  \citenamefont {Mariantoni}, \citenamefont {O'Connell}, \citenamefont {Sank},
  \citenamefont {Wang}, \citenamefont {Weides}, \citenamefont {Wenner},
  \citenamefont {Yin}, \citenamefont {Yamamoto}, \citenamefont {Cleland},\ and\
  \citenamefont {Martinis}}]{Neeley2010}%
  \BibitemOpen
  \bibfield  {author} {\bibinfo {author} {\bibfnamefont {M.}~\bibnamefont
  {Neeley}}, \bibinfo {author} {\bibfnamefont {R.}~\bibnamefont {Bialczak}},
  \bibinfo {author} {\bibfnamefont {M.}~\bibnamefont {Lenander}}, \bibinfo
  {author} {\bibfnamefont {E.}~\bibnamefont {Lucero}}, \bibinfo {author}
  {\bibfnamefont {M.}~\bibnamefont {Mariantoni}}, \bibinfo {author}
  {\bibfnamefont {A.}~\bibnamefont {O'Connell}}, \bibinfo {author}
  {\bibfnamefont {D.}~\bibnamefont {Sank}}, \bibinfo {author} {\bibfnamefont
  {H.}~\bibnamefont {Wang}}, \bibinfo {author} {\bibfnamefont {M.}~\bibnamefont
  {Weides}}, \bibinfo {author} {\bibfnamefont {J.}~\bibnamefont {Wenner}},
  \bibinfo {author} {\bibfnamefont {Y.}~\bibnamefont {Yin}}, \bibinfo {author}
  {\bibfnamefont {T.}~\bibnamefont {Yamamoto}}, \bibinfo {author}
  {\bibfnamefont {A.}~\bibnamefont {Cleland}},\ and\ \bibinfo {author}
  {\bibfnamefont {J.}~\bibnamefont {Martinis}},\ }\bibfield  {title} {\bibinfo
  {title} {Generation of three-qubit entangled states using superconducting
  phase qubits},\ }\href {https://doi.org/10.1038/nature09418} {\bibfield
  {journal} {\bibinfo  {journal} {Nature}\ }\textbf {\bibinfo {volume} {467}},\
  \bibinfo {pages} {570} (\bibinfo {year} {2010})}\BibitemShut {NoStop}%
\bibitem [{\citenamefont {Chen}\ \emph {et~al.}(2014)\citenamefont {Chen},
  \citenamefont {Neill}, \citenamefont {Roushan}, \citenamefont {Leung},
  \citenamefont {Fang}, \citenamefont {Barends}, \citenamefont {Kelly},
  \citenamefont {Campbell}, \citenamefont {Chen}, \citenamefont {Chiaro},
  \citenamefont {Dunsworth}, \citenamefont {Jeffrey}, \citenamefont {Megrant},
  \citenamefont {Mutus}, \citenamefont {O'Malley}, \citenamefont {Quintana},
  \citenamefont {Sank}, \citenamefont {Vainsencher}, \citenamefont {Wenner},
  \citenamefont {White}, \citenamefont {Geller}, \citenamefont {Cleland},\ and\
  \citenamefont {Martinis}}]{Chen2014}%
  \BibitemOpen
  \bibfield  {author} {\bibinfo {author} {\bibfnamefont {Y.}~\bibnamefont
  {Chen}}, \bibinfo {author} {\bibfnamefont {C.}~\bibnamefont {Neill}},
  \bibinfo {author} {\bibfnamefont {P.}~\bibnamefont {Roushan}}, \bibinfo
  {author} {\bibfnamefont {N.}~\bibnamefont {Leung}}, \bibinfo {author}
  {\bibfnamefont {M.}~\bibnamefont {Fang}}, \bibinfo {author} {\bibfnamefont
  {R.}~\bibnamefont {Barends}}, \bibinfo {author} {\bibfnamefont
  {J.}~\bibnamefont {Kelly}}, \bibinfo {author} {\bibfnamefont
  {B.}~\bibnamefont {Campbell}}, \bibinfo {author} {\bibfnamefont
  {Z.}~\bibnamefont {Chen}}, \bibinfo {author} {\bibfnamefont {B.}~\bibnamefont
  {Chiaro}}, \bibinfo {author} {\bibfnamefont {A.}~\bibnamefont {Dunsworth}},
  \bibinfo {author} {\bibfnamefont {E.}~\bibnamefont {Jeffrey}}, \bibinfo
  {author} {\bibfnamefont {A.}~\bibnamefont {Megrant}}, \bibinfo {author}
  {\bibfnamefont {J.~Y.}\ \bibnamefont {Mutus}}, \bibinfo {author}
  {\bibfnamefont {P.~J.~J.}\ \bibnamefont {O'Malley}}, \bibinfo {author}
  {\bibfnamefont {C.~M.}\ \bibnamefont {Quintana}}, \bibinfo {author}
  {\bibfnamefont {D.}~\bibnamefont {Sank}}, \bibinfo {author} {\bibfnamefont
  {A.}~\bibnamefont {Vainsencher}}, \bibinfo {author} {\bibfnamefont
  {J.}~\bibnamefont {Wenner}}, \bibinfo {author} {\bibfnamefont {T.~C.}\
  \bibnamefont {White}}, \bibinfo {author} {\bibfnamefont {M.~R.}\ \bibnamefont
  {Geller}}, \bibinfo {author} {\bibfnamefont {A.~N.}\ \bibnamefont
  {Cleland}},\ and\ \bibinfo {author} {\bibfnamefont {J.~M.}\ \bibnamefont
  {Martinis}},\ }\bibfield  {title} {\bibinfo {title} {Qubit architecture with
  high coherence and fast tunable coupling},\ }\href
  {https://doi.org/10.1103/PhysRevLett.113.220502} {\bibfield  {journal}
  {\bibinfo  {journal} {Phys. Rev. Lett.}\ }\textbf {\bibinfo {volume} {113}},\
  \bibinfo {pages} {220502} (\bibinfo {year} {2014})}\BibitemShut {NoStop}%
\bibitem [{\citenamefont {Barends}\ \emph {et~al.}(2019)\citenamefont
  {Barends}, \citenamefont {Quintana}, \citenamefont {Petukhov}, \citenamefont
  {Chen}, \citenamefont {Kafri}, \citenamefont {Kechedzhi}, \citenamefont
  {Collins}, \citenamefont {Naaman}, \citenamefont {Boixo}, \citenamefont
  {Arute}, \citenamefont {Arya}, \citenamefont {Buell}, \citenamefont
  {Burkett}, \citenamefont {Chen}, \citenamefont {Chiaro}, \citenamefont
  {Dunsworth}, \citenamefont {Foxen}, \citenamefont {Fowler}, \citenamefont
  {Gidney}, \citenamefont {Giustina}, \citenamefont {Graff}, \citenamefont
  {Huang}, \citenamefont {Jeffrey}, \citenamefont {Kelly}, \citenamefont
  {Klimov}, \citenamefont {Kostritsa}, \citenamefont {Landhuis}, \citenamefont
  {Lucero}, \citenamefont {McEwen}, \citenamefont {Megrant}, \citenamefont
  {Mi}, \citenamefont {Mutus}, \citenamefont {Neeley}, \citenamefont {Neill},
  \citenamefont {Ostby}, \citenamefont {Roushan}, \citenamefont {Sank},
  \citenamefont {Satzinger}, \citenamefont {Vainsencher}, \citenamefont
  {White}, \citenamefont {Yao}, \citenamefont {Yeh}, \citenamefont {Zalcman},
  \citenamefont {Neven}, \citenamefont {Smelyanskiy},\ and\ \citenamefont
  {Martinis}}]{Barends2019}%
  \BibitemOpen
  \bibfield  {author} {\bibinfo {author} {\bibfnamefont {R.}~\bibnamefont
  {Barends}}, \bibinfo {author} {\bibfnamefont {C.~M.}\ \bibnamefont
  {Quintana}}, \bibinfo {author} {\bibfnamefont {A.~G.}\ \bibnamefont
  {Petukhov}}, \bibinfo {author} {\bibfnamefont {Y.}~\bibnamefont {Chen}},
  \bibinfo {author} {\bibfnamefont {D.}~\bibnamefont {Kafri}}, \bibinfo
  {author} {\bibfnamefont {K.}~\bibnamefont {Kechedzhi}}, \bibinfo {author}
  {\bibfnamefont {R.}~\bibnamefont {Collins}}, \bibinfo {author} {\bibfnamefont
  {O.}~\bibnamefont {Naaman}}, \bibinfo {author} {\bibfnamefont
  {S.}~\bibnamefont {Boixo}}, \bibinfo {author} {\bibfnamefont
  {F.}~\bibnamefont {Arute}}, \bibinfo {author} {\bibfnamefont
  {K.}~\bibnamefont {Arya}}, \bibinfo {author} {\bibfnamefont {D.}~\bibnamefont
  {Buell}}, \bibinfo {author} {\bibfnamefont {B.}~\bibnamefont {Burkett}},
  \bibinfo {author} {\bibfnamefont {Z.}~\bibnamefont {Chen}}, \bibinfo {author}
  {\bibfnamefont {B.}~\bibnamefont {Chiaro}}, \bibinfo {author} {\bibfnamefont
  {A.}~\bibnamefont {Dunsworth}}, \bibinfo {author} {\bibfnamefont
  {B.}~\bibnamefont {Foxen}}, \bibinfo {author} {\bibfnamefont
  {A.}~\bibnamefont {Fowler}}, \bibinfo {author} {\bibfnamefont
  {C.}~\bibnamefont {Gidney}}, \bibinfo {author} {\bibfnamefont
  {M.}~\bibnamefont {Giustina}}, \bibinfo {author} {\bibfnamefont
  {R.}~\bibnamefont {Graff}}, \bibinfo {author} {\bibfnamefont
  {T.}~\bibnamefont {Huang}}, \bibinfo {author} {\bibfnamefont
  {E.}~\bibnamefont {Jeffrey}}, \bibinfo {author} {\bibfnamefont
  {J.}~\bibnamefont {Kelly}}, \bibinfo {author} {\bibfnamefont {P.~V.}\
  \bibnamefont {Klimov}}, \bibinfo {author} {\bibfnamefont {F.}~\bibnamefont
  {Kostritsa}}, \bibinfo {author} {\bibfnamefont {D.}~\bibnamefont {Landhuis}},
  \bibinfo {author} {\bibfnamefont {E.}~\bibnamefont {Lucero}}, \bibinfo
  {author} {\bibfnamefont {M.}~\bibnamefont {McEwen}}, \bibinfo {author}
  {\bibfnamefont {A.}~\bibnamefont {Megrant}}, \bibinfo {author} {\bibfnamefont
  {X.}~\bibnamefont {Mi}}, \bibinfo {author} {\bibfnamefont {J.}~\bibnamefont
  {Mutus}}, \bibinfo {author} {\bibfnamefont {M.}~\bibnamefont {Neeley}},
  \bibinfo {author} {\bibfnamefont {C.}~\bibnamefont {Neill}}, \bibinfo
  {author} {\bibfnamefont {E.}~\bibnamefont {Ostby}}, \bibinfo {author}
  {\bibfnamefont {P.}~\bibnamefont {Roushan}}, \bibinfo {author} {\bibfnamefont
  {D.}~\bibnamefont {Sank}}, \bibinfo {author} {\bibfnamefont {K.~J.}\
  \bibnamefont {Satzinger}}, \bibinfo {author} {\bibfnamefont {A.}~\bibnamefont
  {Vainsencher}}, \bibinfo {author} {\bibfnamefont {T.}~\bibnamefont {White}},
  \bibinfo {author} {\bibfnamefont {J.}~\bibnamefont {Yao}}, \bibinfo {author}
  {\bibfnamefont {P.}~\bibnamefont {Yeh}}, \bibinfo {author} {\bibfnamefont
  {A.}~\bibnamefont {Zalcman}}, \bibinfo {author} {\bibfnamefont
  {H.}~\bibnamefont {Neven}}, \bibinfo {author} {\bibfnamefont {V.~N.}\
  \bibnamefont {Smelyanskiy}},\ and\ \bibinfo {author} {\bibfnamefont {J.~M.}\
  \bibnamefont {Martinis}},\ }\bibfield  {title} {\bibinfo {title} {Diabatic
  gates for frequency-tunable superconducting qubits},\ }\href
  {https://doi.org/10.1103/PhysRevLett.123.210501} {\bibfield  {journal}
  {\bibinfo  {journal} {Phys. Rev. Lett.}\ }\textbf {\bibinfo {volume} {123}},\
  \bibinfo {pages} {210501} (\bibinfo {year} {2019})}\BibitemShut {NoStop}%
\bibitem [{\citenamefont {Caldwell}\ \emph {et~al.}(2018)\citenamefont
  {Caldwell}, \citenamefont {Didier}, \citenamefont {Ryan}, \citenamefont
  {Sete}, \citenamefont {Hudson}, \citenamefont {Karalekas}, \citenamefont
  {Manenti}, \citenamefont {da~Silva}, \citenamefont {Sinclair}, \citenamefont
  {Acala}, \citenamefont {Alidoust}, \citenamefont {Angeles}, \citenamefont
  {Bestwick}, \citenamefont {Block}, \citenamefont {Bloom}, \citenamefont
  {Bradley}, \citenamefont {Bui}, \citenamefont {Capelluto}, \citenamefont
  {Chilcott}, \citenamefont {Cordova}, \citenamefont {Crossman}, \citenamefont
  {Curtis}, \citenamefont {Deshpande}, \citenamefont {Bouayadi}, \citenamefont
  {Girshovich}, \citenamefont {Hong}, \citenamefont {Kuang}, \citenamefont
  {Lenihan}, \citenamefont {Manning}, \citenamefont {Marchenkov}, \citenamefont
  {Marshall}, \citenamefont {Maydra}, \citenamefont {Mohan}, \citenamefont
  {O'Brien}, \citenamefont {Osborn}, \citenamefont {Otterbach}, \citenamefont
  {Papageorge}, \citenamefont {Paquette}, \citenamefont {Pelstring},
  \citenamefont {Polloreno}, \citenamefont {Prawiroatmodjo}, \citenamefont
  {Rawat}, \citenamefont {Reagor}, \citenamefont {Renzas}, \citenamefont
  {Rubin}, \citenamefont {Russell}, \citenamefont {Rust}, \citenamefont
  {Scarabelli}, \citenamefont {Scheer}, \citenamefont {Selvanayagam},
  \citenamefont {Smith}, \citenamefont {Staley}, \citenamefont {Suska},
  \citenamefont {Tezak}, \citenamefont {Thompson}, \citenamefont {To},
  \citenamefont {Vahidpour}, \citenamefont {Vodrahalli}, \citenamefont
  {Whyland}, \citenamefont {Yadav}, \citenamefont {Zeng},\ and\ \citenamefont
  {Rigetti}}]{Caldwell2018}%
  \BibitemOpen
  \bibfield  {author} {\bibinfo {author} {\bibfnamefont {S.~A.}\ \bibnamefont
  {Caldwell}}, \bibinfo {author} {\bibfnamefont {N.}~\bibnamefont {Didier}},
  \bibinfo {author} {\bibfnamefont {C.~A.}\ \bibnamefont {Ryan}}, \bibinfo
  {author} {\bibfnamefont {E.~A.}\ \bibnamefont {Sete}}, \bibinfo {author}
  {\bibfnamefont {A.}~\bibnamefont {Hudson}}, \bibinfo {author} {\bibfnamefont
  {P.}~\bibnamefont {Karalekas}}, \bibinfo {author} {\bibfnamefont
  {R.}~\bibnamefont {Manenti}}, \bibinfo {author} {\bibfnamefont {M.~P.}\
  \bibnamefont {da~Silva}}, \bibinfo {author} {\bibfnamefont {R.}~\bibnamefont
  {Sinclair}}, \bibinfo {author} {\bibfnamefont {E.}~\bibnamefont {Acala}},
  \bibinfo {author} {\bibfnamefont {N.}~\bibnamefont {Alidoust}}, \bibinfo
  {author} {\bibfnamefont {J.}~\bibnamefont {Angeles}}, \bibinfo {author}
  {\bibfnamefont {A.}~\bibnamefont {Bestwick}}, \bibinfo {author}
  {\bibfnamefont {M.}~\bibnamefont {Block}}, \bibinfo {author} {\bibfnamefont
  {B.}~\bibnamefont {Bloom}}, \bibinfo {author} {\bibfnamefont
  {A.}~\bibnamefont {Bradley}}, \bibinfo {author} {\bibfnamefont
  {C.}~\bibnamefont {Bui}}, \bibinfo {author} {\bibfnamefont {L.}~\bibnamefont
  {Capelluto}}, \bibinfo {author} {\bibfnamefont {R.}~\bibnamefont {Chilcott}},
  \bibinfo {author} {\bibfnamefont {J.}~\bibnamefont {Cordova}}, \bibinfo
  {author} {\bibfnamefont {G.}~\bibnamefont {Crossman}}, \bibinfo {author}
  {\bibfnamefont {M.}~\bibnamefont {Curtis}}, \bibinfo {author} {\bibfnamefont
  {S.}~\bibnamefont {Deshpande}}, \bibinfo {author} {\bibfnamefont {T.~E.}\
  \bibnamefont {Bouayadi}}, \bibinfo {author} {\bibfnamefont {D.}~\bibnamefont
  {Girshovich}}, \bibinfo {author} {\bibfnamefont {S.}~\bibnamefont {Hong}},
  \bibinfo {author} {\bibfnamefont {K.}~\bibnamefont {Kuang}}, \bibinfo
  {author} {\bibfnamefont {M.}~\bibnamefont {Lenihan}}, \bibinfo {author}
  {\bibfnamefont {T.}~\bibnamefont {Manning}}, \bibinfo {author} {\bibfnamefont
  {A.}~\bibnamefont {Marchenkov}}, \bibinfo {author} {\bibfnamefont
  {J.}~\bibnamefont {Marshall}}, \bibinfo {author} {\bibfnamefont
  {R.}~\bibnamefont {Maydra}}, \bibinfo {author} {\bibfnamefont
  {Y.}~\bibnamefont {Mohan}}, \bibinfo {author} {\bibfnamefont
  {W.}~\bibnamefont {O'Brien}}, \bibinfo {author} {\bibfnamefont
  {C.}~\bibnamefont {Osborn}}, \bibinfo {author} {\bibfnamefont
  {J.}~\bibnamefont {Otterbach}}, \bibinfo {author} {\bibfnamefont
  {A.}~\bibnamefont {Papageorge}}, \bibinfo {author} {\bibfnamefont {J.-P.}\
  \bibnamefont {Paquette}}, \bibinfo {author} {\bibfnamefont {M.}~\bibnamefont
  {Pelstring}}, \bibinfo {author} {\bibfnamefont {A.}~\bibnamefont
  {Polloreno}}, \bibinfo {author} {\bibfnamefont {G.}~\bibnamefont
  {Prawiroatmodjo}}, \bibinfo {author} {\bibfnamefont {V.}~\bibnamefont
  {Rawat}}, \bibinfo {author} {\bibfnamefont {M.}~\bibnamefont {Reagor}},
  \bibinfo {author} {\bibfnamefont {R.}~\bibnamefont {Renzas}}, \bibinfo
  {author} {\bibfnamefont {N.}~\bibnamefont {Rubin}}, \bibinfo {author}
  {\bibfnamefont {D.}~\bibnamefont {Russell}}, \bibinfo {author} {\bibfnamefont
  {M.}~\bibnamefont {Rust}}, \bibinfo {author} {\bibfnamefont {D.}~\bibnamefont
  {Scarabelli}}, \bibinfo {author} {\bibfnamefont {M.}~\bibnamefont {Scheer}},
  \bibinfo {author} {\bibfnamefont {M.}~\bibnamefont {Selvanayagam}}, \bibinfo
  {author} {\bibfnamefont {R.}~\bibnamefont {Smith}}, \bibinfo {author}
  {\bibfnamefont {A.}~\bibnamefont {Staley}}, \bibinfo {author} {\bibfnamefont
  {M.}~\bibnamefont {Suska}}, \bibinfo {author} {\bibfnamefont
  {N.}~\bibnamefont {Tezak}}, \bibinfo {author} {\bibfnamefont {D.~C.}\
  \bibnamefont {Thompson}}, \bibinfo {author} {\bibfnamefont {T.-W.}\
  \bibnamefont {To}}, \bibinfo {author} {\bibfnamefont {M.}~\bibnamefont
  {Vahidpour}}, \bibinfo {author} {\bibfnamefont {N.}~\bibnamefont
  {Vodrahalli}}, \bibinfo {author} {\bibfnamefont {T.}~\bibnamefont {Whyland}},
  \bibinfo {author} {\bibfnamefont {K.}~\bibnamefont {Yadav}}, \bibinfo
  {author} {\bibfnamefont {W.}~\bibnamefont {Zeng}},\ and\ \bibinfo {author}
  {\bibfnamefont {C.}~\bibnamefont {Rigetti}},\ }\bibfield  {title} {\bibinfo
  {title} {Parametrically activated entangling gates using transmon qubits},\
  }\href {https://doi.org/10.1103/PhysRevApplied.10.034050} {\bibfield
  {journal} {\bibinfo  {journal} {Phys. Rev. Appl.}\ }\textbf {\bibinfo
  {volume} {10}},\ \bibinfo {pages} {034050} (\bibinfo {year}
  {2018})}\BibitemShut {NoStop}%
\bibitem [{\citenamefont {Paolo}\ \emph {et~al.}(2022)\citenamefont {Paolo},
  \citenamefont {Leroux}, \citenamefont {Hazard}, \citenamefont {Serniak},
  \citenamefont {Gustavsson}, \citenamefont {Blais},\ and\ \citenamefont
  {Oliver}}]{dipaolo2022extensible}%
  \BibitemOpen
  \bibfield  {author} {\bibinfo {author} {\bibfnamefont {A.~D.}\ \bibnamefont
  {Paolo}}, \bibinfo {author} {\bibfnamefont {C.}~\bibnamefont {Leroux}},
  \bibinfo {author} {\bibfnamefont {T.~M.}\ \bibnamefont {Hazard}}, \bibinfo
  {author} {\bibfnamefont {K.}~\bibnamefont {Serniak}}, \bibinfo {author}
  {\bibfnamefont {S.}~\bibnamefont {Gustavsson}}, \bibinfo {author}
  {\bibfnamefont {A.}~\bibnamefont {Blais}},\ and\ \bibinfo {author}
  {\bibfnamefont {W.~D.}\ \bibnamefont {Oliver}},\ }\href@noop {} {\bibinfo
  {title} {Extensible circuit-qed architecture via amplitude- and
  frequency-variable microwaves}} (\bibinfo {year} {2022}),\ \Eprint
  {https://arxiv.org/abs/2204.08098} {arXiv:2204.08098 [quant-ph]} \BibitemShut
  {NoStop}%
\bibitem [{\citenamefont {Barends}\ \emph {et~al.}(2014)\citenamefont
  {Barends}, \citenamefont {Kelly}, \citenamefont {Megrant}, \citenamefont
  {Veitia}, \citenamefont {Sank}, \citenamefont {Jeffrey}, \citenamefont
  {White}, \citenamefont {Mutus}, \citenamefont {Fowler}, \citenamefont
  {Campbell}, \citenamefont {Chen}, \citenamefont {Chen}, \citenamefont
  {Chiaro}, \citenamefont {Dunsworth}, \citenamefont {Neill}, \citenamefont
  {O’Malley}, \citenamefont {Roushan}, \citenamefont {Vainsencher},
  \citenamefont {Wenner}, \citenamefont {Korotkov}, \citenamefont {Cleland},\
  and\ \citenamefont {Martinis}}]{barends_superconducting_2014}%
  \BibitemOpen
  \bibfield  {author} {\bibinfo {author} {\bibfnamefont {R.}~\bibnamefont
  {Barends}}, \bibinfo {author} {\bibfnamefont {J.}~\bibnamefont {Kelly}},
  \bibinfo {author} {\bibfnamefont {A.}~\bibnamefont {Megrant}}, \bibinfo
  {author} {\bibfnamefont {A.}~\bibnamefont {Veitia}}, \bibinfo {author}
  {\bibfnamefont {D.}~\bibnamefont {Sank}}, \bibinfo {author} {\bibfnamefont
  {E.}~\bibnamefont {Jeffrey}}, \bibinfo {author} {\bibfnamefont {T.~C.}\
  \bibnamefont {White}}, \bibinfo {author} {\bibfnamefont {J.}~\bibnamefont
  {Mutus}}, \bibinfo {author} {\bibfnamefont {A.~G.}\ \bibnamefont {Fowler}},
  \bibinfo {author} {\bibfnamefont {B.}~\bibnamefont {Campbell}}, \bibinfo
  {author} {\bibfnamefont {Y.}~\bibnamefont {Chen}}, \bibinfo {author}
  {\bibfnamefont {Z.}~\bibnamefont {Chen}}, \bibinfo {author} {\bibfnamefont
  {B.}~\bibnamefont {Chiaro}}, \bibinfo {author} {\bibfnamefont
  {A.}~\bibnamefont {Dunsworth}}, \bibinfo {author} {\bibfnamefont
  {C.}~\bibnamefont {Neill}}, \bibinfo {author} {\bibfnamefont
  {P.}~\bibnamefont {O’Malley}}, \bibinfo {author} {\bibfnamefont
  {P.}~\bibnamefont {Roushan}}, \bibinfo {author} {\bibfnamefont
  {A.}~\bibnamefont {Vainsencher}}, \bibinfo {author} {\bibfnamefont
  {J.}~\bibnamefont {Wenner}}, \bibinfo {author} {\bibfnamefont {A.~N.}\
  \bibnamefont {Korotkov}}, \bibinfo {author} {\bibfnamefont {A.~N.}\
  \bibnamefont {Cleland}},\ and\ \bibinfo {author} {\bibfnamefont {J.~M.}\
  \bibnamefont {Martinis}},\ }\bibfield  {title} {\bibinfo {title}
  {Superconducting quantum circuits at the surface code threshold for fault
  tolerance},\ }\href {https://doi.org/10.1038/nature13171} {\bibfield
  {journal} {\bibinfo  {journal} {Nature}\ }\textbf {\bibinfo {volume} {508}},\
  \bibinfo {pages} {500} (\bibinfo {year} {2014})}\BibitemShut {NoStop}%
\bibitem [{\citenamefont {Rol}\ \emph {et~al.}(2019)\citenamefont {Rol},
  \citenamefont {Battistel}, \citenamefont {Malinowski}, \citenamefont
  {Bultink}, \citenamefont {Tarasinski}, \citenamefont {Vollmer}, \citenamefont
  {Haider}, \citenamefont {Muthusubramanian}, \citenamefont {Bruno},
  \citenamefont {Terhal},\ and\ \citenamefont {DiCarlo}}]{rol_fast_2019}%
  \BibitemOpen
  \bibfield  {author} {\bibinfo {author} {\bibfnamefont {M.~A.}\ \bibnamefont
  {Rol}}, \bibinfo {author} {\bibfnamefont {F.}~\bibnamefont {Battistel}},
  \bibinfo {author} {\bibfnamefont {F.~K.}\ \bibnamefont {Malinowski}},
  \bibinfo {author} {\bibfnamefont {C.~C.}\ \bibnamefont {Bultink}}, \bibinfo
  {author} {\bibfnamefont {B.~M.}\ \bibnamefont {Tarasinski}}, \bibinfo
  {author} {\bibfnamefont {R.}~\bibnamefont {Vollmer}}, \bibinfo {author}
  {\bibfnamefont {N.}~\bibnamefont {Haider}}, \bibinfo {author} {\bibfnamefont
  {N.}~\bibnamefont {Muthusubramanian}}, \bibinfo {author} {\bibfnamefont
  {A.}~\bibnamefont {Bruno}}, \bibinfo {author} {\bibfnamefont {B.~M.}\
  \bibnamefont {Terhal}},\ and\ \bibinfo {author} {\bibfnamefont
  {L.}~\bibnamefont {DiCarlo}},\ }\bibfield  {title} {\bibinfo {title} {Fast,
  high-fidelity conditional-phase gate exploiting leakage interference in
  weakly anharmonic superconducting qubits},\ }\href
  {https://doi.org/10.1103/PhysRevLett.123.120502} {\bibfield  {journal}
  {\bibinfo  {journal} {Phys. Rev. Lett.}\ }\textbf {\bibinfo {volume} {123}},\
  \bibinfo {pages} {120502} (\bibinfo {year} {2019})}\BibitemShut {NoStop}%
\bibitem [{\citenamefont {Kjaergaard}\ \emph {et~al.}(2022)\citenamefont
  {Kjaergaard}, \citenamefont {Schwartz}, \citenamefont {Greene}, \citenamefont
  {Samach}, \citenamefont {Bengtsson}, \citenamefont {O'Keeffe}, \citenamefont
  {McNally}, \citenamefont {Braum\"uller}, \citenamefont {Kim}, \citenamefont
  {Krantz}, \citenamefont {Marvian}, \citenamefont {Melville}, \citenamefont
  {Niedzielski}, \citenamefont {Sung}, \citenamefont {Winik}, \citenamefont
  {Yoder}, \citenamefont {Rosenberg}, \citenamefont {Obenland}, \citenamefont
  {Lloyd}, \citenamefont {Orlando}, \citenamefont {Marvian}, \citenamefont
  {Gustavsson},\ and\ \citenamefont {Oliver}}]{PhysRevX.12.011005}%
  \BibitemOpen
  \bibfield  {author} {\bibinfo {author} {\bibfnamefont {M.}~\bibnamefont
  {Kjaergaard}}, \bibinfo {author} {\bibfnamefont {M.~E.}\ \bibnamefont
  {Schwartz}}, \bibinfo {author} {\bibfnamefont {A.}~\bibnamefont {Greene}},
  \bibinfo {author} {\bibfnamefont {G.~O.}\ \bibnamefont {Samach}}, \bibinfo
  {author} {\bibfnamefont {A.}~\bibnamefont {Bengtsson}}, \bibinfo {author}
  {\bibfnamefont {M.}~\bibnamefont {O'Keeffe}}, \bibinfo {author}
  {\bibfnamefont {C.~M.}\ \bibnamefont {McNally}}, \bibinfo {author}
  {\bibfnamefont {J.}~\bibnamefont {Braum\"uller}}, \bibinfo {author}
  {\bibfnamefont {D.~K.}\ \bibnamefont {Kim}}, \bibinfo {author} {\bibfnamefont
  {P.}~\bibnamefont {Krantz}}, \bibinfo {author} {\bibfnamefont
  {M.}~\bibnamefont {Marvian}}, \bibinfo {author} {\bibfnamefont
  {A.}~\bibnamefont {Melville}}, \bibinfo {author} {\bibfnamefont {B.~M.}\
  \bibnamefont {Niedzielski}}, \bibinfo {author} {\bibfnamefont
  {Y.}~\bibnamefont {Sung}}, \bibinfo {author} {\bibfnamefont {R.}~\bibnamefont
  {Winik}}, \bibinfo {author} {\bibfnamefont {J.}~\bibnamefont {Yoder}},
  \bibinfo {author} {\bibfnamefont {D.}~\bibnamefont {Rosenberg}}, \bibinfo
  {author} {\bibfnamefont {K.}~\bibnamefont {Obenland}}, \bibinfo {author}
  {\bibfnamefont {S.}~\bibnamefont {Lloyd}}, \bibinfo {author} {\bibfnamefont
  {T.~P.}\ \bibnamefont {Orlando}}, \bibinfo {author} {\bibfnamefont
  {I.}~\bibnamefont {Marvian}}, \bibinfo {author} {\bibfnamefont
  {S.}~\bibnamefont {Gustavsson}},\ and\ \bibinfo {author} {\bibfnamefont
  {W.~D.}\ \bibnamefont {Oliver}},\ }\bibfield  {title} {\bibinfo {title}
  {Demonstration of density matrix exponentiation using a superconducting
  quantum processor},\ }\href {https://doi.org/10.1103/PhysRevX.12.011005}
  {\bibfield  {journal} {\bibinfo  {journal} {Phys. Rev. X}\ }\textbf {\bibinfo
  {volume} {12}},\ \bibinfo {pages} {011005} (\bibinfo {year}
  {2022})}\BibitemShut {NoStop}%
\bibitem [{\citenamefont {Neg\^{\i}rneac}\ \emph {et~al.}(2021)\citenamefont
  {Neg\^{\i}rneac}, \citenamefont {Ali}, \citenamefont {Muthusubramanian},
  \citenamefont {Battistel}, \citenamefont {Sagastizabal}, \citenamefont
  {Moreira}, \citenamefont {Marques}, \citenamefont {Vlothuizen}, \citenamefont
  {Beekman}, \citenamefont {Zachariadis}, \citenamefont {Haider}, \citenamefont
  {Bruno},\ and\ \citenamefont {DiCarlo}}]{negirneac_high-fidelity_2021}%
  \BibitemOpen
  \bibfield  {author} {\bibinfo {author} {\bibfnamefont {V.}~\bibnamefont
  {Neg\^{\i}rneac}}, \bibinfo {author} {\bibfnamefont {H.}~\bibnamefont {Ali}},
  \bibinfo {author} {\bibfnamefont {N.}~\bibnamefont {Muthusubramanian}},
  \bibinfo {author} {\bibfnamefont {F.}~\bibnamefont {Battistel}}, \bibinfo
  {author} {\bibfnamefont {R.}~\bibnamefont {Sagastizabal}}, \bibinfo {author}
  {\bibfnamefont {M.~S.}\ \bibnamefont {Moreira}}, \bibinfo {author}
  {\bibfnamefont {J.~F.}\ \bibnamefont {Marques}}, \bibinfo {author}
  {\bibfnamefont {W.~J.}\ \bibnamefont {Vlothuizen}}, \bibinfo {author}
  {\bibfnamefont {M.}~\bibnamefont {Beekman}}, \bibinfo {author} {\bibfnamefont
  {C.}~\bibnamefont {Zachariadis}}, \bibinfo {author} {\bibfnamefont
  {N.}~\bibnamefont {Haider}}, \bibinfo {author} {\bibfnamefont
  {A.}~\bibnamefont {Bruno}},\ and\ \bibinfo {author} {\bibfnamefont
  {L.}~\bibnamefont {DiCarlo}},\ }\bibfield  {title} {\bibinfo {title}
  {High-fidelity controlled-$z$ gate with maximal intermediate leakage
  operating at the speed limit in a superconducting quantum processor},\ }\href
  {https://doi.org/10.1103/PhysRevLett.126.220502} {\bibfield  {journal}
  {\bibinfo  {journal} {Phys. Rev. Lett.}\ }\textbf {\bibinfo {volume} {126}},\
  \bibinfo {pages} {220502} (\bibinfo {year} {2021})}\BibitemShut {NoStop}%
\bibitem [{\citenamefont {Collodo}\ \emph {et~al.}(2020)\citenamefont
  {Collodo}, \citenamefont {Herrmann}, \citenamefont {Lacroix}, \citenamefont
  {Andersen}, \citenamefont {Remm}, \citenamefont {Lazar}, \citenamefont
  {Besse}, \citenamefont {Walter}, \citenamefont {Wallraff},\ and\
  \citenamefont {Eichler}}]{Collodo2020}%
  \BibitemOpen
  \bibfield  {author} {\bibinfo {author} {\bibfnamefont {M.~C.}\ \bibnamefont
  {Collodo}}, \bibinfo {author} {\bibfnamefont {J.}~\bibnamefont {Herrmann}},
  \bibinfo {author} {\bibfnamefont {N.}~\bibnamefont {Lacroix}}, \bibinfo
  {author} {\bibfnamefont {C.~K.}\ \bibnamefont {Andersen}}, \bibinfo {author}
  {\bibfnamefont {A.}~\bibnamefont {Remm}}, \bibinfo {author} {\bibfnamefont
  {S.}~\bibnamefont {Lazar}}, \bibinfo {author} {\bibfnamefont {J.-C.}\
  \bibnamefont {Besse}}, \bibinfo {author} {\bibfnamefont {T.}~\bibnamefont
  {Walter}}, \bibinfo {author} {\bibfnamefont {A.}~\bibnamefont {Wallraff}},\
  and\ \bibinfo {author} {\bibfnamefont {C.}~\bibnamefont {Eichler}},\
  }\bibfield  {title} {\bibinfo {title} {Implementation of conditional phase
  gates based on tunable $zz$ interactions},\ }\href
  {https://doi.org/10.1103/PhysRevLett.125.240502} {\bibfield  {journal}
  {\bibinfo  {journal} {Phys. Rev. Lett.}\ }\textbf {\bibinfo {volume} {125}},\
  \bibinfo {pages} {240502} (\bibinfo {year} {2020})}\BibitemShut {NoStop}%
\bibitem [{\citenamefont {Chu}\ and\ \citenamefont
  {Yan}(2021)}]{chu_coupler-assisted_2021}%
  \BibitemOpen
  \bibfield  {author} {\bibinfo {author} {\bibfnamefont {J.}~\bibnamefont
  {Chu}}\ and\ \bibinfo {author} {\bibfnamefont {F.}~\bibnamefont {Yan}},\
  }\bibfield  {title} {\bibinfo {title} {Coupler-{Assisted}
  {Controlled}-{Phase} {Gate} with {Enhanced} {Adiabaticity}},\ }\href
  {https://doi.org/10.1103/PhysRevApplied.16.054020} {\bibfield  {journal}
  {\bibinfo  {journal} {Physical Review Applied}\ }\textbf {\bibinfo {volume}
  {16}},\ \bibinfo {pages} {054020} (\bibinfo {year} {2021})}\BibitemShut
  {NoStop}%
\bibitem [{\citenamefont {Johansson}\ \emph {et~al.}(2013)\citenamefont
  {Johansson}, \citenamefont {Nation},\ and\ \citenamefont
  {Nori}}]{JOHANSSON20131234}%
  \BibitemOpen
  \bibfield  {author} {\bibinfo {author} {\bibfnamefont {J.}~\bibnamefont
  {Johansson}}, \bibinfo {author} {\bibfnamefont {P.}~\bibnamefont {Nation}},\
  and\ \bibinfo {author} {\bibfnamefont {F.}~\bibnamefont {Nori}},\ }\bibfield
  {title} {\bibinfo {title} {Qutip 2: A python framework for the dynamics of
  open quantum systems},\ }\href
  {https://doi.org/https://doi.org/10.1016/j.cpc.2012.11.019} {\bibfield
  {journal} {\bibinfo  {journal} {Computer Physics Communications}\ }\textbf
  {\bibinfo {volume} {184}},\ \bibinfo {pages} {1234} (\bibinfo {year}
  {2013})}\BibitemShut {NoStop}%
\bibitem [{\citenamefont {{Google Quantum AI}}\ and\ \citenamefont
  {Collaborators}(2025)}]{google_2025}%
  \BibitemOpen
  \bibfield  {author} {\bibinfo {author} {\bibnamefont {{Google Quantum AI}}}\
  and\ \bibinfo {author} {\bibnamefont {Collaborators}},\ }\bibfield  {title}
  {\bibinfo {title} {Quantum error correction below the surface code
  threshold},\ }\href {https://www.nature.com/articles/s41586-024-08449-y}
  {\bibfield  {journal} {\bibinfo  {journal} {Nature}\ }\textbf {\bibinfo
  {volume} {638}},\ \bibinfo {pages} {920–926} (\bibinfo {year}
  {2025})}\BibitemShut {NoStop}%
\bibitem [{\citenamefont {Ding}(2025)}]{ding2025data}%
  \BibitemOpen
  \bibfield  {author} {\bibinfo {author} {\bibfnamefont {Q.}~\bibnamefont
  {Ding}},\ }\href@noop {} {\bibinfo {title} {Pulse design of baseband flux
  control for adiabatic controlled-phase gates in superconducting circuits
  [data set]}},\ \bibinfo {howpublished}
  {\url{https://doi.org/10.5281/zenodo.15454126}} (\bibinfo {year} {2025}),\
  \bibinfo {note} {zenodo}\BibitemShut {NoStop}%
\bibitem [{\citenamefont {Zener}(1932)}]{zener1932}%
  \BibitemOpen
  \bibfield  {author} {\bibinfo {author} {\bibfnamefont {C.}~\bibnamefont
  {Zener}},\ }\bibfield  {title} {\bibinfo {title} {{Non-adiabatic crossing of
  energy levels}},\ }\href@noop {} {\bibfield  {journal} {\bibinfo  {journal}
  {Proc. R. Soc. Lond. A 137}\ } (\bibinfo {year} {1932})}\BibitemShut
  {NoStop}%
\bibitem [{\citenamefont {Landau}(1932)}]{landau1932}%
  \BibitemOpen
  \bibfield  {author} {\bibinfo {author} {\bibfnamefont {L.~D.}\ \bibnamefont
  {Landau}},\ }\bibfield  {title} {\bibinfo {title} {{Theory of energy
  transfer. II}},\ }\href@noop {} {\bibfield  {journal} {\bibinfo  {journal}
  {Physikalische Zeitschrift der Sowjetunion}\ } (\bibinfo {year}
  {1932})}\BibitemShut {NoStop}%
\bibitem [{\citenamefont {Rubbmark}\ \emph {et~al.}(1981)\citenamefont
  {Rubbmark}, \citenamefont {Kash}, \citenamefont {Littman},\ and\
  \citenamefont {Kleppner}}]{rubbmark_dynamical_1981}%
  \BibitemOpen
  \bibfield  {author} {\bibinfo {author} {\bibfnamefont {J.~R.}\ \bibnamefont
  {Rubbmark}}, \bibinfo {author} {\bibfnamefont {M.~M.}\ \bibnamefont {Kash}},
  \bibinfo {author} {\bibfnamefont {M.~G.}\ \bibnamefont {Littman}},\ and\
  \bibinfo {author} {\bibfnamefont {D.}~\bibnamefont {Kleppner}},\ }\bibfield
  {title} {\bibinfo {title} {Dynamical effects at avoided level crossings: {A}
  study of the {Landau}-{Zener} effect using {Rydberg} atoms},\ }\href
  {https://doi.org/10.1103/PhysRevA.23.3107} {\bibfield  {journal} {\bibinfo
  {journal} {Physical Review A}\ }\textbf {\bibinfo {volume} {23}},\ \bibinfo
  {pages} {3107} (\bibinfo {year} {1981})}\BibitemShut {NoStop}%
\bibitem [{\citenamefont {Ding}(2023)}]{QD_2023}%
  \BibitemOpen
  \bibfield  {author} {\bibinfo {author} {\bibfnamefont {Q.}~\bibnamefont
  {Ding}},\ }\emph {\bibinfo {title} {Pulse Design for Two-Qubit Gates in
  Superconducting Circuits}},\ \href@noop {} {Master's thesis},\ \bibinfo
  {school} {Massachusetts Institute of Technology} (\bibinfo {year}
  {2023})\BibitemShut {NoStop}%
\bibitem [{\citenamefont {Slepian}\ and\ \citenamefont
  {Pollak}(1961)}]{slepian_prolate_1961}%
  \BibitemOpen
  \bibfield  {author} {\bibinfo {author} {\bibfnamefont {D.}~\bibnamefont
  {Slepian}}\ and\ \bibinfo {author} {\bibfnamefont {H.~O.}\ \bibnamefont
  {Pollak}},\ }\bibfield  {title} {\bibinfo {title} {Prolate spheroidal wave
  functions, fourier analysis and uncertainty — {I}},\ }\href
  {https://doi.org/10.1002/j.1538-7305.1961.tb03976.x} {\bibfield  {journal}
  {\bibinfo  {journal} {The Bell System Technical Journal}\ }\textbf {\bibinfo
  {volume} {40}},\ \bibinfo {pages} {43} (\bibinfo {year} {1961})}\BibitemShut
  {NoStop}%
\bibitem [{\citenamefont {Landau}\ and\ \citenamefont
  {Pollak}(1961)}]{landau_prolate_1961}%
  \BibitemOpen
  \bibfield  {author} {\bibinfo {author} {\bibfnamefont {H.~J.}\ \bibnamefont
  {Landau}}\ and\ \bibinfo {author} {\bibfnamefont {H.~O.}\ \bibnamefont
  {Pollak}},\ }\bibfield  {title} {\bibinfo {title} {Prolate spheroidal wave
  functions, fourier analysis and uncertainty — {II}},\ }\href
  {https://doi.org/10.1002/j.1538-7305.1961.tb03977.x} {\bibfield  {journal}
  {\bibinfo  {journal} {The Bell System Technical Journal}\ }\textbf {\bibinfo
  {volume} {40}},\ \bibinfo {pages} {65} (\bibinfo {year} {1961})}\BibitemShut
  {NoStop}%
\bibitem [{\citenamefont {Landau}\ and\ \citenamefont
  {Pollak}(1962)}]{landau_prolate_1962}%
  \BibitemOpen
  \bibfield  {author} {\bibinfo {author} {\bibfnamefont {H.~J.}\ \bibnamefont
  {Landau}}\ and\ \bibinfo {author} {\bibfnamefont {H.~O.}\ \bibnamefont
  {Pollak}},\ }\bibfield  {title} {\bibinfo {title} {Prolate spheroidal wave
  functions, fourier analysis and uncertainty — {III}: {The} dimension of the
  space of essentially time- and band-limited signals},\ }\href
  {https://doi.org/10.1002/j.1538-7305.1962.tb03279.x} {\bibfield  {journal}
  {\bibinfo  {journal} {The Bell System Technical Journal}\ }\textbf {\bibinfo
  {volume} {41}},\ \bibinfo {pages} {1295} (\bibinfo {year}
  {1962})}\BibitemShut {NoStop}%
\bibitem [{\citenamefont {Slepian}(1964)}]{slepian_prolate_1964}%
  \BibitemOpen
  \bibfield  {author} {\bibinfo {author} {\bibfnamefont {D.}~\bibnamefont
  {Slepian}},\ }\bibfield  {title} {\bibinfo {title} {Prolate spheroidal wave
  functions, {Fourier} analysis and uncertainty — {IV}: {Extensions} to many
  dimensions; generalized prolate spheroidal functions},\ }\href
  {https://doi.org/10.1002/j.1538-7305.1964.tb01037.x} {\bibfield  {journal}
  {\bibinfo  {journal} {The Bell System Technical Journal}\ }\textbf {\bibinfo
  {volume} {43}},\ \bibinfo {pages} {3009} (\bibinfo {year}
  {1964})}\BibitemShut {NoStop}%
\bibitem [{\citenamefont {Slepian}(1978)}]{slepian_prolate_1978}%
  \BibitemOpen
  \bibfield  {author} {\bibinfo {author} {\bibfnamefont {D.}~\bibnamefont
  {Slepian}},\ }\bibfield  {title} {\bibinfo {title} {Prolate spheroidal wave
  functions, fourier analysis, and uncertainty — {V}: the discrete case},\
  }\href {https://doi.org/10.1002/j.1538-7305.1978.tb02104.x} {\bibfield
  {journal} {\bibinfo  {journal} {The Bell System Technical Journal}\ }\textbf
  {\bibinfo {volume} {57}},\ \bibinfo {pages} {1371} (\bibinfo {year}
  {1978})}\BibitemShut {NoStop}%
\bibitem [{\citenamefont {Slepian}(1983)}]{slepian_comments_1983}%
  \BibitemOpen
  \bibfield  {author} {\bibinfo {author} {\bibfnamefont {D.}~\bibnamefont
  {Slepian}},\ }\bibfield  {title} {\bibinfo {title} {Some {Comments} on
  {Fourier} {Analysis}, {Uncertainty} and {Modeling}},\ }\href
  {https://www.jstor.org/stable/2029386} {\bibfield  {journal} {\bibinfo
  {journal} {SIAM Review}\ }\textbf {\bibinfo {volume} {25}},\ \bibinfo {pages}
  {379} (\bibinfo {year} {1983})}\BibitemShut {NoStop}%
\bibitem [{\citenamefont {Dolph}(1946)}]{dolph_current_1946}%
  \BibitemOpen
  \bibfield  {author} {\bibinfo {author} {\bibfnamefont {C.}~\bibnamefont
  {Dolph}},\ }\bibfield  {title} {\bibinfo {title} {A {Current} {Distribution}
  for {Broadside} {Arrays} {Which} {Optimizes} the {Relationship} between
  {Beam} {Width} and {Side}-{Lobe} {Level}},\ }\href
  {https://doi.org/10.1109/JRPROC.1946.225956} {\bibfield  {journal} {\bibinfo
  {journal} {Proceedings of the IRE}\ }\textbf {\bibinfo {volume} {34}},\
  \bibinfo {pages} {335} (\bibinfo {year} {1946})}\BibitemShut {NoStop}%
\bibitem [{\citenamefont {Parks}\ and\ \citenamefont
  {McClellan}(1972{\natexlab{a}})}]{parks_chebyshev_1972}%
  \BibitemOpen
  \bibfield  {author} {\bibinfo {author} {\bibfnamefont {T.}~\bibnamefont
  {Parks}}\ and\ \bibinfo {author} {\bibfnamefont {J.}~\bibnamefont
  {McClellan}},\ }\bibfield  {title} {\bibinfo {title} {Chebyshev
  {Approximation} for {Nonrecursive} {Digital} {Filters} with {Linear}
  {Phase}},\ }\href {https://doi.org/10.1109/TCT.1972.1083419} {\bibfield
  {journal} {\bibinfo  {journal} {IEEE Transactions on Circuit Theory}\
  }\textbf {\bibinfo {volume} {19}},\ \bibinfo {pages} {189} (\bibinfo {year}
  {1972}{\natexlab{a}})}\BibitemShut {NoStop}%
\bibitem [{\citenamefont {Parks}\ and\ \citenamefont
  {McClellan}(1972{\natexlab{b}})}]{parks_program_1972}%
  \BibitemOpen
  \bibfield  {author} {\bibinfo {author} {\bibfnamefont {T.}~\bibnamefont
  {Parks}}\ and\ \bibinfo {author} {\bibfnamefont {J.}~\bibnamefont
  {McClellan}},\ }\bibfield  {title} {\bibinfo {title} {A program for the
  design of linear phase finite impulse response digital filters},\ }\href
  {https://doi.org/10.1109/TAU.1972.1162381} {\bibfield  {journal} {\bibinfo
  {journal} {IEEE Transactions on Audio and Electroacoustics}\ }\textbf
  {\bibinfo {volume} {20}},\ \bibinfo {pages} {195} (\bibinfo {year}
  {1972}{\natexlab{b}})}\BibitemShut {NoStop}%
\bibitem [{\citenamefont {McClellan}\ and\ \citenamefont
  {Parks}(1973)}]{mcclellan_unified_1973}%
  \BibitemOpen
  \bibfield  {author} {\bibinfo {author} {\bibfnamefont {J.}~\bibnamefont
  {McClellan}}\ and\ \bibinfo {author} {\bibfnamefont {T.}~\bibnamefont
  {Parks}},\ }\bibfield  {title} {\bibinfo {title} {A unified approach to the
  design of optimum {FIR} linear-phase digital filters},\ }\href
  {https://doi.org/10.1109/TCT.1973.1083764} {\bibfield  {journal} {\bibinfo
  {journal} {IEEE Transactions on Circuit Theory}\ }\textbf {\bibinfo {volume}
  {20}},\ \bibinfo {pages} {697} (\bibinfo {year} {1973})}\BibitemShut
  {NoStop}%
\bibitem [{\citenamefont {Rabiner}(1972)}]{Rabiner1972}%
  \BibitemOpen
  \bibfield  {author} {\bibinfo {author} {\bibfnamefont {L.}~\bibnamefont
  {Rabiner}},\ }\bibfield  {title} {\bibinfo {title} {Linear program design of
  finite impulse response (fir) digital filters},\ }\href
  {https://doi.org/10.1109/TAU.1972.1162395} {\bibfield  {journal} {\bibinfo
  {journal} {IEEE Transactions on Audio and Electroacoustics}\ }\textbf
  {\bibinfo {volume} {20}},\ \bibinfo {pages} {280} (\bibinfo {year}
  {1972})}\BibitemShut {NoStop}%
\bibitem [{\citenamefont {Rabiner}\ \emph {et~al.}(1975)\citenamefont
  {Rabiner}, \citenamefont {Mcclellan},\ and\ \citenamefont
  {Parks}}]{Rabiner1975}%
  \BibitemOpen
  \bibfield  {author} {\bibinfo {author} {\bibfnamefont {L.~R.}\ \bibnamefont
  {Rabiner}}, \bibinfo {author} {\bibfnamefont {J.~P.}\ \bibnamefont
  {Mcclellan}},\ and\ \bibinfo {author} {\bibfnamefont {T.~W.}\ \bibnamefont
  {Parks}},\ }\bibfield  {title} {\bibinfo {title} {Fir digital filter design
  techniques using weighted chebyshev approximation},\ }\href
  {https://api.semanticscholar.org/CorpusID:12579115} {\bibfield  {journal}
  {\bibinfo  {journal} {Proceedings of the IEEE}\ }\textbf {\bibinfo {volume}
  {63}},\ \bibinfo {pages} {595} (\bibinfo {year} {1975})}\BibitemShut
  {NoStop}%
\bibitem [{\citenamefont {Rabiner}\ and\ \citenamefont
  {Gold}(1975)}]{rabiner_theory_1975}%
  \BibitemOpen
  \bibfield  {author} {\bibinfo {author} {\bibfnamefont {L.~R.}\ \bibnamefont
  {Rabiner}}\ and\ \bibinfo {author} {\bibfnamefont {B.}~\bibnamefont {Gold}},\
  }\href@noop {} {\emph {\bibinfo {title} {Theory and {Application} of
  {Digital} {Signal} {Processing}}}}\ (\bibinfo  {publisher} {Prentice-Hall},\
  \bibinfo {year} {1975})\BibitemShut {NoStop}%
\bibitem [{\citenamefont {Ding}\ \emph {et~al.}(2024)\citenamefont {Ding},
  \citenamefont {Oppenheim}, \citenamefont {Boufounos}, \citenamefont
  {Gustavsson}, \citenamefont {Grover}, \citenamefont {Baran},\ and\
  \citenamefont {Oliver}}]{Ding2024}%
  \BibitemOpen
  \bibfield  {author} {\bibinfo {author} {\bibfnamefont {Q.}~\bibnamefont
  {Ding}}, \bibinfo {author} {\bibfnamefont {A.~V.}\ \bibnamefont {Oppenheim}},
  \bibinfo {author} {\bibfnamefont {P.~T.}\ \bibnamefont {Boufounos}}, \bibinfo
  {author} {\bibfnamefont {S.}~\bibnamefont {Gustavsson}}, \bibinfo {author}
  {\bibfnamefont {J.~A.}\ \bibnamefont {Grover}}, \bibinfo {author}
  {\bibfnamefont {T.~A.}\ \bibnamefont {Baran}},\ and\ \bibinfo {author}
  {\bibfnamefont {W.~D.}\ \bibnamefont {Oliver}},\ }\bibfield  {title}
  {\bibinfo {title} {Application of weighted chebyshev approximation in pulse
  design for quantum gates},\ }in\ \href
  {https://doi.org/10.1109/SiPS62058.2024.00024} {\emph {\bibinfo {booktitle}
  {2024 IEEE Workshop on Signal Processing Systems (SiPS)}}}\ (\bibinfo {year}
  {2024})\ pp.\ \bibinfo {pages} {89--94}\BibitemShut {NoStop}%
\bibitem [{\citenamefont {Wood}\ and\ \citenamefont
  {Gambetta}(2018)}]{Christopher2018}%
  \BibitemOpen
  \bibfield  {author} {\bibinfo {author} {\bibfnamefont {C.~J.}\ \bibnamefont
  {Wood}}\ and\ \bibinfo {author} {\bibfnamefont {J.~M.}\ \bibnamefont
  {Gambetta}},\ }\bibfield  {title} {\bibinfo {title} {Quantification and
  characterization of leakage errors},\ }\href
  {https://doi.org/10.1103/PhysRevA.97.032306} {\bibfield  {journal} {\bibinfo
  {journal} {Phys. Rev. A}\ }\textbf {\bibinfo {volume} {97}},\ \bibinfo
  {pages} {032306} (\bibinfo {year} {2018})}\BibitemShut {NoStop}%
\bibitem [{\citenamefont {Nielsen}(2002)}]{Nielsen2002}%
  \BibitemOpen
  \bibfield  {author} {\bibinfo {author} {\bibfnamefont {M.~A.}\ \bibnamefont
  {Nielsen}},\ }\bibfield  {title} {\bibinfo {title} {A simple formula for the
  average gate fidelity of a quantum dynamical operation},\ }\href
  {https://doi.org/https://doi.org/10.1016/S0375-9601(02)01272-0} {\bibfield
  {journal} {\bibinfo  {journal} {Physics Letters A}\ }\textbf {\bibinfo
  {volume} {303}},\ \bibinfo {pages} {249} (\bibinfo {year}
  {2002})}\BibitemShut {NoStop}%
\end{thebibliography}%
\end{document}